\documentclass[journal=jctcce,manuscript=article,layout=twocolumn]{achemso}
\usepackage{url}
\usepackage{graphicx}
\usepackage{color}
\usepackage{amsmath}
\usepackage{amsfonts}
\usepackage{amssymb}
\usepackage{rotating}
\usepackage{longtable}

\title{Electronic Spectra of Ytterbium Fluoride from Relativistic Electronic Structure Calculations}

\author{Johann V. Pototschnig}
\email{j.v.pototschnig@vu.nl}
\affiliation{Amsterdam Center for Multiscale Modeling, Department of Theoretical 
Chemistry, Faculty of Sciences, VU University Amsterdam, De Boelelaan 1083, 
NL-1081 HV Amsterdam, The Netherlands}
\author{Kenneth G. Dyall}
\email{diracsolutions@gmail.com}
\affiliation{Dirac Solutions, 10527 NW Lost Park Drive,
Portland, OR 97229, U.S.A.}
\author{Lucas Visscher}
\email{visscher@chem.vu.nl}
\affiliation{Amsterdam Center for Multiscale Modeling, Department of Theoretical 
Chemistry, Faculty of Sciences, VU University Amsterdam, De Boelelaan 1083, 
NL-1081 HV Amsterdam, The Netherlands}
\author{Andr\'e Severo Pereira Gomes}
\email{andre.gomes@univ-lille.fr}
\affiliation{Universit\'e de Lille, CNRS, 
UMR 8523 -- PhLAM -- Physique des Lasers, 
Atomes et Mol\'ecules, F-59000 Lille, France}

\keywords{Gaussian basis sets -- relativistic basis sets -- lanthanide elements
 -- double zeta -- triple zeta -- quadruple zeta -- correlating functions -- 
 Ytterbium Fluoride -- spectroscopic constants -- electronic spectra -- Fock-space
 coupled cluster}

\begin{document}

\def\wn{~cm$^{-1}$}

\voffset 1 true cm

\begin{abstract}
We report an investigation of the low-lying excited states of the YbF molecule--a candidate molecule for experimental measurements of the electron electric dipole moment--with 2-component based multi-reference configuration interaction (MRCI), equation of motion coupled cluster (EOM-CCSD) and the extrapolated intermediate Hamiltonian Fock-space coupled cluster (XIHFS-CCSD). Specifically, we address the question of the nature of these low-lying states in terms of configurations containing filled or partially-filled Yb $4f$ shells. We show that while it does not appear possible to carry out calculations with both kinds of configurations contained in the same active space, reliable information can be extracted from different sectors of Fock space--that is, by performing electron attachment and detachment IHFS-CCSD and EOM-CCSD calculation on the closed-shell YbF$^+$ and YbF$^-$ species, respectively. From these we observe $\Omega = 1/2, 3/2$ states that arise from the $4f^{13}\sigma_{6s}^2$,  $4f^{14}5d$/$6p$, and $4f^{13}5d\sigma_{6s}$ configurations appear in the same energy range around the ground-state equilibrium geometry and they are therefore able to interact. As these states are generated from different sectors of Fock space, they are almost orthogonal and provide complementary descriptions of parts of the excited state manifold. To obtain a comprehensive picture, we introduce a simple adiabatization model to extract energies of interacting $\Omega = 1/2, 3/2$ states that can be compared to experimental observations.

\end{abstract}

\section{Introduction}

In a previous paper\cite{Gomes2010}, we introduced all-electron relativistic basis sets for the lanthanides (La -- Lu) and discussed their performance for the determination of spectroscopic constants for the ground state of ytterbium fluoride (YbF), an open-shell molecule with a $^2\Sigma^+$ ground state. This molecule has received a fair amount of experimental  and theoretical attention because of its potential application in the observation of parity-violating interactions \cite{Sapirstein2002,Berger2004,Sunaga2016} via determination of the electric dipole moment of the electron (eEDM)---see, for instance refs.
\citenum{Parpia1998,Nayak2006,Titov2005,Hudson2002,Steimle2007, Fukuda2016,Sunaga2016,Fitch2020} 
and references therein). There is also some interest in the Yb atom, cation and dimer in connection to ultracold physics.\cite{Takasu2003, Fukuhara2009,tecmer-yb2-2019} 
An example\cite{Tsigutkin2009} is the parity violation observed in the Yb atom.

A noteworthy finding in our previous work was the sensitivity of coupled cluster calculations to the basis set in use and, indirectly, to the amount of electron correlation recovered in the calculations. 
We observed a spike in the values of the $T_1$ diagnostic around the ground state equilibrium geometry, so it appears that the perturbative treatment of triple excitations in the CCSD(T) calculations breaks down in this region of the potential energy curves unless there is enough flexibility in the correlation treatment. 
The same was recently observed by~\citeauthor{Pasteka2016}~\cite{Pasteka2016} for the nuclear quadruple coupling constant.
This suggests the existence of a low lying perturbing state, which we want to investigate further in the current work. 

Experimental\cite{NistDiatomic,Dunfield1995} and previous theoretical \cite{DOLG1992a,Liu1998,Su2009} investigations suggest that in the ground state the unpaired electron is located in a $\sigma_s$ orbital with dominant contributions from the 6s orbital of Yb, corresponding to a Yb($4f^{14}\sigma_{6s}$)F configuration. This $^2\Sigma_{1/2}$ state ground state was studied in greater detail by combining microwave and optical spectroscopy for the odd $^{171}$Yb isotope.\cite{Glassman2014}

Experimentally\cite{Dunfield1995}, the lowest excited state observed is assigned as $^2\Pi_{1/2}$, with an energy of 18106~\wn{}, while the $^2\Pi_{3/2}$ component is found at 19471~\wn{}, yielding a spin-orbit splitting of 1365~\wn{} of this spin-orbit split A~$^2\Pi$ state. The lower component will be denoted $3_{1/2}$ in the current work. Experiments indicate a perturbation of its vibrational levels~\cite{Dunfield1995, Sauer1999, Lim2017, Lim2018}, which was attributed to the presence of a perturbing state (denoted by $4_{1/2}$ here) found at 18705 \wn{}\cite{Dunfield1995}. This perturbing $4_{1/2}$ state is sometimes referred to as [18.6]$_{1/2}$ by experimentalists\cite{Dunfield1995, Sauer1999} (energy in \wn{} divided by $10^3$ in the square brackets, and $\Omega$-value as subscript). The mixing of these two $\Omega=1/2$ states gives rise to states designated as [557] and [561] (the values in square brackets referring to transition energies in Thz from the vibronic ground state) with transition energies of 18574 and 18699 \wn, respectively\cite{Lim2017,Lim2018}. These two states are of importance for laser cooling schemes that have been investigated\cite{Smallman2014} and tested\cite{Lim2018} with the purpose of realizing high-accuracy measurements of YbF at very low temperatures. 
Besides these first excited states, Smallman\cite{Smallman2014} investigated also two not yet fully characterized mixed states, [574]($\approx$19150~\wn{}) and [578]($\approx$19280~\wn{}) at higher energies. 
These can be compared with the $\Omega=3/2$ state at 19471~\wn{} found earlier by Dunfield\cite{Dunfield1995}, which will be denoted $2_{3/2}$ in the current work. Uttam \textit{et al.} furthermore measured additional unidentified higher bands at about 23035, 23256 and 26015 \wn{}.\cite{Uttam1995,Lee1977} 

Theoretically, excited states arising from the Yb($4f^{14}6p$)F and Yb($4f^{14}5d$)F configurations were considered by~ \citeauthor{Nayak2006}\cite{Nayak2006} with RAS-CI based on 4-component spinors, yielding the $A^2\Pi_{1/2}$ ($3_{1/2}$),  $A^2\Pi_{3/2}$ ($2_{3/2}$), and a $^2\Sigma_{1/2}$ state.
Earlier multireference CI calculations by~\citeauthor{DOLG1992a}\cite{DOLG1992a} furthermore indicate the possibility of low-lying $\Omega = 1/2$ states arising from the Yb($4f^{13}$[$F_{7/2}^\circ$]$\sigma_{6s}^2$)F or Yb($4f^{13}$[$F_{7/2}^\circ$]$5d\sigma_{6s}$)F configurations, to be lying below or close to the Yb($4f^{14}6p$)F states. 
This was also found in the DFT calculations of~\citeauthor{Liu1998}\cite{Liu1998} who place excited states arising from the Yb($4f^{13}$[$F_{7/2}^\circ$]$\sigma_{6s}^2$)F configuration in the range from 9000 to 15000 \wn{} relative to the Yb($4f^{14}\sigma_{6s}$)F ground state. These findings make it of interest to explicitly consider the configuration interaction between the $f^{13}$ and $f^{14}$ configurations in the Yb atom\cite{Dzuba2010}.

The vibronic states are additionally split due to hyperfine interactions. In atomic experiments they were measured for the ground and excited states\cite{Dickinson2001, Steimle2007, Steimle2008, Steimle2012, Glassman2014, Wang2019} also using Zeeman spectroscopy.\cite{Ma2009} 
The hyperfine interaction of the atom\cite{Porsev1999, Mani2011,Nayak2006b} and molecule\cite{Naik2014} were studied theoretically, and should have similar uncertainties to the contribution of the eEDM to the spectrum due to the similarity of the  matrix elements. 
Recently, uncertainties of the hyperfine constants arising in relativistic coupled cluster computations have been studied.\cite{Haase2020}

It is clear from the above that a proper description of the Yb atom and the YbF molecule requires an accurate treatment of both spin-orbit coupling and electron correlation, for ground as well as excited states. A popular approach is the so-called two-step approach to spin-orbit coupling (SOC), in which electron correlation methods based on non-relativistic or scalar relativistic Hamiltonians are used to obtain excited state energies, that are in turn used to dress a spin-orbit configuration interaction (SOCI) matrix. This approach can yield quite accurate spin-orbit coupled states, but results are particularly sensitive to the the number of spin-free states serving as a basis for the SOCI step.\cite{Vallet2000, Marian2001, Danilo2010, Farhat2015,Chmaisani2018,Chmaisani2019,Kervazo2019} An alternative is to include SOC already at the mean-field level, and use fully SO-coupled molecular spinors to construct the correlated wave functions.~\cite{relativity:liu2020}. 
This can be done with four-component Hamiltonians, as done for the ground~\cite{Gomes2010} and excited states~\cite{Nayak2006,Nayak2006b,Nayak2009,Nayak2011} of YbF, or with more computationally efficient two-component Hamiltonians based on the eXact 2-Component (X2C) approach~\cite{jensen:rehe2005,x2c:kutzelnigg2005,x2c:atom2mol:liu2006,x2c:atom2mol:peng2007,Ilias2007,Sikkema2009,x2c:liu2009,Konecny2016}, 
in which a transformation to decouple the positive and negative energy states of the Dirac Hamiltonian can be carried out in matrix form, yielding  the same  positive energy spectrum as the original 4-component Hamiltonian. More details can be found in the recent review by~\citeauthor{relativity:liu2020}\cite{relativity:liu2020}. Among the different X2C flavors, we can distinguish two main strategies for the decoupling, which is performed based on : (i) the one-electron Dirac Hamiltonian prior to the mean-field step~\cite{Ilias2007,Konecny2016}, and for which two-electron spin-orbit contributions due to the untransformed two-electron potential are included via atomic mean-field contributions calculated with the \textsc{AMFI} code~\cite{HE1996365,Marian2001,prog:amfi} (X2C-AMFI); (ii) after a converged 4-component mean-field calculation on atoms~\cite{x2c:atom2mol:liu2006,x2c:atom2mol:peng2007,x2c:liu2009} or molecules~\cite{Sikkema2009} (${^2}$DC$^{M}$). Recent benchmarks show that ${^2}$DC$^{M}$ calculations closely reproduce equivalent 4-component ones for valence~\cite{Shee2018} or core~\cite{halbert2021} states.

Moreover, the aforementioned calculations for the excited states of YbF have mostly employed multireference CI (MRCI) approaches. While these can provide great flexibility in capturing static correlation, it remains the case that dynamical correlation is better accounted for with coupled cluster approaches. Among the coupled cluster singles and doubles (CCSD) approaches for excited states, we have the equation of motion (EOM-CCSD) method as well as Fock-space (IHFS-CCSD) methods~\cite{Bartlett2007}, of which the single electron attachment, detachment, and singly excited states variants are the most commonly used. The two approaches have been found to yield very accurate results in general and in particular for calculations with relativistic Hamiltonians as discussed elsewhere (see ref.~\citenum{Shee2018} and references therein).

The first goal of this work is therefore to go beyond the investigations performed to date in the literature, and apply the relativistic EOM-CCSD and IHFS-CCSD approaches to describe the low-lying excited states of YbF. For such states, where the most important excited state configurations appear have a single open-shell character ($4f^{14}\sigma_{6s}$, $4f^{13}$[$F_{7/2}^\circ$]$\sigma_{6s}^2$, $4f^{14}5d$, $4f^{13}$[$F_{7/2}^\circ$]$5d\sigma_{6s}$, $4f^{14}6p$), these coupled cluster approaches are in principle applicable, provided one starts from closed shell configurations such as Yb($4f^{14}\sigma_{6s}^2$)F$^-$ or Yb($4f^{14})$F$^+$. Additionally, we assess the performance of relativistic MRCI with respect to the coupled cluster methods. Our second goal is to confirm whether any low-lying state is close enough to the ground state to perturb the latter, and explain the anomalous behavior observed in the open-shell ground-state calculations in the literature.


\section{Computational Details}

All relativistic electronic structure calculations were performed with a development version of the \textsc{Dirac} program suite\cite{DIRAC} (revision \texttt{6e10c5d3}), employing for Yb the valence double-zeta (\texttt{24s19p13d9f2g}),  triple-zeta (\texttt{30s24p18d14f4g2h}) and quadruple-zeta (\texttt{35s30p19d16f6g4h2i}) basis sets from the previous work~\cite{Gomes2010}, 
along with the matching augmented correlation-consistent (aug-cc-pVnZ, $n=2,3,4$) basis sets of Dunning \cite{Dunning1989} for F. All basis sets were kept uncontracted, with the small component basis generated by restricted kinetic balance. In addition to these individual basis sets, we have used the calculations with triple- and quadruple-zeta sets to construct extrapolations to the complete basis set limit ($E_\infty$) for the underlying potential energy curves, using the relation \cite{HelgakerBook} 
\begin{equation}
E_\infty (\mathbf{R}) = \frac{ 4^3 E_4(\mathbf{R})   - 3^3 E_3(\mathbf{R})}{4^3 - 3^3}
\end{equation}
where the subscripts denote the cardinal numbers for the basis sets and
$E_n(\mathbf{R})$ the energy for a given geometry and electronic structure 
method for a basis of cardinal number $n (=2,3,4)$.

In the coupled cluster computations the ${^2}$DC$^{M}$ Hamiltonian\cite{Ilias2007, Sikkema2009} was applied, all two-electron integrals over small component (S) basis sets (i.e.\ the so-called (SS$|$SS)-type integrals) appearing in the SCF step have been replaced by a simple correction\cite{Visscher1997}. In order to account for spin-orbit coupling and other relativistic effects the X2C-AMFI Hamiltonian was employed for the Kramers-restricted configuration interaction (KRCI) method.  

Spectroscopic constants ($r_e$, $D_e$, $\omega_e$ and $\omega_e x_e$) were determined from a Morse potential fit in the vicinity of the potential energy minima. 
The potential energy curves were determined for bond lengths between 1.6 \AA\  and  2.3 \AA\ spaced by 0.02 \AA\ and additional points with larger spacing up to 3.5 \AA\ .
In the calculation of $D_e$ the asymptotic dissociation limit is calculated from the energies of the isolated neutral atoms, F in the $^2P_{3/2}$ state and Yb in the $^1S_0$ state.

The dataset associated with this manuscript (outputs from calculations, codes to extract and process information from these, and code to obtain the spectroscopic constants) is provided in ref.~\citenum{pototschnig:ybf:dataset}.

\subsection{Kramers-restricted configuration interaction}

For YbF we first consider Kramers-restricted configuration interaction (KRCI) based on an average-of-configuration Hartree-Fock approach (AOC-SCF)\cite{Thyssen2001}.
This method was employed in order to treat the open shells, where one or two valence electrons were distributed over the s- and d-orbitals and the f-shell was either completely filled or contained one hole, depending on the states of interest. 
The AOC-SCF reference wave function in the KRCI computation is occupied according to a definition given by a generalized active space (GAS)\cite{Fleig2006}. 
In this approach the Hamiltonian is computed for all allowed configurations and then diagonalized.
The GAS space was defined by a f-shell which was completely filled or contained one hole and one or two electrons distributed over 29 orbitals.


\subsection{Equation-of-motion coupled cluster}

The first approach we use to describe the dynamical correlation that is largely missing in KRCI is EOM-CCSD, which can give access to electronic states of different kinds, depending on the single determinant wave function that is chosen as the starting point. In it, the CCSD amplitudes are determined for the chosen ground state in the first step, and subsequently the similarity transformed Hamiltonian is constructed using these amplitudes and the desired states are generated by an operator that either removes or adds an electron. 



The first set of states was obtained by electron attachment on Yb($4f^{14})$F$^+$ ion, where the HOMO ($\sigma_{6s,1/2}$) of YbF was initially empty.
This computation on the (0h,1p) sector of Fock space yielded states with $4f^{14}$ and a valence electron in the $\sigma_{6s}$, d or p orbital. 
This means that, in the process of obtaining the potentials for the ground and excited states of YbF, we immediately obtain energies of CCSD quality for YbF$^+$, and therefore vertical ionization potentials (IP) at each geometry. 

Another set of states was obtained by ionizing the Yb($4f^{14}\sigma_{6s}^2$)F$^-$ anion, where the HOMO ($\sigma_{6s,1/2}$) of YbF was initially doubly occupied.
States arising from the Yb($4f^{14}\sigma_{6s}$)F, Yb($4f^{13}$[$F_{7/2}^\circ$]$\sigma_{6s}^2$)F, and Yb($4f^{13}$[$F_{5/2}^\circ$]$\sigma_{6s}^2$)F configurations were obtained by considering the (1h,0p) sector of Fock space.
This means that, in the process of obtaining the potentials for the ground and excited states of YbF, we immediately obtain energies of CCSD quality for YbF$^-$, and therefore vertical electron affinities (EA) at each geometry. We note that states arising from the (2h,1p) and (1h,2p) manifolds are also accessible from EOM-IP and EOM-EA calculations, though the energy of electronic states determined by such configurations will be less accurate than states dominated by single detachment or attachment configurations.

The EOM-CCSD electronic states are obtained by an iterative diagonalization (Davidson) procedure in which only the energies of a certain number of the lowest states are determined. For the IP-EOM-CCSD we obtained 16 $\Omega = 1/2$, 8 $\Omega = 3/2$, 6 $\Omega = 5/2$ and 2 $\Omega = 7/2$ states, whereas for EA-EOM-CCSD we obtained 8 $\Omega = 1/2$, 6 $\Omega = 3/2$, 4 $\Omega = 5/2$ and 2 $\Omega = 7/2$ states.

As transition moments are not yet available for the EOM-CCSD implementation in \textsc{Dirac}, we have only obtained the potential energy curves. These are nevertheless useful since, by not requiring the definition of model spaces or the use of an extrapolation procedure, they serve as a cross validation of the IHFS-CCSD calculations below. 

\subsection{Fock-space coupled cluster}

%

Fock-space coupled cluster\cite{Visscher2001} (FS-CCSD) is our second approach to include dynamical correlation in the electronically excited states. Here it was employed in a similar fashion to EOM-CCSD, starting from YbF$^+$ or YbF$^-$ and proceeding to the (0h,1p) and (1h,0p) sectors of Fock space, respectively. For FS-CCSD a model space is defined by selecting a number of occupied and virtual orbitals and how many electrons are added and removed. The matrix for this subspace is constructed and subsequently diagonalized, thus yielding all states within the chosen model space, in this case states arising from single electron attachment (EA) or single electron detachment (IP). This method requires solving first for the underlying sectors, starting with (0h,0p), which corresponds to CCSD. Due to computational constraints, we have truncated the virtual space so that 117, 230 and 296 orbitals were used in the double-, triple- and quadruple-zeta CCSD calculations, respectively. 

The separation into a model and external space leads to the appearance of the so-called intruder states, a well-known difficulty with Fock-space coupled cluster and other effective Hamiltonian approaches, that can be dealt with in many cases by the intermediate Hamiltonian (IH) Fock-space coupled cluster (IHFS-CCSD) method\cite{Landau2000,Landau2001a}. 

The IH approach was employed to compute Yb($4f^{14} \{\sigma_{6s},6p,5d\ldots\}^1$)F states starting from YbF$^+$. 
The active $P$ space in such calculations contained about 50 spinors varying slightly with bond distance and basis set. 
Of these 26 spinors are always present in the model ($P_m$) space, whereas the remaining active spinors are placed in the intermediate ($P_I$) space. 
Due to using the $(0h,1p)$ sector for the cation, states arising from configurations where the Yb $4f$ shell is partially filled (such as Yb($4f^{13}$[$F_{7/2}^\circ$]$\sigma_{6s}^2$)F, Yb($4f^{13}$[$F_{7/2}^\circ$]$5d\sigma_{6s}$)F or Yb($4f^{13}$[$F_{7/2}^\circ$]$\sigma_{6s}6p$)F) are not 
accessible in this  calculation.

The approach outlined above was, however, not enough to avoid divergence in the computation of the (1h,0p) sector using the anion as a reference. Therefore, the extrapolated intermediate Hamiltonian (XIH) Fock-space coupled-cluster approach\cite{Eliav2005} (XIHFS-CCSD) was applied using the same shifts as in reference \citenum{Eliav2005}. 
Values of 0.1 and 0.2 Hartree were selected if one of the holes is not in the model space. These shifts were doubled for two holes outside the model space. Using the determined energies an extrapolation to the system without shifts was performed. 
The model ($P_m$) space in these computations contained 22 spinors, the intermediate ($P_I$) space about 24 spinors depending on the bond distance and basis set. 
Since we start out from the anion and only allow holes, only Yb($4f^{13}$[$F_{7/2}^\circ$]$\sigma_{6s}^2$)F, Yb($4f^{14}\sigma_{6s}$)F, and  Yb($4f^{14}\sigma_{6s}^2$)F($2p^{5}$) configurations are accessible in this computation. 

Combining the two sectors allows us to get different excited states of YbF, although there are limitations. Firstly, the interactions between configurations with open f-shell and the ones with an electron in the p- or d-shell are not included, since they will be obtained for different sectors of Fock space. This interaction will nevertheless be treated with a simple adiabatization approach, described in sections~\ref{compdet:adiabatization} and~\ref{sec:adiabatization}. Secondly, configurations such as  Yb($4f^{13}$[$F_{7/2}^\circ$]$\sigma_{6s}6p$)F or Yb($4f^{13}$[$F_{7/2}^\circ$]$5d\sigma_{6s}$)F are not included in the current treatment. This limitations is not as significant because these states have higher energies than the ones we are interested in. Both of these problems could be dealt with by using the (1h,1p) sector, but this goes beyond the current work as convergence is very unstable for this sector and it requires the use of an open-shell reference. 

\subsection{Adiabatization of electronic states\label{compdet:adiabatization}}

As we separated the computations of states with 
$4f^{13}$ and $4f^{14}$ character, these states cannot interact with each other, and states with the same $\Omega$ value cross although they should have an avoided crossing. In order to correct this deficiency we considered a simple adiabatization model, in which we set up and diagonalize the following matrix for each $\Omega$ value: 
\begin{align}
M=\begin{pmatrix}
E\left(f^{13}\right) & \mathbf{C} \\
\mathbf{C} & E\left(f^{14}\right) \\
\end{pmatrix}
\label{eq:adiabatization}
\end{align}
where $\mathbf{C}$ is a matrix where every entry is a coupling constant (whose value is kept constant for all states and geometries considered; we have investigated values of 0.01, 0.001 and 0.0001 a.u.), $E$ are matrices with the eigenvalues of the different electronic states on the diagonal. The potential curves were computed for different coupling constants and the results are shown in section~\ref{sec:adiabatization}. We note that since the ground state energy, associated with the Yb($4f^{14}\sigma_{6s}$)F configuration, appears in both coupled cluster approaches for the coupled cluster methods, we have only considered one such energy. As we shall see in the discussion, this is  valid in the region between 1.8~\AA~and 2.5~\AA, since for these distances the ground-state energies from IP and EA calculations are nearly identical.   


\section{Results and Discussion}

We start our discussion with the electronic transitions of the atomic Yb$^+$ cation, before moving on to the YbF molecule. This is because because the atom's  electronic structure is similar to the Yb in YbF since, due to the large electronegativity of fluorine, one electron is almost completely removed from the Yb atom.

Subsequently, the potential energy curves for Kramers-restricted configuration interaction are presented. A discussion of the coupled cluster approaches follows, with a focus on the comparison of the coupled cluster results for the Fock space and equation-of-motion approach. 
This section is followed by a presentation of the spectroscopic data for the ground and excited states. 
In the last part we take a closer look at the mixing of states at around 18000~\wn{} and apply the adiabatization procedure. 

\subsection{Ytterbium cation}
\label{sec:Yb_cation}

As discussed in the introduction, states from both the $4f^{13}$ and the $4f^{14}$ configurations are of importance. This is difficult to realize in a balanced manner when using one set of orbitals to describe all states. Any change in the occupation of the $4f$-shell will alter the screening of the $5s$ and $5p$ orbitals of Yb, resulting in differences between orbital sets optimized for a $4f^{13}$ or a $4f^{14}$ configuration. Additionally, the $4f$ orbitals are very compact and since they are the first $f$-shell there are no orthogonality conditions limiting the radial expansion or contraction of the orbital. Depictions of the orbitals for both configurations can be found in the supporting information. 

These observations help to understand why it turned out to be very difficult to treat both sets of states in the same calculation, which we attempted to do from AOC-SCF on the Yb$^+$. We started out by performing AOC-SCF computations on the atom, based on the $4f$ closed shell configuration. While we obtained the correct ground state configuration, the $^2$F$^\circ_{7/2}$ has an energy of about 46000~\wn{} (over two times higher than the experimental value), and the wrong order for the hole states is observed. If the wave function is optimized for a $4f^{13}$ configuration, one obtains the  $^2$F$^\circ_{7/2}$ as the lowest state and finds the true ground state more than 20000 \wn{} higher. Because of these difficulties, the KRCI calculations discussed below all follow the strategy of different orbital sets that is also employed in the subsequent coupled cluster calculations.

Table~\ref{tab:Yb_cation_KRCI} contains KRCI values of electronic transitions for the cation. The transition energies show deviations of about 10~\% and the spin-orbit splitting is underestimated for states with a $4f^{14}$ configuration.  The transition dipole moment (TDM) of the $^2$P$^\circ_{1/2}$ state is underestimated by about 13~\%, the one for the $^2$P$^\circ_{3/2}$ state overestimated by about  47~\%.
The second set of states with a hole in the f-shell and different distributions of the 2 valence electrons are given in the lower part of table~\ref{tab:Yb_cation_KRCI}, 
the energies are relative to the $^2$F$^\circ_{7/2}$ state. 
In this case the two valence electrons are distributed over the $s$- and $d$-shell. The lowest state with a $4f^{13}$[$F_{7/2}^\circ$]$6s6p$ configuration has a transition energy of 47912.31~\wn{} and was not included in the current treatment.

\begin{table*}[hbtp]
\caption{Kramers-restricted configuration interaction transition energies (in \wn{}), transition dipole moments (TDM), and line strength (S) for the Yb$^+$ cation, the later two are in atomic units($e^2 a_0^2$).  Reference values and notation have been taken from the NIST database\cite{NistDiatomic}. For the $4f^{13}$ configurations, energies relative to the $^2$F$^\circ_{7/2}$ state are also given. 2z,3z, 4z, and extr. indicate double, triple, quadruple zeta and extrapolated results, respectively.
}
\label{tab:Yb_cation_KRCI}
\begin{tabular}{ c | l | r | r | r | r | r | r | r | r | r}
&&\multicolumn{2}{c |}{ NIST\cite{NistDiatomic} }&\multicolumn{2}{c |}{ 2z }&\multicolumn{2}{c |}{3z }&\multicolumn{2}{c |}{4z} & extr.
\\
\hline
state&conf&E&S&E&TDM&E&TDM&E&TDM & E
\\
\hline
$^2$S$_{1/2}$       & $4f^{14}6s$ &     0 &      &     0 &      &     0 &     &     0  & 	 &  0
\\
$^2$D$_{3/2}$       & $4f^{14}5d$ & 22961 &      & 23322 & 0.0  & 22802 & 0.0  & 23606 & 0.0 & 24192
\\
$^2$D$_{5/2}$       & $4f^{14}5d$ & 24333 &      & 23882 & 0.0 & 23321 & 0.0  & 24117 & 0.0  & 24698
\\
$^2$P$^\circ_{1/2}$ & $4f^{14}6p$ & 27062 & 6.1  & 25210 & 3.5  & 24533 & 3.8  & 25331 & 3.6 & 25914
\\
$^2$P$^\circ_{3/2}$ & $4f^{14}6p$ & 30392 & 11.4 & 28104 & 16.9 & 27385 & 18.9 & 28153 & 17.4 & 28712
\\
\hline
state&conf&E&$\Delta$ E &$\Delta$E& &$\Delta$E& &$\Delta$E& & $\Delta$E
\\
\hline
$^2$F$^\circ_{7/2}$ & $4f^{13}6s^2$              & 21419 &    0 &     0  &&     0 &&     0 && 0\\
$^3\left[3/2\right]^\circ_{5/2}$ & $4f^{13}5d 6s$  & 26759 &  5340 &  4260 &&  5538 &&  4618 && 3946 \\
$^3\left[3/2\right]^\circ_{3/2}$ & $4f^{13}5d 6s$  & 28758 &  7339 &  6387 &&  7822 &&  7123 && 6613 \\
$^3\left[11/2\right]^\circ_{9/2} $& $4f^{13}5d 6s$ & 30224 &  8806 &  8214 &&  9325 &&  8314 && 7576 \\
$^3\left[11/2\right]^\circ_{11/2}$& $4f^{13}5d 6s$ & 30563 &  9144 &  8320 &&  9431 &&  8447 && 7729 \\
\end{tabular}
\end{table*}

The excited states of the Yb atom for the $4f^{14}$ configuration have already been investigated by relativistic Fock-space coupled cluster\cite{Eliav1995, Nayak2006b, Sur2007, Mani2011} as well as for the cation\cite{Eliav1995}, including the transition moment of magnetic transitions\cite{Sur2007}. With our current calculations we can go beyond these studies and investigate the f$^{13}$ configurations as well. Before discussing our IHFS-CCSD calculations for Yb$^+$, we focus on the EOM-CCSD excitation energies, shown in table~\ref{tab:Yb_cation_EOM}. 
The EOM-IP-CCSD energies of $4f^{13}$ states obtained from the extrapolation to the complete basis set limit underestimate the experimental transition energies by around 3000~\wn{}, whereas the values for $4f^{14}$ states, obtained with EOM-EA-CCSD are within 1000~\wn{} of the experimental values, which yields a quantitative improvement over the KRCI ones for both configurations, even though qualitatively the two methods provide a similar picture. From that and the preceeding discussion, we attribute the relatively lower accuracy for the $4f^{13}$ to arise from the incomplete account of the relaxation of the wave function upon the creation of the hole in the f shell. Beyond the states presented in table~\ref{tab:Yb_cation_EOM}, which are dominated by single electron attachment and detachment, we are able to access states with significant (1h,2p) and (2h,1p) character with EOM-CCSD. These states, available in the supplementary information, are about 10000~\wn{} higher in energy than the experimental ones. 

\begin{table*}[hbtp]
\caption{Transition energies (in \wn{}) for the Yb$^+$ cation,  obtained for different basis set with EOM-IP-CCSD ($4f^{13}$) and EOM-EA-CCSD ($4f^{14}$), except for the ground state, for which both methods yield the same configuration and total energy. 2z,3z, 4z, and extr. indicate double, triple, quadruple zeta and extrapolated results, respectively. Reference values were obtained from the NIST database.\cite{NIST_ASD}}
\label{tab:Yb_cation_EOM}
\begin{tabular}{ c | l | r | r | r | r | r  }
state&conf& NIST\cite{NistDiatomic} &2z&3z&4z& extr.
\\
\hline
$^2$S$_{1/2}$       & $4f^{14}6s$     &  0 & 0 & 0 & 0 & 0 
\\
\hline
$^2$F$^\circ_{7/2}$ & $4f^{13}6s^2$  & 21419 & 12054 & 13524 & 16092 & 17966
\\
$^2$F$^\circ_{5/2}$  & $4f^{13}6s^2$ & 31568 & 22629 & 24139 & 26655 & 28491
\\
\hline
$^2$D$_{3/2}$       & $4f^{14}5d$     & 22961 & 24073 & 24209 & 24060 & 23951
\\
$^2$D$_{5/2}$       & $4f^{14}5d$     & 24333 & 25351 & 25457 & 25341 & 25257
\\
$^2$P$^\circ_{1/2}$  & $4f^{14}6p$    & 27062 & 27539 & 27780 & 27857 & 27913
\\
$^2$P$^\circ_{3/2}$  & $4f^{14}6p$     & 30392 & 30954 & 31246 & 31323 & 31380
\\
\end{tabular}
\end{table*}

Finally, our IHFS-CCSD results are presented in table~\ref{tab:Yb_cation_FSCC}. The transition energies for $4f^{14}$ configuration reproduce well the experimental ones, with errors below 6~\%, and only show a small dependence on the basis set. 
The states arising from the $4f^{13}$ configuration ($4f^{13}$[$F_{7/2}^\circ$]$s^2$ etc.), in contrast,  show a significant dependence on the basis and a rather slow convergence and underestimate the value by about 30~\%, which makes them less accurate than the EOM-CCSD ones. This lower accuracy is a consequence of the reduced flexibility in the model spaces, due to the need of adding the $5p$-shell just below the $4f$-shell to the intermediate space, in order to achieve convergence. These results are in line with the observations of~\citeauthor{Shee2018}~\cite{Shee2018}, in that the formal equivalence between EOM-CCSD and IHFS-CCSD for the sectors of Fock space considered depends, in fact, on the flexibility of the main model space.

Furthermore, the removal of the $5p$ spinors from the main model space underscores the importance of the $5p$ for the energetics of the states with a hole in the $4f$ shell, since 
by doing so, we undress the contributions from the $5p$ configurations, and thus prevent them from interacting effectively with $4f^{13}$ determinants. 

\begin{table*}[hbtp]
\caption{Transition energies for the Yb$^+$ cation. Reference values have been obtained from the NIST database\cite{NistDiatomic}, the computed values were obtained for different basis set with Fock-space coupled cluster.}
\label{tab:Yb_cation_FSCC}
\begin{tabular}{ c | l | r | r | r | r | r | r  }
state&conf& NIST\cite{NistDiatomic} & 2z & 3z & 4z & extr. & DCB\cite{Eliav1995}
\\
\hline
$^2$S$_{1/2}$        & $4f^{14}6s$   &     0 &    0  &   0   &   0&  0 & 0
\\
\hline
$^2$F$^\circ_{7/2}$  & $4f^{13}6s^2$ & 21419 & 11087 & 12390 & 13618 & 14514
\\
$^2$F$^\circ_{5/2}$  & $4f^{13}6s^2$ & 31568 & 21631 & 22976 & 24170 & 25042

\\
\hline
$^2$D$_{3/2}$        & $4f^{14}5d$   & 22961 & 24058 & 24223 & 24059 & 23938 & 23720

\\
$^2$D$_{5/2}$        & $4f^{14}5d$   & 24333 & 25336 & 25469 & 25340 & 25246 & 24998

\\
$^2$P$^\circ_{1/2}$  & $4f^{14}6p$   & 27062 & 27518 & 27774 & 27851 & 27907 & 27870
\\
$^2$P$^\circ_{3/2}$  & $4f^{14}6p$   & 30392 & 30934 & 31241 & 31316 & 31371 & 31312
\end{tabular}
\end{table*}

\subsection{Kramers-restricted configuration interaction potential energy curves}

\begin{figure*}[hbtp]
\centering
\includegraphics[width=0.99\textwidth]{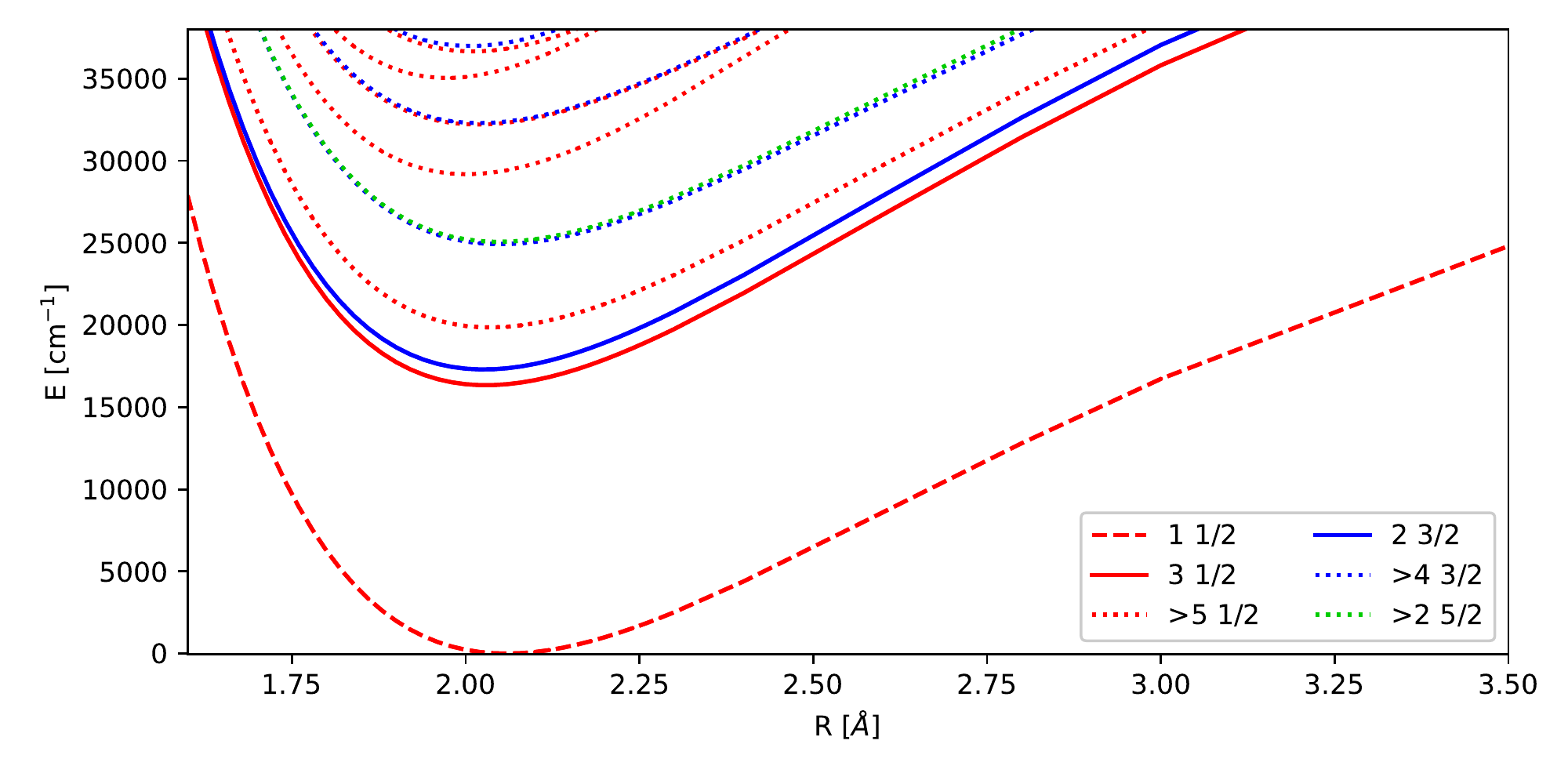}
\includegraphics[width=0.99\textwidth]{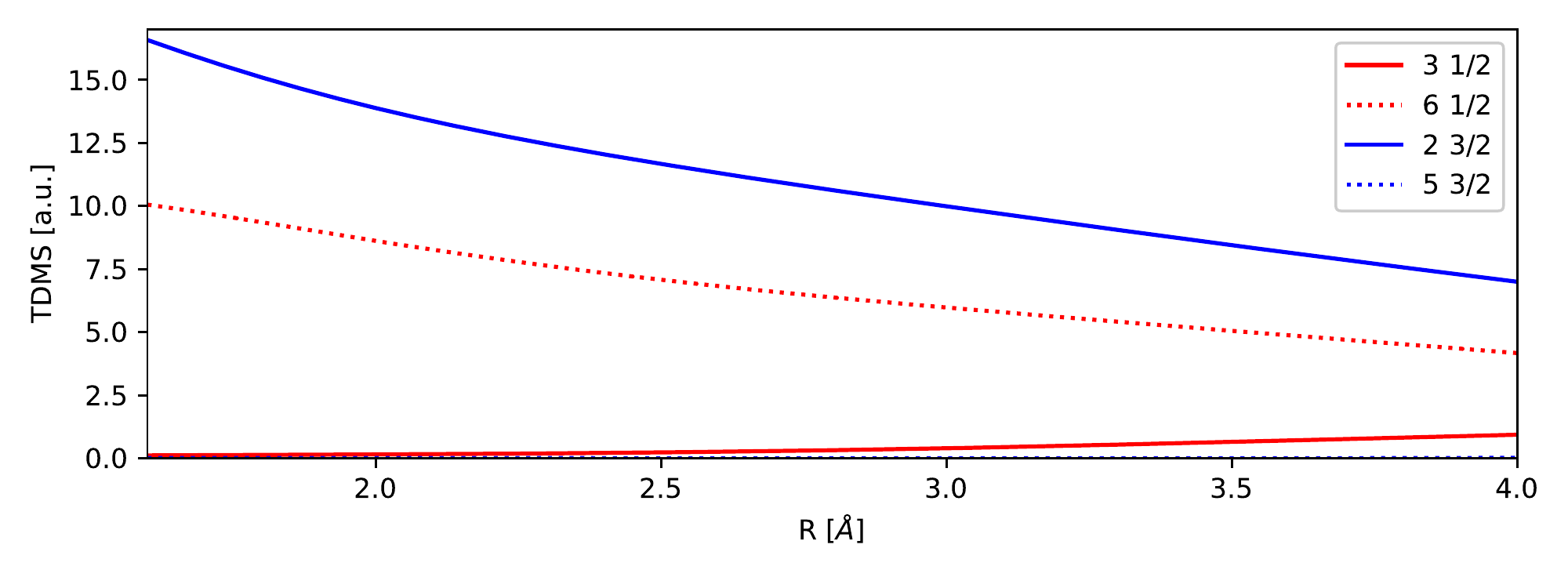}
\caption{Kramers-restricted configuration interaction potential energy curves (PECs) and transition dipole moments (TDMs) for states with $4f^{14}$ for a quadruple zeta basis set.}
\label{fig:PES_CI_zoom_f14}
\end{figure*}

The configuration interaction results for excited states of YbF were obtained by an approach corresponding to the one used for Yb$^+$, the results for a closed f-shell are depicted in figure~\ref{fig:PES_CI_zoom_f14}.
The lowest two excited states in the upper part of the figure belong to the Yb($4f^{14}6p$)F configuration, but approach asymptotically the $^2$D$_{3/2}$ state.  The asymptote of the next three states is $^2$D$_{5/2}$ corresponding to the Yb($4f^{14}5d$)F configuration for smaller internuclear separations. 
For $\Omega=5/2$ the transition dipole moment with the ground state is zero, for the other four the values are shown in the lower part of figure~\ref{fig:PES_CI_zoom_f14}. 
The first $\Omega=3/2$ and the third $\Omega=1/2$ state have a larger transition dipole moment close to equilibrium, but get close to each other at the largest internuclear separations.

\begin{figure*}[hbtp]
\centering
\includegraphics[width=0.99\textwidth]{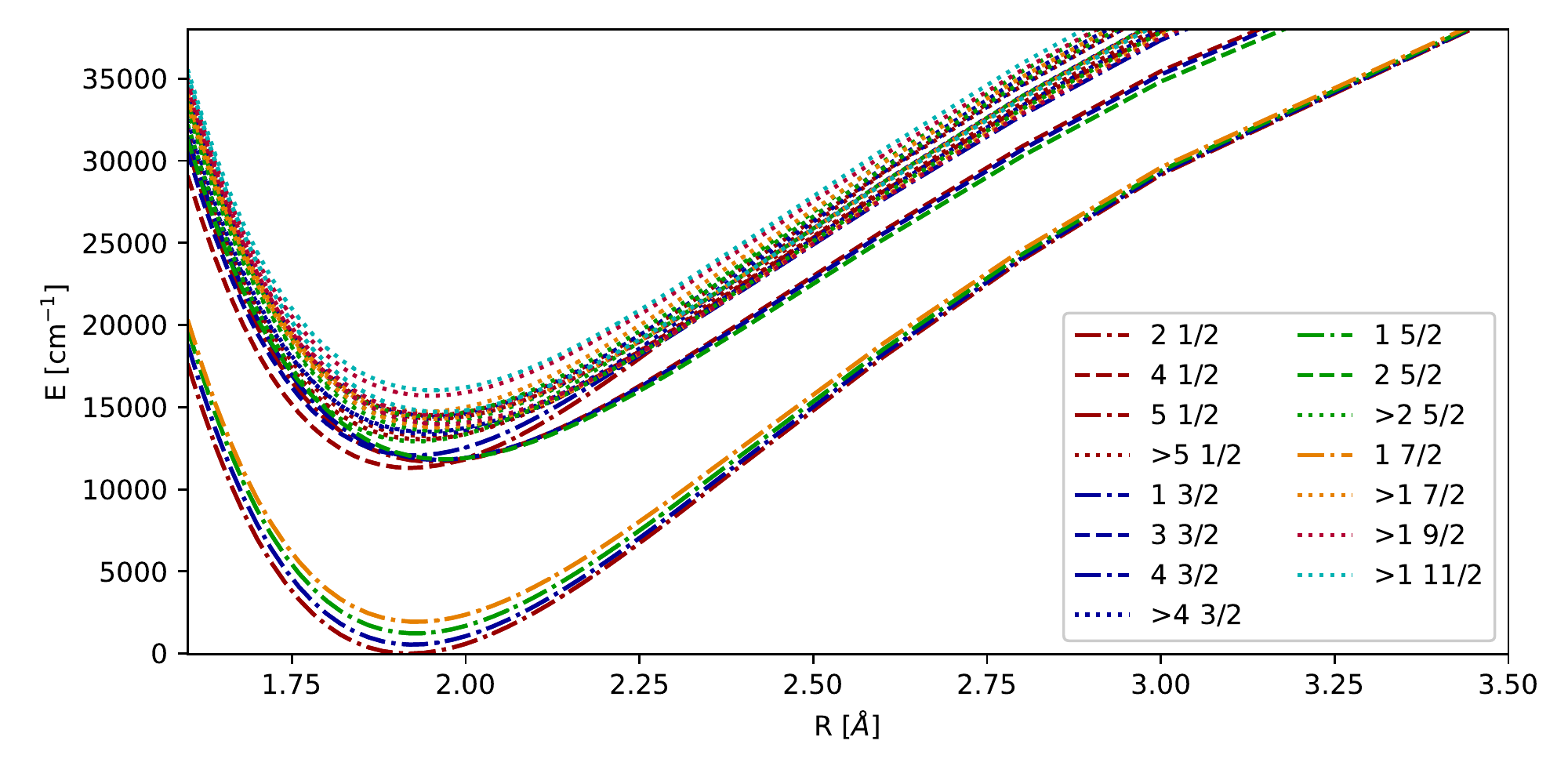}
\includegraphics[width=0.99\textwidth]{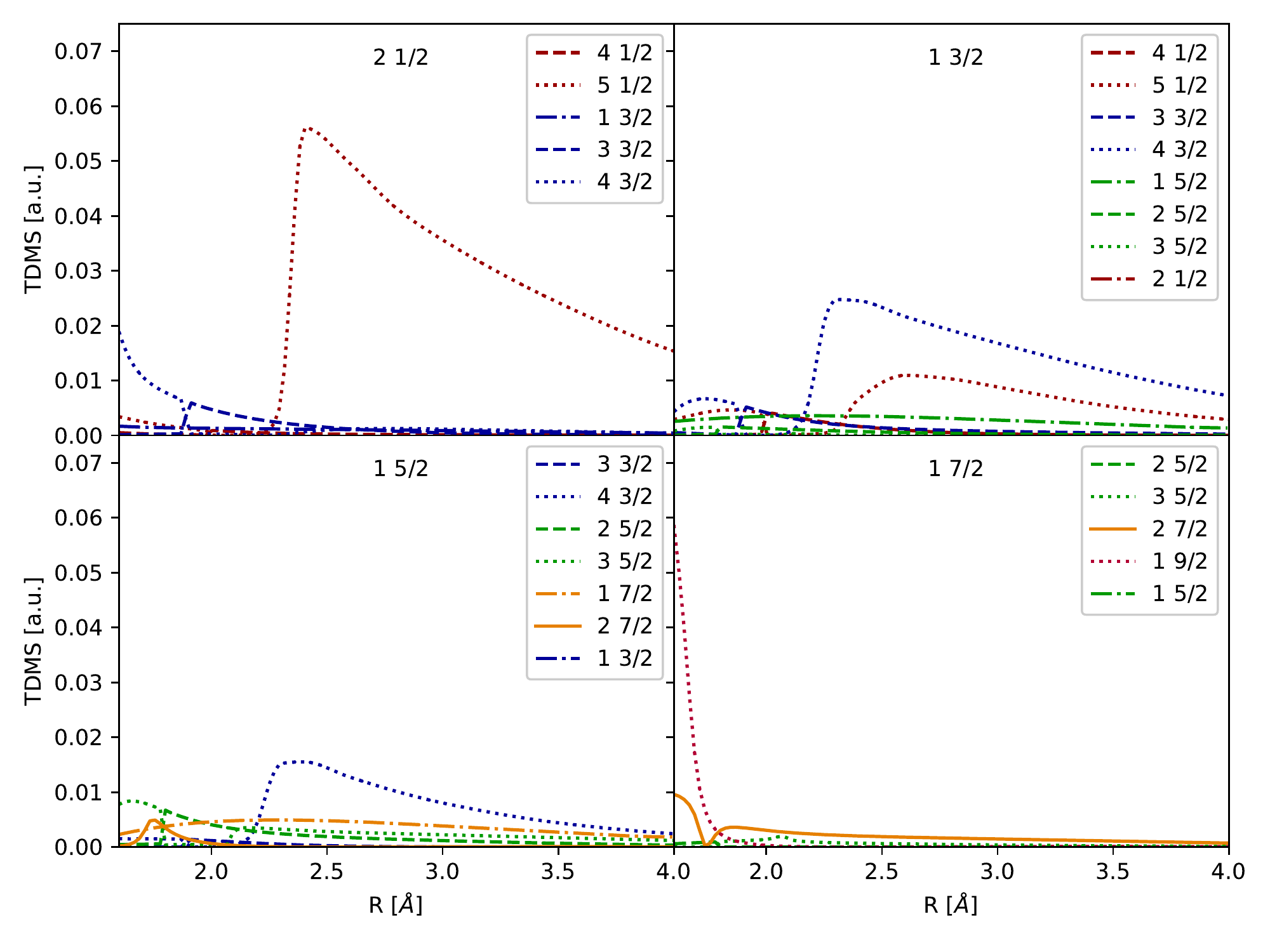}
\caption{Kramers-restricted configuration interaction PECs and TDMs for states with $4f^{13}$ for a quadruple zeta basis set. The TDMs are provided for the different lowest states with $\Omega = 1/2$, $3/2$, $5/2$, and $7/2$.}
\label{fig:PES_CI_zoom_f13}
\end{figure*}

The states of the $4f^{13}$ manifold are depicted in figure~\ref{fig:PES_CI_zoom_f13}. The lowest four states belonging to the Yb($4f^{13}$[$F_{7/2}^\circ$]$\sigma_{6s}^2$)F configuration are well separated from a dense region with a lot of states about 12000~\wn{} higher. Most of these states are of the Yb($4f^{13}$[$F_{7/2}^\circ$]$5d\sigma_{6s}$)F configuration, with the Yb($4f^{13}$[$F_{5/2}^\circ$]$\sigma_{6s}^2$)F state slightly higher in energy asymptotically and more strongly bound, resulting in several avoided crossings. 
For each of the four Yb($4f^{13}$[$F_{7/2}^\circ$]$\sigma_{6s}^2$)F states the transition dipole moments with higher excited states of varying $\Omega$ are plotted in the lower part of figure~\ref{fig:PES_CI_zoom_f13}. 
The transition dipole moments are substantially smaller than the ones for the closed $4f$-shells but some of them are non-zero. 

\begin{figure*}[hbtp]
\centering
\includegraphics[width=0.99\textwidth]{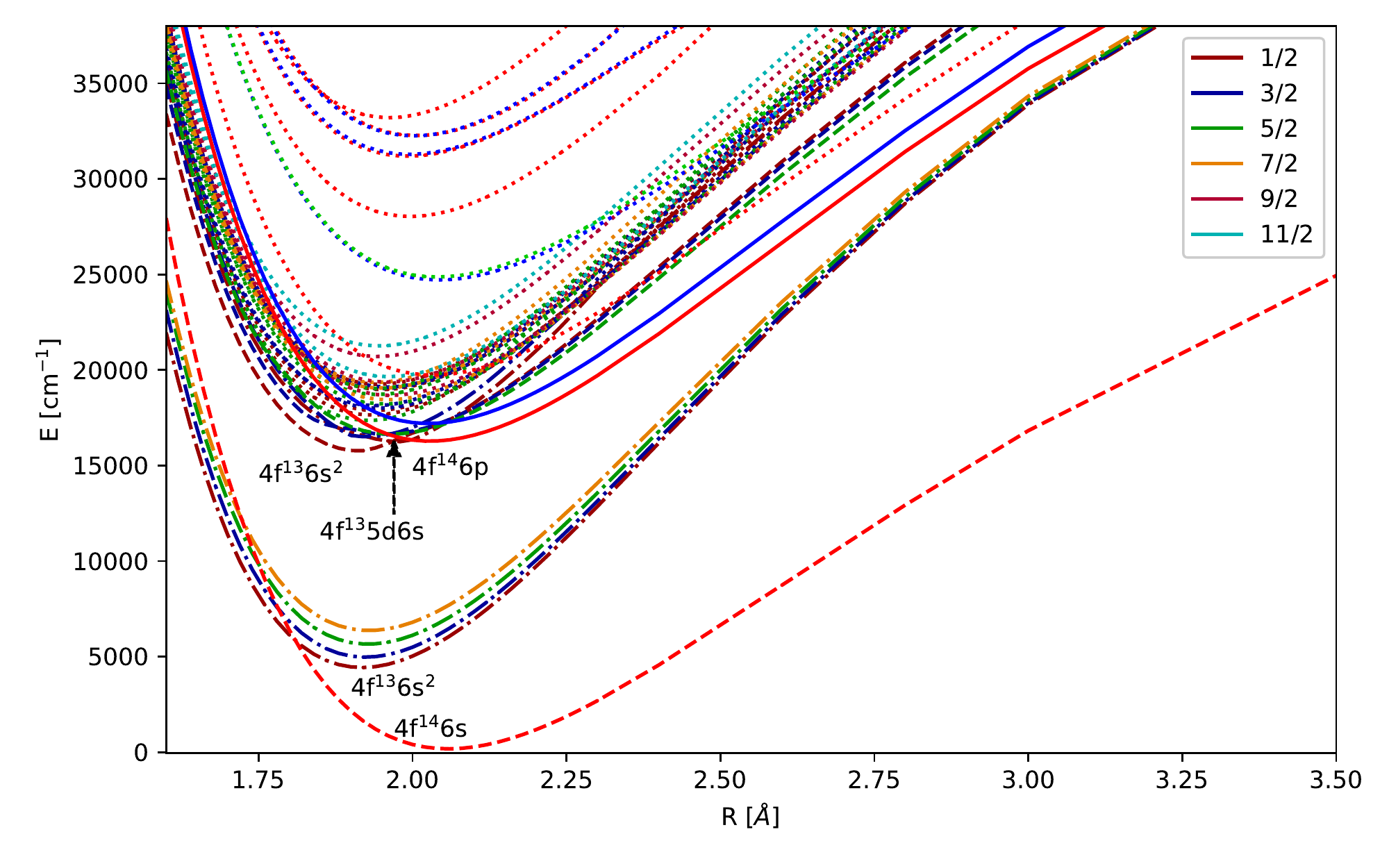}
\caption{Combination of the sets of KRCI potentials for the CBS limit. The lowest $\Omega=1/2$ are denoted by their dominant configuration.}
\label{fig:PES_CI}
\end{figure*}

The two separate sets of potential energy curves can now be combined and figure~\ref{fig:PES_CI} is obtained.
The potential energy curves have been determined up to 11~\AA{} and for the closed-shell case they were shifted so that the lowest state is at 0~\wn{} at the largest distance. Accordingly, the PECs for the hole states were shifted to 21418.75~\wn{} at this distance. There is still some interaction between ytterbium and fluorine at this distance, but the long range behaviour can be expected to be similar for the two configurations (this assumption was checked, see  supplementary information for further details). Taking into account the position of the minima, the curvatures, spin-orbit splitting, the avoided crossings and asymptotes the states can be assigned to a dominant configuration, shown in figure~\ref{fig:PES_CI}. 

An alternative to AOC-SCF for obtaining orbitals for several configurations is multiconfigurational SCF, but similar difficulties as for AOC-SCF in obtaining a balaced description of the $4f^{14}$ and $4f^{13}$ states are observed: either the wrong ground state is obtained (if only the hole states optimized in MCSCF), or the hole states are too high in energy by about 20000~\wn{} (if the ground state is optimized). 
We also made attempts using state-averaged MCSCF in a non-relativistic quantum chemistry code and observed the same difficulties (see dataset~\cite{pototschnig:ybf:dataset}). If the $4f^{13}$ configurations are excluded one obtains meaningful results, but at the expense of obtaining a Yb($4f^{13}$[$F_{7/2}^\circ$]$\sigma_{6s}^2$)F states too high in energy. 
If all the states are included, the wrong ground state is obtained. 

\subsection{Coupled cluster potential energy curves}

\begin{figure*}
\includegraphics[width=\textwidth]{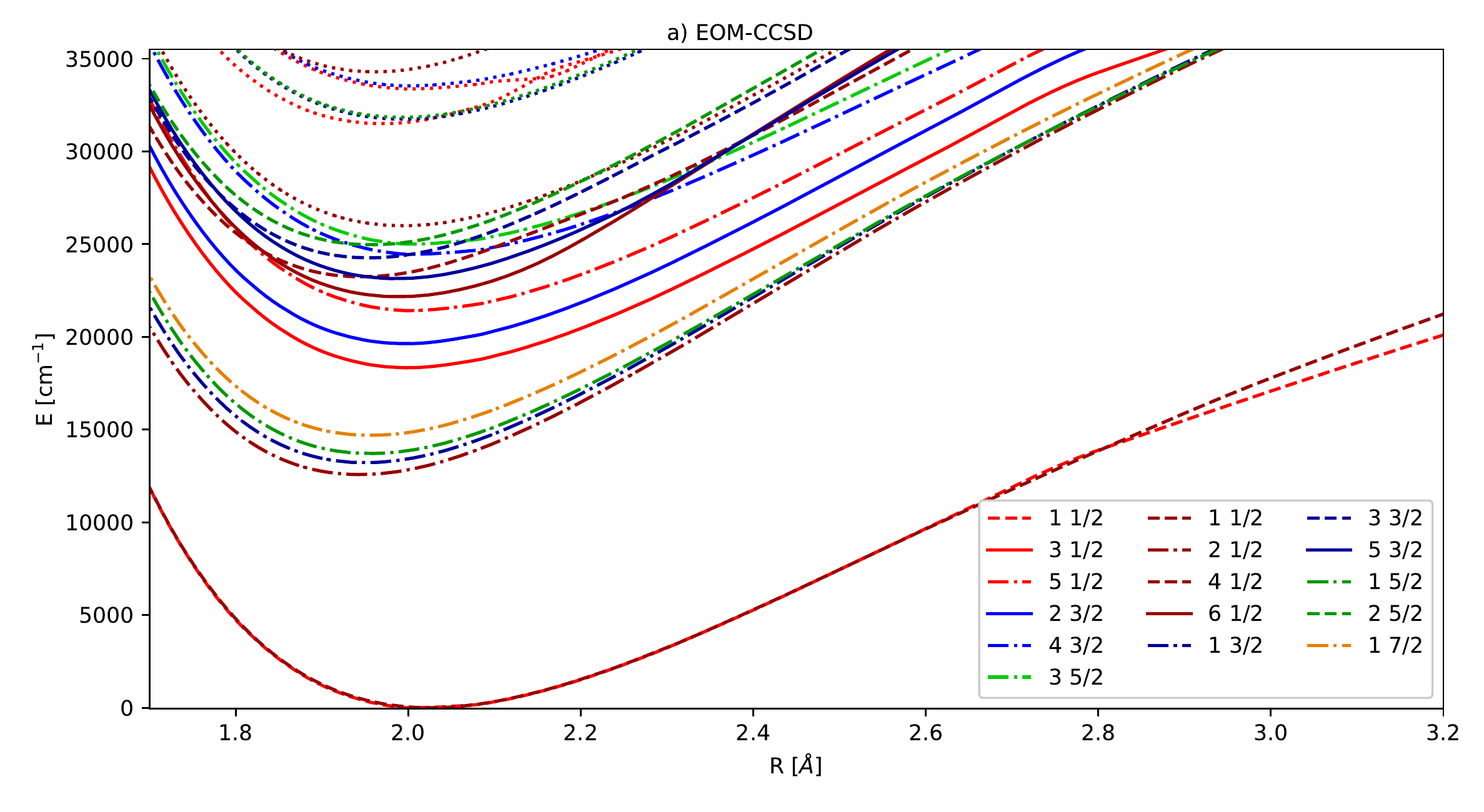}
\includegraphics[width=\textwidth]{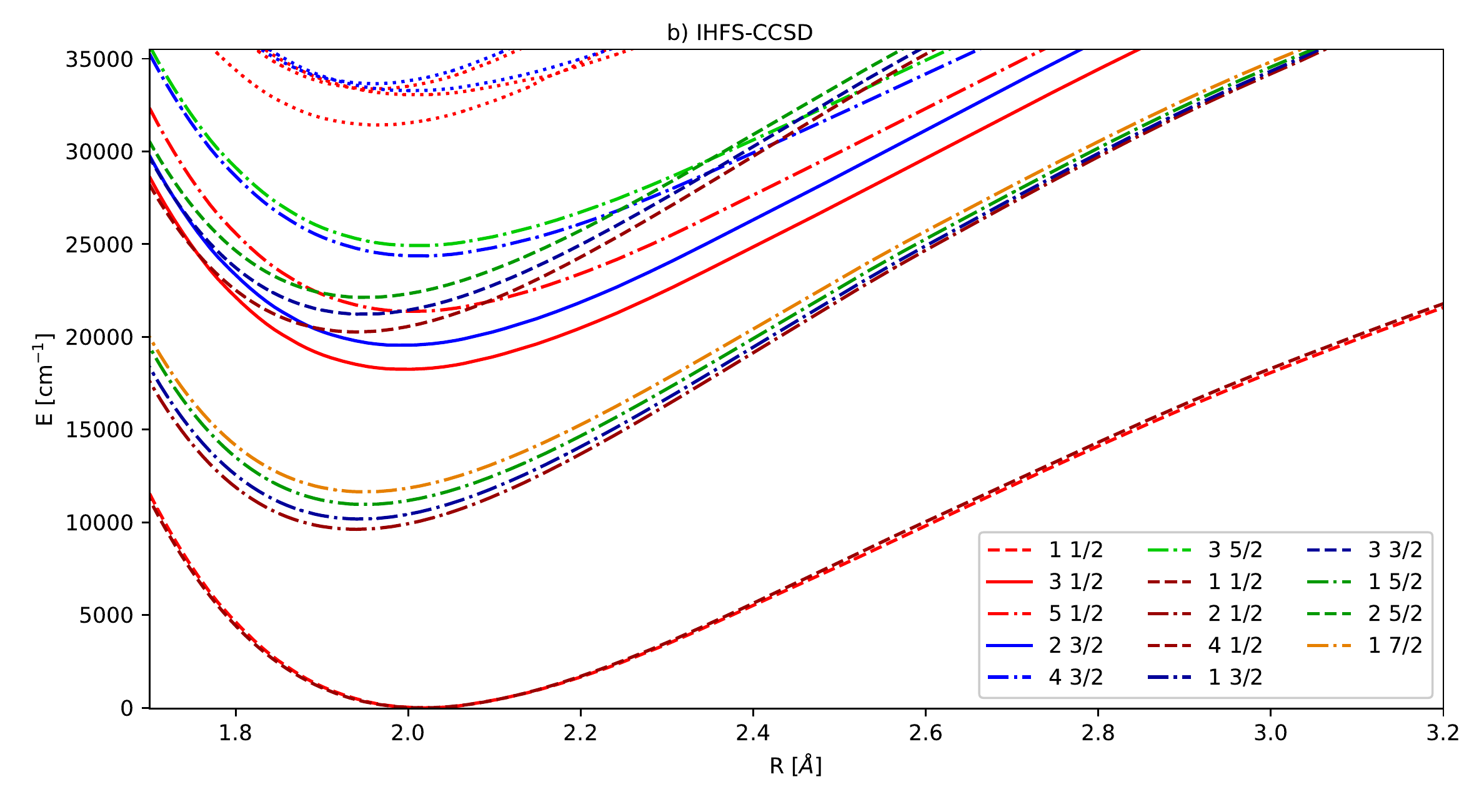}
\caption{Potential energy curves obtained by extrapolating triple and quadruple zeta basis sets.
EOM-CCSD results in the upper part, IHFS-CCSD in the lower part. The $\Omega$ values of $1/2$, $3/2$, $5/2$ and $7/2$ are indicated by the colors red, blue, green and orange. The light colors are used for the (0h,1p) sector, dark ones for (1h,0p). For the states below 30000~\wn{} we employ the same color coding and state notation as in figure~\ref{fig:PES_CI}.}
\label{fig:PECs_CC_comb}
\end{figure*}

The potential energy curves of excited states obtained by the equation-of-motion and Fock space methods are displayed in figure~\ref{fig:PECs_CC_comb}, the values for the complete basis set limit are shown.
 The basis set dependence in the molecule is similar to the one observed for Yb$^+$: energies for $4f^{14}$ states depend only weakly on the basis set, while the gap between the ground state and the excited states corresponding to the Yb($4f^{13}$[$F_{7/2}^\circ$]$\sigma_{6s}^2$)F configuration increases upon improving the basis sets.

While the EOM-CCSD excitations energies of Yb$^+$ are closer to the experimental ones, the Yb($4f^{13}$[$F_{5/2}^\circ$]$\sigma_{6s}^2$)F states are too high to perturb the Yb($4f^{14}5d$[$_{3/2}$])F PECs. From the extended potential energy curves provided in the supporting information, we can observe that the ground state of the non-interacting system (Yb($4f^{14}\sigma_{6s}^2$)F($2p^{5}$)) is repulsive and has a high energy at the equilibrium distance. This results in several avoided crossings being observed at 3, 3.5, and 5~\AA{}.

Since the Yb($4f^{14}\sigma_{6s}$)F ground state is accessible for both sectors employed in the coupled cluster calculations (Fock space as well as EOM), we can assess the compatibility of the two separate sets of calculations (in the sense of having comparable accuracies) by looking more closely at the differences between the ground states in figure~\ref{fig:PECs_CC_comb}. From that, we can see that the EOM-CCSD and IHFS-CCSD approaches the curves are on top of each other from the smallest considered internuclear separation up to about 2.8~\AA{}. This assures us that there should not be artifacts in putting together and comparing the calculations on the two sectors.

\subsection{Dissociation and ionization energies}

Since the (1h,0p) and (0h,1p) sectors have been considered in our EOM-CCSD and IHFS-CCSD calculations, 
we have as a by-product of our calculations the ionization potentials (IP) and electron affinities (EA) for YbF for all computed distances. Therefore these quantities are presented first in table~\ref{tab:EA_IP_DE}, before proceeding to the spectroscopic constants.

\begin{table*}[hbtp]
\caption{Ionization potential (IP), electron affinity (EA) and dissociation energy (D$_e$) of Yb, F, and YbF. All values in~\wn{}. They are listed for a quadrupole zeta basis set and a basis set extrapolation.  The minimum of the potentials were determined using a Morse fit and used to compute the adiabatic values listed here. }
\label{tab:EA_IP_DE}
\begin{tabular}{ l | c | r | r | r | r | r | r | r  r }
 quant. & system & \multicolumn{2}{c |}{ KRCI } & \multicolumn{2}{c |}{ EOM-CCSD } & \multicolumn{2}{c |}{ IHFS-CCSD } & experiment
  \\
  \hline
 & &4z&extr.&4z&extr.&4z&extr.&
 \\ 
 IP  & Yb & 38406 & 39128 & 50735 & 50822 & 50740 & 50837 & 50443\cite{NIST_ASD}
 \\ 
 IP2  & Yb & 90581 & 90926 & 97919 & 98035 & 97918 & 98040 & 98232\cite{NIST_ASD}
 \\ 
 IP  & F & 127131 & 126617 & 144153 & 143321 & 144076 & 144703 & 140525\cite{NIST_ASD}
 \\ 
 EA  &  F  & 13326 & 11978 & 27279 & 27740 & 27246 & 27759 & 27432\cite{Blondel2001}
 \\ 
 IP  & YbF & 58478 & 56884 & 48471 & 48578 & 48426 & 49901 & 47700\cite{Kaledin1999}
 \\ 
 EA & YbF & 7423 & 7326 & 9713 & 9876 & 9579 & 8197 &
 \\ 
 D$_e$(IP) & YbF & 26059 & 25887 & 43824 &  47782 & 40591 & 40931 & 43260\cite{Yokozeki1976}
 \\ 
 D$_e$(EA) & YbF & 40394 & 39660 & 45534 & 49629 & 40430 & 49053 & 43260\cite{Yokozeki1976}
\\
\end{tabular}
\end{table*}

Unlike coupled cluster calculations, for KRCI a consistent definition of active spaces is difficult, and its lack of size-consistency results in large deviations from experiment and from the coupled cluster values. 
For adiabatic electron affinities, for which to the best of our knowledge there are no experimental values, the extrapolated values are 8393 and 8197 cm$^{-1}$ for EOM-CCSD and IHFS-CCSD, respectively. For a distance of 6.5~\AA\ a value of 28651 cm$^{-1}$ was obtained, which is reasonably close to the electron affinity of fluorine (27432 cm$^{-1}$).\cite{Blondel2001}
Corresponding results for the atoms are listed in the table, which allow to calculate the dissociation energies (D$_e$).
They deviate from the experimental values of $43600 \pm 800$ cm$^{-1}$ by \citeauthor{Kaledin1999}\cite{Kaledin1999} and $43260\pm 800$~\wn{} by \citeauthor{Yokozeki1976}\cite{Yokozeki1976}. 
The ionization potentials in table~\ref{tab:EA_IP_DE} show acceptable agreement with experimental values. 

\subsection{Spectroscopic constants}

The spectroscopic constants for the ground state are now considered. In table~\ref{tab:ground_state} our results are summarized, along those from the literature.
\begin{table*}[hbtp]
\caption{Spectroscopic constants for  ground state parameters for different approaches. Dissociation energies (D$_e$), harmonic frequencies ($\omega_e$) and anharmonicity constants ($\omega_e\chi_e$) are given in \wn{}, the equilibrium bond distances (r$_e$) in \AA{}.
For the theoretical results we listed the CBS values. }
\label{tab:ground_state}
\begin{tabular}{ r | r | r | r | r | r | r}
method&ref.&r$_e$&$\omega_e$&$\omega_e\chi_e$&D$_e$
\\
\hline
KRCI & YbF &  2.0829 &         465 &        2.40  &            0\\
EOM-CCSD & YbF$^+$  & 2.0230 & 511 & 2.80 & 49629 \\
         & YbF$^-$  & 2.0250 & 508 & 2.53 & 47782 \\
IHFS-CCSD & YbF$^+$ & 2.0176 & 515 & 2.82 & 49053 \\
        & YbF$^-$   & 2.0159 & 513 & 2.42 & 40931 \\
&&&&&\\
CCSD\cite{Gomes2010}&YbF&2.0174&507.6&2.357&40904 \\
CCSD(T)\cite{Gomes2010}&YbF&2.0289&528.2&1.939&41156 \\
CCSD\cite{Su2009}&&2.0127&566.8&3.7885&55650 \\
RASCI\cite{Nayak2006}&&2.051&529&& \\
CCSD(T)\cite{Heiberg2003}&&2.03&&&38900 \\
CISD\cite{Cao2002a}&&2.034&502&&42100 \\
DFT\cite{Liu1998}&&1.987&532&&45000 \\
&&&&&\\
exp.\cite{Yokozeki1976}&&&&&43260 \\
exp.\cite{Dunfield1995}&&2.0158&506.6674&2.2452& \\
exp.\cite{Uttam1995}&&&505.5&1.9& \\
exp.\cite{Kaledin1999}&&&&&43600 \\
exp.\cite{Dickinson2001}&&2.016514&&& \\
exp.\cite{Lim2017}&&2.0195&506.616&2.235& \\
\end{tabular}
\end{table*}
We observe that the extrapolated KRCI bond distances, at about 2.058~\AA{}, are significantly longer (by around 0.04~\AA{}) than experiment~\cite{Dickinson2001}, whereas the coupled cluster calculations show differences from experiment smaller than 0.01~\AA{}, with EOM-CCSD showing slightly larger discrepancies than IHFS-CCSD. Between the extrapolated EOM-CCSD and IHFS-CCSD, we also see small differences between the $4f^{14}$ and $4f^{13}$: for EOM-CCSD these differ by around 0.001\AA{} whereas for IHFS-CCSD the difference is slightly under 0.002\AA{}, with the $4f^{13}$ configuration yielding a slightly underestimated value, compared to experiment, something that can be traced back to the differences in model spaces for this configuration. 

Our results for harmonic frequencies further indicate that KRCI seems to underestimate the bonding strength in YbF, as the harmonic vibrational frequency is smaller (491 \wn{}) than experiment (between 505.5 and 506.7 \wn{} depending on the experiment). The coupled cluster results, on the other hand, show the typical 5-6 \wn{} overestimation of the harmonic frequencies with respect to experiment (something also seen for the anharmonic constants), which can be attributed to lack of triples in the EOM or FS treatment, that would introduce further orbital relaxation. This can be seen in comparison to the unrestricted coupled cluster calculations of~\citeauthor{Gomes2010}\cite{Gomes2010}, which in spite of the large value of the $T_1$ diagnostic, reproduce well the experimental bond lengths, harmonic frequencies and anharmonic constants. 

Taken together, our 2-component CCSD-based calculations and the 4-component ones of ~\citeauthor{Gomes2010}\cite{Gomes2010} compare consistently better to experiment than the other theoretical works for bond lengths, vibrational frequencies and anharmonic constants. For the dissociation energies, on the other hand, the extrapolated calculations presented here do not provide a significant improvement over the results of prior theoretical investigations (quadruple zeta values are closer to the experimental ones for this quantity, see table \ref{tab:EA_IP_DE}). Especially, electron attachment values are off, which might be related to the absence of the configuration with a hole in the p orbitals of fluorine. 



Moving now to excited states, we start by considering the four lowest excited states, which belong to the Yb($4f^{13}$[$F_{7/2}^\circ$]$\sigma_{6s}^2$)F configuration. These states are well separated from the ground state (the lowest excited state is about 10000~\wn{} above the ground state) and higher excited states. That such states are quite well separated from the ground state would, in our view, tend to exclude the interaction with a low-lying excited state as an explanation for the appearance of the large $T_1$ diagnostic values observed by~\citeauthor{Gomes2010}\cite{Gomes2010}. From their spectroscopic constants, presented in table~\ref{tab:low_state}, we see that with the exception of DFT all methods yield similar level splittings of about 500, 1200, and 2000~\wn{}. To the best of the authors knowledge there is no experimental data available for these states, due to their negligible  transition dipole moments for dipole excitations (see for instance figure~\ref{fig:PES_CI_zoom_f13}) and small Franck-Condon factors due to the difference in bond distances between these states and the ground state. 

\begin{table*}[hbtp]
\caption{Spectroscopic constants for the lowest excited states Yb($4f^{13}$[$F_{7/2}^\circ$]$\sigma_{6s}^2$)F for different wave function methods using the potential energy curves extrapolated to the basis set limit. 
In the case of KRCI and MRCI\cite{DOLG1992a} the ground state is not included in the computation and absolute transition energies are not available.  
The transition energy (T$_e$), level splitting (T$_{rel}$, energy relative to $2_{1/2}$), harmonic frequencies ($\omega_e$) and anharmonicity constants ($\omega_e\chi_e$) are given in \wn{}, the equilibrium bond distances (r$_e$) in \AA{}.
}
\label{tab:low_state}
\begin{tabular}{ r | r | r | r | r | r | r } 
state & method & T$_e$ & r$_e$&$\omega_e$&$\omega_e\chi_e$&T$_{rel}$ \\ 
\hline
$2_{1/2}$ & KRCI   &  & 1.9200 & 631 & 2.51 & \\ 
   & EOM-CCSD & 12568 & 1.9432 & 591 & 2.59 & \\ 
 & IHFS-CCSD  &  9627 & 1.9396 & 599 & 2.79 & \\ 
 & DFT\cite{Liu1998} & 3790 & 1.9570 & 561 & & \\ 
 & MRCI\cite{DOLG1992a} & & 1.9480 & 600 & & \\ 
  $1_{3/2}$ & KRCI &  & 1.9253 & 628 & 2.50 & 540 \\ 
   & EOM-CCSD & 13211 & 1.9494 & 588 & 2.61 & 643 \\ 
  & IHFS-CCSD & 10180 & 1.9438 & 595 & 2.79 & 553 \\ 
 & DFT\cite{Liu1998} & 9520 & 1.9440 & 597 & & 5730 \\ 
& MRCI\cite{DOLG1992a} & & 1.951 0 & 598 & & 428 \\ 
    $1_{5/2}$&KRCI &  & 1.9296 & 622 & 2.45 & 1223 \\ 
   & EOM-CCSD & 13703 & 1.9553 & 582 & 2.61 & 1135 \\ 
  & IHFS-CCSD & 10968 & 1.9493 & 589 & 2.78 & 1341 \\ 
& DFT\cite{Liu1998} & 10970 & 1.9360 & 598 & & 7180 \\ 
& MRCI\cite{DOLG1992a} & & 1.9540 & 594 & & 1021 \\ 
  $1_{7/2}$ & KRCI &  & 1.9315 & 616 & 2.43 & 1933 \\ 
   & EOM-CCSD & 14685 & 1.9556 & 577 & 2.62 & 2117 \\ 
  & IHFS-CCSD & 11645 & 1.9496 & 583 & 2.77 & 2018 \\ 
& DFT\cite{Liu1998} & 16530 & 1.936 & 592 & & 12740 \\ 
& MRCI\cite{DOLG1992a} & & 1.954 & 589 & & 1709 
\end{tabular}
\end{table*}

The smallest equilibrium distance was obtained for the $2_{1/2}$ state with 1.94~\AA{} for the coupled cluster methods and 0.02~\AA{} less for KRCI.
The vibrational frequencies are between 570 and 600~\wn{} for the coupled cluster methods and about 30~\wn{} higher for KRCI.  

For higher excited states, as apparent from the figures in the previous section, the identification and assignment of states gets more difficult and there are differences between the methods. We have nevertheless provided in table~\ref{tab:omega_135_2} the spectroscopic constants for excited with $\Omega$ values of $1/2$, $3/2$, and $5/2$, respectively.

\onecolumn
\begin{longtable}{llll|rrrr}
\caption{Spectroscopic constants for excited states with $\Omega=1/2, 3/2, 5/2$, starting from 18000~\wn{} for different methods using the values after extrapolation to the basis set limit.
Transition energy (T$_e$), vibrational constant ($\omega_e$), and anharmonicty constant ($\omega_e\chi_e$) are given in \wn{}, the equilibrium bond distance (r$_e$) in \AA{}. Experimental transitions that were not assigned(n.a.) are also listed.
\label{tab:omega_135_2}}\\
\hline\hline
$\Omega$ & 
method   & 
state      &
configuration  & 
T$_e$    &
r$_e$    & 
$\omega_e$    &
$\omega_e\chi_e$ \\
\hline
\endfirsthead
\multicolumn{8}{c}%
{\tablename\ \thetable\ -- \textit{Continued from previous page}} \\
\hline
\hline
$\Omega$ & 
method   & 
state       &
configuration  & 
T$_e$    &
r$_e$    & 
$\omega_e$    &
$\omega_e\chi_e$ \\
\hline
\endhead
\hline \multicolumn{8}{r}{\textit{Continued on next page}} \\
\endfoot
\hline
\multicolumn{8}{l}
{\textsuperscript{a} KRCI transition energies for the 4f$^{13}$ sector were obtained by adding 4144~\wn{},} \\
\multicolumn{8}{l}
{an estimate for the energy of the lowest state in this manifold.} \\
\endlastfoot

%
1/2 & KRCI\textsuperscript{a} 
  & 3 & $4f^{13}\sigma_{6s}^2$ & 15572 & 1.9038 & 655 & 13.57 \\ 
 && 4 & $4f^{13}5d\sigma_{6s}$& 16020 & 1.9603 & 620 & 0.08 \\ 
 && 5 & $4f^{14}6p$ & 16189 & 2.0504 & 496 & 2.38 \\ 
 && 6 & $4f^{13}5d\sigma_{6s}$ & 17350 & 1.9360 & 598 & 2.50 \\ 
 && 7 & $4f^{13}5d\sigma_{6s}$ & 17587 & 1.9417 & 592 & 2.56 \\ 
 && 8 & $4f^{13}5d\sigma_{6s}$ & 18675 & 1.9539 & 582 & 2.71 \\ 
 && 9 & $4f^{13}5d\sigma_{6s}$ & 18952 & 1.9479 & 578 & 2.71 \\
 && 10 & $4f^{14}5d$ & 19631 & 2.0552 & 490 & 2.49 \\ 
&EOM-CCSD & 3 & $4f^{14}6p$ & 18373 & 2.0004 & 536 & 2.72 \\ 
 && 4 & $4f^{14}5d$ & 21448 & 2.0079 & 532 & 2.78 \\ 
 && 5 & $4f^{14}6p$ & 22147 & 1.9886 & 573 & 1.18\\ 
 && 6 &$4f^{13}\sigma_{6s}^2$ & 23241 & 1.9432 & 582 & 4.06 \\ 
&IHFS-CCSD & 3 & $4f^{14}6p$ & 18249 & 1.9953 & 539 & 2.63 \\ 
 && 4 & $4f^{13}\sigma_{6s}^2$ & 20267 & 1.9397 & 597 & 2.78 \\ 
 && 5 & $4f^{14}5d$ & 21375 & 2.0032 & 533 & 2.73 \\ 
&MRCI\cite{DOLG1992a} &  & $4f^{13}\sigma_{6s}^2$ & & 1.948 & 600 & \\ 
&exp.\cite{Dunfield1995} \textsuperscript{a} & 3 & & 18106.20 & & 537 & 3 \\ 
&exp.\cite{Dunfield1995} & 4 & [18.6]$_{1/2}$ & 18705.06 & & &  \\ 
&exp.\cite{Lim2017} & & [557] & 18574 & 1.9656 & 502.15 & \\ 
&exp.\cite{Lim2017} & & [561] & 18699 & 1.9571 & & \\ 
%
%
&&&&&&&\\
3/2 & KRCI\textsuperscript{a}
  & 2 & $4f^{13}\sigma_{6s}^2$ & 16206 & 1.9331 & 711 & 8.45 \\ 
 && 3 & $4f^{13}5d\sigma_{6s}$ & 16346 & 1.9505 & 516 & 0.80 \\ 
 && 4 & $4f^{14}6p$ & 17123 & 2.0473 & 499 & 2.37 \\ 
 && 5 & $4f^{13}5d\sigma_{6s}$ & 17693 & 1.9473 & 585 & 2.47 \\ 
 && 6 & $4f^{13}5d\sigma_{6s}$ & 17843 & 1.9408 & 584 & 3.30\\ 
&EOM-CCSD & 2 & $4f^{14}6p$ & 19672 & 1.9971 & 540 & 2.72 \\ 
 && 3 & $4f^{14}6p$ & 23137 & 1.9857 & 557 & 2.28 \\ 
 && 4 & $4f^{13}\sigma_{6s}^2$ & 24251 & 1.9537 & 584 & 2.64 \\ 
 && 5 & $4f^{14}5d$ & 24468 & 2.0177 & 509 & 2.78 \\ 
&IHFS-CCSD & 2 & $4f^{14}6p$  & 19543 & 1.9920 & 542 & 2.63 \\ 
 && 3 & $4f^{13}\sigma_{6s}^2$ & 21222 & 1.9480 & 591 & 2.80 \\ 
 && 4 & $4f^{14}5d$ & 24363 & 2.0120 & 512 & 2.73 \\ 
&MRCI\cite{DOLG1992a} & & $4f^{13}\sigma_{6s}^2$ & & 1.953 & 596 & \\ 
&exp.\cite{Dunfield1995} & 2 & & 19471.49 & &  &  \\ 
%
%
    &      &   &             &       &        &     &      \\
5/2 & KRCI\textsuperscript{a}
           & 2 & $4f^{13}5d\sigma_{6s}$ & 16340 & 1.9677 & 564 & 2.42 \\ 
    &      & 3 & $4f^{13}\sigma_{6s}^2$ & 17063 & 1.9302 & 635 & 1.41 \\ 
    &      & 4 & $4f^{13}5d\sigma_{6s}$ & 17700 & 1.9627 & 571 & 0.93\\ 
    &      & 5 & $4f^{13}5d\sigma_{6s}$ & 18422 & 1.9451 & 584 & 3.07 \\ 
    & EOM-CCSD & 2 & $4f^{13}\sigma_{6s}^2$ & 24957 & 1.95359 & 577 & 2.62 \\ 
    & & 3 & $4f^{14}5d$ & 25023 & 2.0146 & 513 & 2.80 \\ 
    & IHFS-CCSD & 2 & $4f^{13}\sigma_{6s}^2$ & 22127 & 1.9499 & 584 & 2.77 \\ 
    & & 3 & $4f^{14}5d$ & 24919 & 2.0089 & 515 & 2.77 \\ 
    & MRCI\cite{DOLG1992a} & & $4f^{13}\sigma_{6s}^2$ & & 1.954 & 590 & \\ 
&&&&&&&\\
n.a.&exp.\cite{Smallman2014} & & [574] & 19150 & \\ 
&exp.\cite{Smallman2014} & & [578] & 19280 & \\ 
&exp.\cite{Uttam1995} & & Q & 23035.3 & & 523 & 2 \\ 
&exp.\cite{Uttam1995} & & Q & 23256.0 & & 507 & 2 \\ 
&exp.\cite{Uttam1995} & & P & 26014.8 & & 574.6 &2 .8 \\ 
\hline
\hline
\end{longtable}
\twocolumn

The comparison with experimental results allows assignment of the lowest excited state reported in experiments and give some indications for higher states. 
The lowest $\Omega=1/2$ state observed in experiment can be identified as the $3_{1/2}$ state. 
Spectroscopic parameters agree well with the ones obtained by fitting to the A$^2\Pi_{1/2}$ in experiments. 
A bond distance of 1.9935~\AA{} obtained by fitting to the same states in ref.~\citenum{Ma2009} agrees well with the coupled cluster values for the  $3_{1/2}$ state, the vibrational constant of about 540~\wn{} is close to the experimental value of ref.~\citenum{Dunfield1995}. 
Similarly, the lowest $\Omega=3/2$ state reported by~\citeauthor{Dunfield1995}\cite{Dunfield1995} can be identified as the $2_{3/2}$ state, see table \ref{tab:omega_135_2}. 

The lowest states with $\Omega=1/2$ and $\Omega=3/2$ for this energy range approach asymptotically a state with a Yb($4f^{14}5d$)F configuration, but if one analyses the EOM-CCSD and IHFS-CCSD orbital composition, significant contributions of the atomic $6p$ are identified.
The  $\Omega=3/2$ state is dominated (97\%) by a single configuration, corresponding to a HOMO($\sigma_{6s, 1/2}$)$\rightarrow$LUMO+1 where the latter is made up of a mixture of $6p_{\pi}$ and $5d_{\pi}$ orbitals (the $6p_{\pi}$ contributions being the dominant -- $\simeq 80\%$ -- in the reference YbF$^+$ orbitals).
The few other significant configurations arise from excitations to higher-lying orbitals with increasingly large ($\ge 50\%\, 5d_{\pi}$) contributions.
The $\Pi_{1/2}$ state is also dominated by a single configuration, now corresponding to a HOMO($\sigma_{1/2}$)$\rightarrow$LUMO transition, and shows a rather similar picture in terms of the relative weights of the $6p_{\pi}$ and $5d_{\pi}$ orbitals, with very small contributions from the ground-state mixing due to spin-orbit coupling.
The splitting in Yb$^+$ of $^2$D$_{3/2}$ and $^2$D$_{5/2}$ is 1372~\wn{}, for $^2$P$_{1/2}$ and $^2$P$_{3/2}$ it is 3330~\wn{}. 
The separation between the lowest excited $\Omega = 1/2$ states in the closed shell computation is 3779 and 3126~\wn{} for EOM-CCSD and IHFS-CCSD, respectively. 
This is an indication that the two states must be regarded as a fairly strong admixture of 6p and 5d orbitals of j={1/2} or {3/2}, as one can expect a much smaller spin-orbit splitting in the axial field of the molecule (about $1/3$ of the atomic spin-orbit splitting for the P state). 
This picture also finds experimental support in recent measurements of hyperfine constants ($d$ and $eQ_0q$) for the ground and $\Pi_{1/2}$ excited state of YbF\cite{Steimle2007}, where a simple ligand-field model disregarding the contributions from $5d_\pi$ orbitals predicted values of $d$ a factor of 2 larger than the measurements.
For bond distances much larger that the equilibrium one the system gets closer to the configurations in Yb$^+$ with a dominating 5d contribution.

As already mentioned this energy range above 18000~\wn{} is dense with a large number of excited states that can mix with each other and result in new mixed states, like the [557] and [561] ones\cite{Lim2017}. 
These will be addressed in section \ref{sec:adiabatization}. 

Uttam \textit{et al.}\cite{Uttam1995} reported three unidentified states with energies above 22000~\wn{} which are listed in table \ref{tab:omega_135_2} and cannot uniquely be identified with the current results. 
The one at 26014.8~\wn{} has a larger vibrational constant indicating a more strongly bound state, possibly of the  $4f^{13}\sigma_{6s}^2$  configuration. The vibrational spacing of the two states at 23000~\wn{} rather points to states with a closed f shell.



\subsection{Perturbation of the $3_{1/2}$ excited state}
\label{sec:adiabatization}

Due to the use of different sectors of Fock space to obtain the $4f^{14}$ and $4f^{13}$ configurations, the excited states with the same $\Omega$ values cannot interact among themselves, as is the case within each sector. However, from the discussion above, it is clear that dealing with states which are artificially prevented from interaction makes it difficult to establish a comparison to experiment, for states from about 18000~\wn{} to about 26000~\wn{}, which is where these configurations should be the most entangled. In order to remedy that, in the following we introduce a simple adiabatization model (equation~\ref{eq:adiabatization}) that allows us to investigate how coupling such states would affect the overall spectra in the aforementioned energy region. 

\begin{figure*}[hbtp]
\centering
\includegraphics[width=0.95\textwidth]{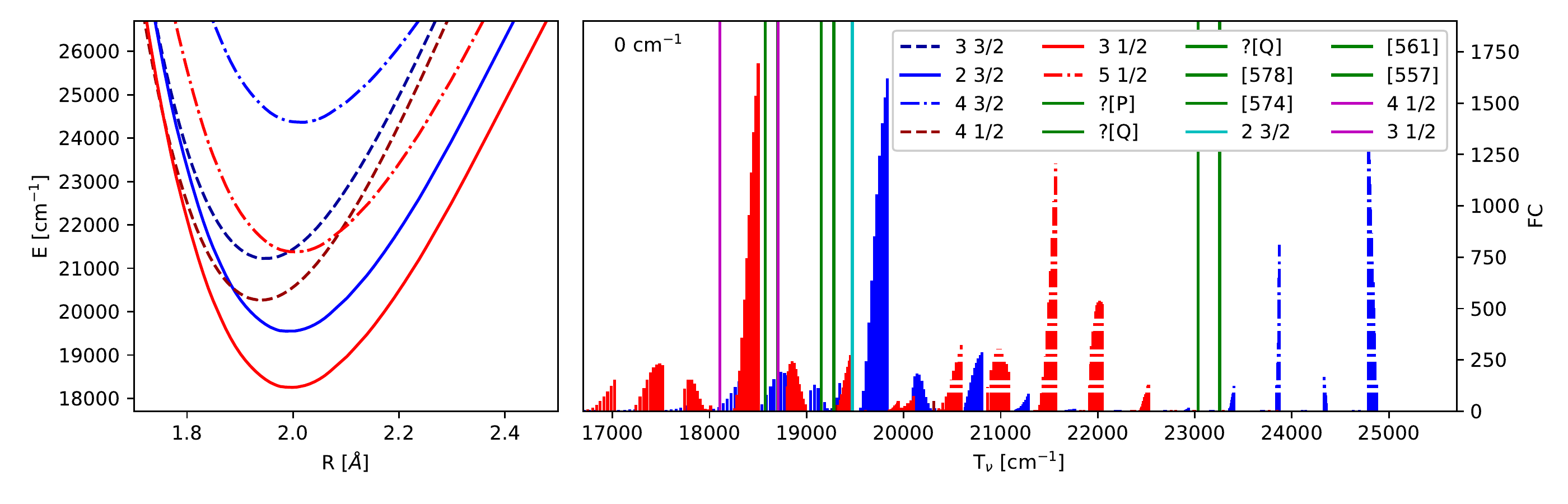}
\\
\includegraphics[width=0.95\textwidth]{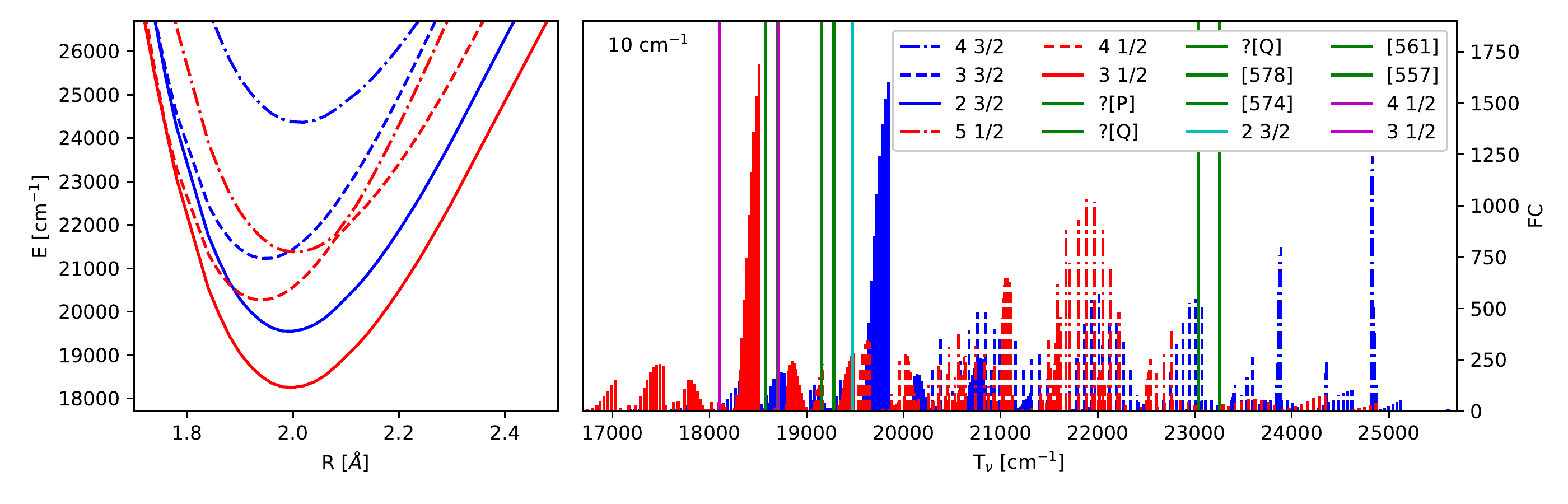}
\\
\includegraphics[width=0.95\textwidth]{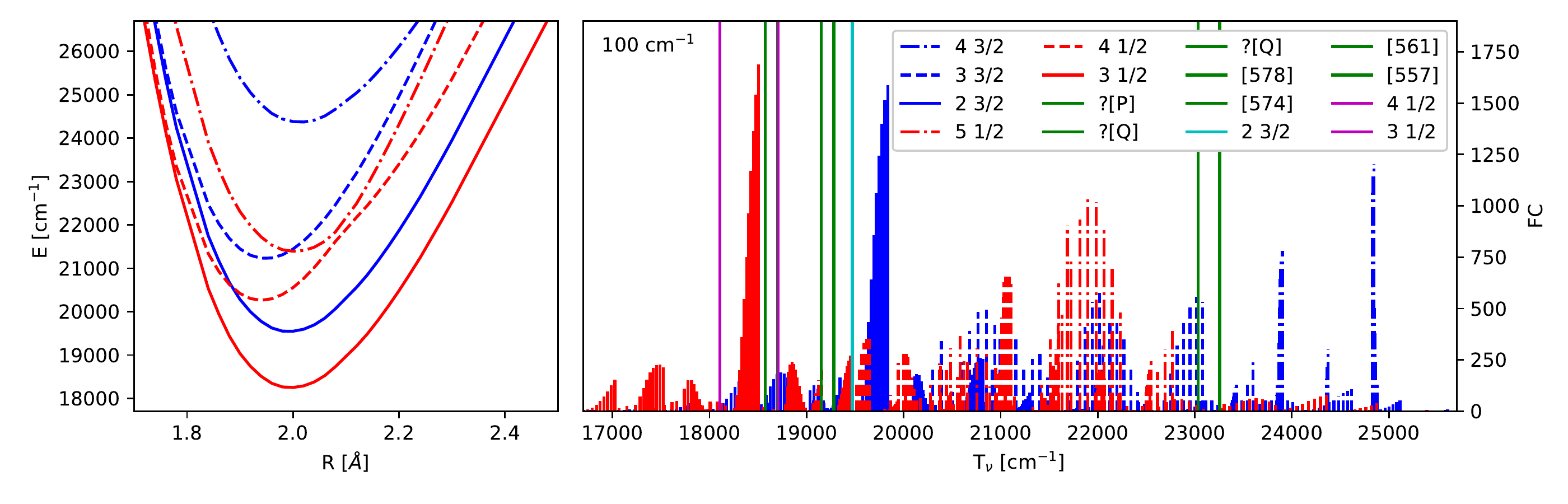} 
\\
\includegraphics[width=0.95\textwidth]{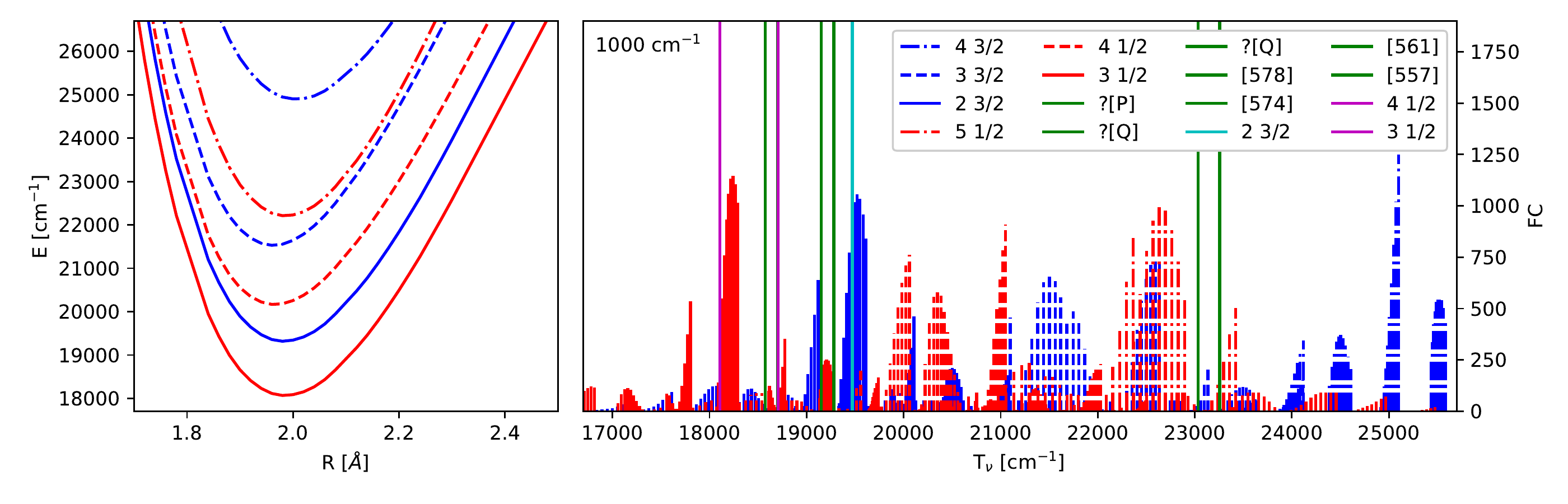}
\caption{Frank-Condon factors before and after adiabatization for the IHFS-CCSD potential energy curves. C is the coupling strength in Hartree. 
The lowest 10 vibrational levels of the ground state were used as well as the lowest 60 vibrational levels of the excited state. The experimental values have been added as straight lines.\cite{Dunfield1995,Lim2017,Smallman2014}}
\label{fig:coupling_strength}
\end{figure*}

In the following we only consider the IHFS-CCSD potential energy curves, as the spectroscopic parameters are more reliable for CCSD than for KRCI. The coupled cluster results for the two methods are quite similar, and FS-CCSD was selected (because it does not include the (2h,1p) and (1h,2p) transitions with rather large uncertainties). 
Figure~\ref{fig:coupling_strength} contains the original FS-CCSD curves as well as the ones obtained after adiabatzation with three different coupling constants. 
Looking at the potential energy curves for this energetic region, there are two $\Omega=1/2$ and two $\Omega=3/2$ states of Yb($4f^{14}6p$)F and Yb($4f^{14}5d$)F configurations originating from the (0h,1p) sector. 
For both $\Omega$ values there is an additional state with a Yb($4f^{13}$[$F_{5/2}^\circ$]$\sigma_{6s}^2$)F configuration stemming from the (1h,0p) sector. 
By looking at the KRCI results one expects additional states belonging to the Yb($4f^{13}$[$F_{7/2}^\circ$]$5d\sigma_{6s}$)F configuration for this energy range, which will not be included in the current considerations. 

As already mentioned earlier the lowest $\Omega=1/2$ and $\Omega=3/2$ states can be identified clearly and assigned to experimental observations. 
There are several experimental states in this energy region attributed to the mixing of states. 
The [557] and [561] ones\cite{Lim2017} are assumed to arise from a mixing of the $3_{1/2}$ and $4_{1/2}$. 
The vibrational constant of the perturbing state ($4_{1/2}$) was estimated to be 605~\wn{} in ref.~\citenum{Dunfield1995}. 
This agrees with the $4_{1/2}$ state in figure~\ref{fig:coupling_strength} with a Yb($4f^{13}$[$F_{5/2}^\circ$]$\sigma_{6s}^2$)F configuration, see also table~\ref{tab:omega_135_2}.
[574] and [578]\cite{Smallman2014} have not been identified and since their $\Omega$ value is unknown, we were not able to assign them to a configuration.

Next we take a look at the changes introduced by adiabatization. 
For small and intermediate coupling strengths there are no major differences in the potential energy curves, although close to the crossing points the potentials are deformed. 
Intermediate coupling strengths with slightly deformed potentials close to the avoided crossings will be the most realistic description. For very large coupling strengths one obtains parallel potential energy curves due to the strong repulsion. This also results in a major change of the spectra above 19000~\wn{}.
One of the differences between the adiabatic spectrum and the upermost one in figure~\ref{fig:coupling_strength} is that the Frank-Condon factors of the $4_{1/2}$ state, which is of the Yb($4f^{13}\sigma_{6s}^2$)F configuration, are now noticeable and the spacing of the energy levels of the  5~$1/2$ is changed. Similarly, transitions belonging to the 5~$1/2$ appear. 

The influence of adiabatization on spectroscopic parameters can be investigated by comparing spectroscopic constants calculated for the IHFS-CCSD curves without and with a coupling of 100~\wn{} (table~\ref{tab:FSCC_compare}). We observe that for this coupling strength, there are small but non-negligible changes for the excitation energies, harmonic frequencies and anharmonicity constants, for all but the fourth and fifth $\Omega = 1/2$ states; there, the coupling does seem to significantly change the anharmonicity constants. Equilibrium distances, on the other hand, are largely unperturbed in all cases. Furthermore, as expected from the preceding discussion, no changes are observed for the ground-state, since it is too separated in energy from the other electronic states.

\begin{table*}[hbtp]
\caption{Spectroscopic data obtained by fitting Morse potentials to the lowest points of the potential energy curves obtained with FSCC for the extrapolated basis set(CBS). This table combines results from both sectors starting either with a closed (f$^{14}$) or open (f$^{13}$) f-shell. Additionally, the table contains spectroscopic parameters after adiabatization with a specific coupling constant (C). 
The transition energy (T$_e$), vibrational constant ($\omega_e$), and anharmonicty constant ($\omega_e$x$_e$) are given in \wn{}, the equilibrium bond distance (r$_e$) in \AA{}.}
\label{tab:FSCC_compare}
\begin{tabular}{ r | r | r | r | r | r | r | r | r | r | r  }
 $\Omega$ &    \multicolumn{5}{c |}{ CBS } & \multicolumn{5}{ c }{C = 100~\wn{} } \\
 \hline
  &  state &  r$_e$    &     $\omega_e$   &   $\omega_e\chi_e$ & T$_e$  
  &  state &  r$_e$    &     $\omega_e$   &   $\omega_e\chi_e$ & T$_e$\\
 \hline
   1/2    &  f$^{14}$ - 1 &      2.018 &         515 &        2.9  &            0  
   &    1 &      2.018 &         515 &        2.8  &            0\\ 
          &  f$^{13}$ - 2 &      1.940 &         599 &        2.8  &         9627  
   &    2 &      1.940 &         599 &        2.8  &         9617\\ 
          &  f$^{14}$ - 2 &      1.995 &         539 &        2.6  &        18249
  &    3 &      1.995 &         538 &        2.6  &        18247\\ 
          &  f$^{13}$ - 3 &      1.940 &         597 &        2.8  &        20267
  &    4 &      1.935 &         603 &        8.6  &        20258 \\ 
          &  f$^{14}$ - 3 &      2.003 &         533 &        2.7  &        21375
  &    5 &      2.002 &         586 &        0.4  &        21359\\ 
          &  f$^{14}$ - 4 &      1.964 &         581 &        1.8  &        31416
  &    6 &      1.964 &         581 &        1.8  &        31419\\ 
\hline
   3/2    &  f$^{13}$ - 1 &      1.944 &         595 &        2.8  &        10180
   &    1 &      1.944 &         594 &        2.8  &        10170 \\ 
          &  f$^{14}$ - 1 &      1.992 &         542 &        2.6  &        19543
  &    2 &      1.992 &         542 &        2.6  &        19540\\ 
          &  f$^{13}$ - 2 &      1.948 &         591 &        2.8  &        21222
  &    3 &      1.948 &         591 &        2.8  &        21217\\ 
          &  f$^{14}$ - 2 &      2.012 &         512 &        2.7  &        24363
  &    4 &      2.012 &         512 &        2.7  &        24369\\ 
\hline
   5/2    &  f$^{13}$ - 1 &      1.949 &         589 &        2.8  &        10967 
   &    1 &      1.949 &         589 &        2.8  &        10960\\ 
          &  f$^{13}$ - 2 &      1.950 &         584 &        2.8  &        22127
  &    2 &      1.950 &         583 &        2.8  &        22117\\ 
          &  f$^{14}$ - 1 &      2.009 &         515 &        2.8  &        24919
  &    3 &      2.009 &         516 &        2.7  &        24926\\ 
\end{tabular}
\end{table*}

\section{Conclusion}

In this manuscript we have presented a study of the ground and excited states of the YbF molecule, with 2-component multireference CI, equation of motion and Fock space coupled cluster approaches (in all cases, performing extrapolations to the complete basis set limit). In particular, we have focused on obtaining electronic states up to around 24000~\wn{} arising from configurations which differ in the occupation of the $4f$ shell ($4f^{14}$ and $4f^{14}$), which are very difficult to treat on the same footing due to a number of subtle correlation and relaxation effects. 

In order to achieve such a balanced description, our strategy consisted of starting from 
YbF$^+$ and YbF$^-$, in order to arrive at the wavefunctions for YbF through the (1h,0p) and (0h,1p) sectors of Fock space. Once obtained, electronic states with same $\Omega$ values coming from these different sectors are further coupled through a simple adiabatization model in which the coupling strength is taken as a constant. 

As a general rule we find that the CI calculations do capture the essential physics of the system, though they are not as reliable as the coupled cluster approaches for excitation energies, bond lengths, harmonic vibrational frequencies and anharmonic constants. In effect, the coupled cluster calculations for the (1h,0p) and (0h,1p) sectors yield the same potential energy curves for the ground state, for internuclear distances up to around 2.8\AA{}, which is sufficient to capture the bound regions of all states under consideration, 

We have determined that the lowest lying excited states arise from the Yb($4f^{13}$[$F_{7/2}^\circ$]$\sigma_{6s}^2$)F configuration, with transition energies of around 10000~\wn{}, and a splitting about 2000~\wn{}.  These states are, however, not generally accessible in experiment due to their low dipolar intensity and significantly shifted minima of the potential energy curve resulting in small Frank-Condon factors. 

The next set of states, coming above 18000~\wn{}, arise from the
Yb($4f^{14}6p$)F, Yb($4f^{14}5d$)F,  Yb($4f^{13}$[$F_{5/2}^\circ$]$\sigma_{6s}^2$)F, and \newline 
Yb($4f^{13}$[$F_{7/2}^\circ$]$5d\sigma_{6s}$)F configurations. 
Among these, the Yb($4f^{13}$[$F_{7/2}^\circ$]$\sigma_{6s}^2$)F configurations generally display the shortest equilibrium distances  and deepest potential well, while the Yb($4f^{14}5d$)F and Yb($4f^{14}6p$)F configurations exhibits the largest bond distances and smallest harmonic frequencies, with the other configurations falling somewhere in between. 
The lowest $\Omega=1/2$ and $\Omega=3/2$ states of this group show a  Yb($4f^{14}6p$)F orbital composition around the ground-state equilibrium structure, though for longer bond lengths they asymptotically approach the Yb($4f^{14}5d$)F configuration. 

We note that configurations with three unpaired electrons, such as Yb($4f^{13}$[$F_{7/2}^\circ$]$5d\sigma_{6s}$)F, were only considered with the KRCI method, which has larger uncertainties. 
This only allows us to make some qualitative statements, e.g. that their bond distances and vibrational constant should be between the values for the other configurations and that they should be higher in energy than the lowes excited Yb($4f^{14}6p$)F and Yb($4f^{13}$[$F_{5/2}^\circ$]$\sigma_{6s}^2$)F states.

A simple method was applied in order to adiabatize the curves obtained for different sectors and reference wave functions. 
It was applied to potential energy curves between 18000~and 26000~\wn{} and small changes of the Franck-Condon factors were observed. The influence on spectroscopic constant was minor, with the exception of the asymmetry constant for two states. 
However, the approximation introduced (same coupling strength for all states and all geometries) is perhaps not flexible enough, and more sophisticated models should be investigated.

\section{Acknowledgements}

LV and JVP wish to thank The Netherlands Organization for Scientific Research (NWO) for financial support via the ECHO and computer time. JVP acknowledges funding by the Austrian Science Fund(FWF):J 4177-N36.
ASPG acknowledges support from PIA ANR project CaPPA (ANR-11-LABX-0005-01), the Franco-German project CompRIXS (Agence nationale de la recherche ANR-19-CE29-0019, Deutsche Forschungsgemeinschaft JA 2329/6-1), I-SITE ULNE projects OVERSEE, the French Ministry of Higher Education and Research, region Hauts de France council and European Regional Development Fund (ERDF) project CPER CLIMIBIO, and the French national supercomputing facilities (grants DARI A0070801859 and A0090801859). ASPG, LH, JP and LV acknowledge support from MESONM International Associated Laboratory (LAI) (ANR-16-IDEX-0004).

\bibliography{ybf}

\end{document}


\def\wn{~cm$^{-1}$}

This is the supporting information for the manuscript entitled: "Electronic Spectra of Ytterbium Fluoride from Relativistic Electronic Structure Calculations" by J. V. Pototschnig, K. G. Dyall, L. Visscher and A. S. P. Gomes.

In section~\ref{sec:orbitals} a closer look is taken at the orbitals and their energies for the two different references, either with a closed or open f-shell.
Orbital energies for different diatomic distances are depicted alongside orbitals for the Yb atom and the YbF diatomic(for two internuclear separations). 
Section~\ref{sec:KRCI} contains additional results obtained for Kramers restricted configuration interaction. 
Firstly, we compare results obtained with the X2C-AMFI code to full 4-component computations. 
Next, potential energy curves, including spectroscopic parameters, and dipole moments--both transition dipole moments (TDM) and permanent electric dipole moments (PEDM)--for different basis set sizes are presented. Subsequently, the problem of the relative position of the f$^{13}$ and f$^{14}$ states is addressed and its basis set dependence. Lastly, 
transition dipole moments are listed. 
Section~\ref{sec:FSCC} contains the potential energy curves and spectroscopic parameters for different basis set sizes applying the Fock space coupled cluster method and spectroscopic parameters for the states after adiabatic mixing as presented in the main text. 
An extended list of equation-of-motion coupled cluster transition energies for the Yb$^+$ is listed in section~\ref{sec:EOMCC}. This section also contains EOM-CCSD potential energy curves for different basis set sizes. 
Finally, in section~\ref{sec:Spectra}  Franck-Condon factors for potentials stemming from the different methods are depicted and compared.

\section{Orbitals}
\label{sec:orbitals}

\subsection{Orbital energies}

In this part the orbital energies obtained by AOC-SCF for several internuclear separations are shown. 

\begin{figure}[hbp]
\centering
\includegraphics[width=0.9\textwidth]{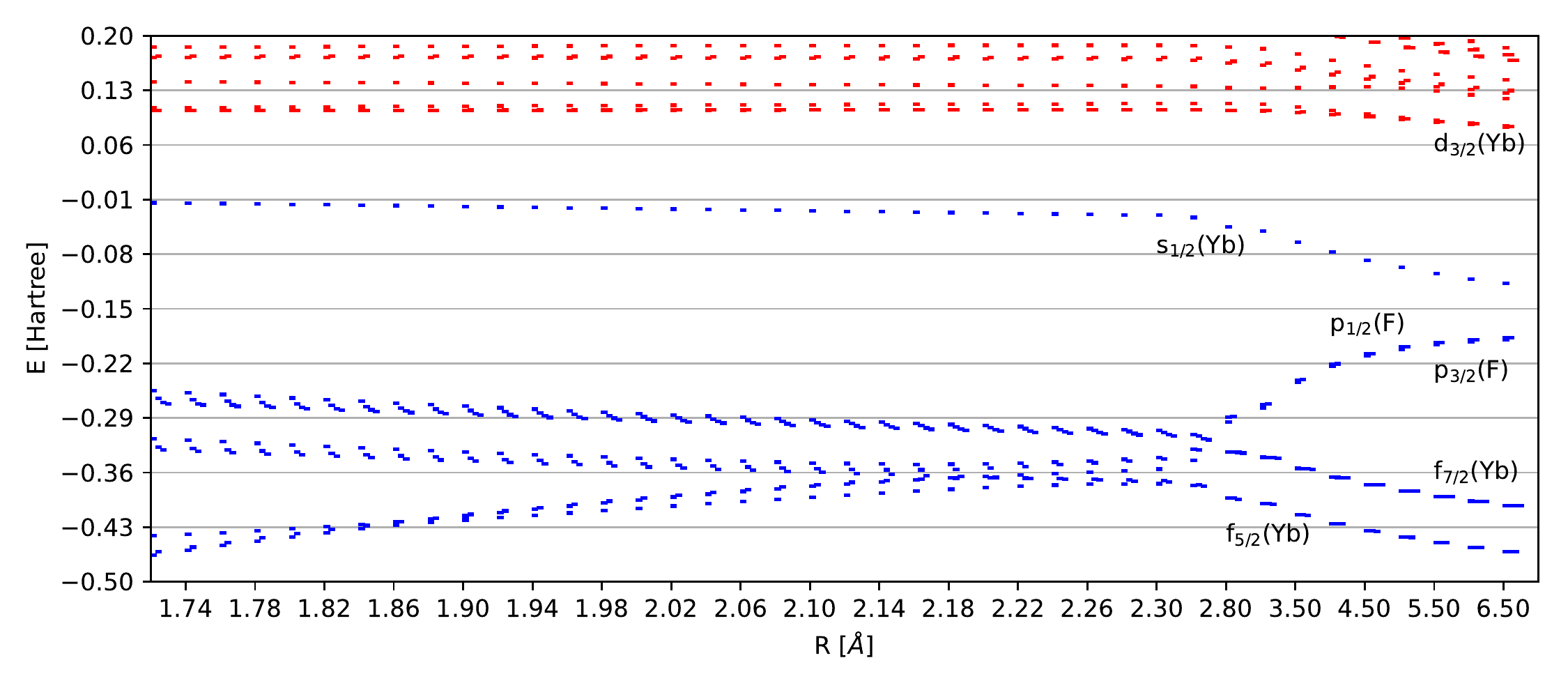}
\caption{Orbital energies of YbF$^-$ for different radii. Blue and red are occupied and virtual orbitals, respectively. }
\label{fig:ORB_ANION4_NEW}
\end{figure}

\begin{figure}[hbtp]
\centering
\includegraphics[width=0.9\textwidth]{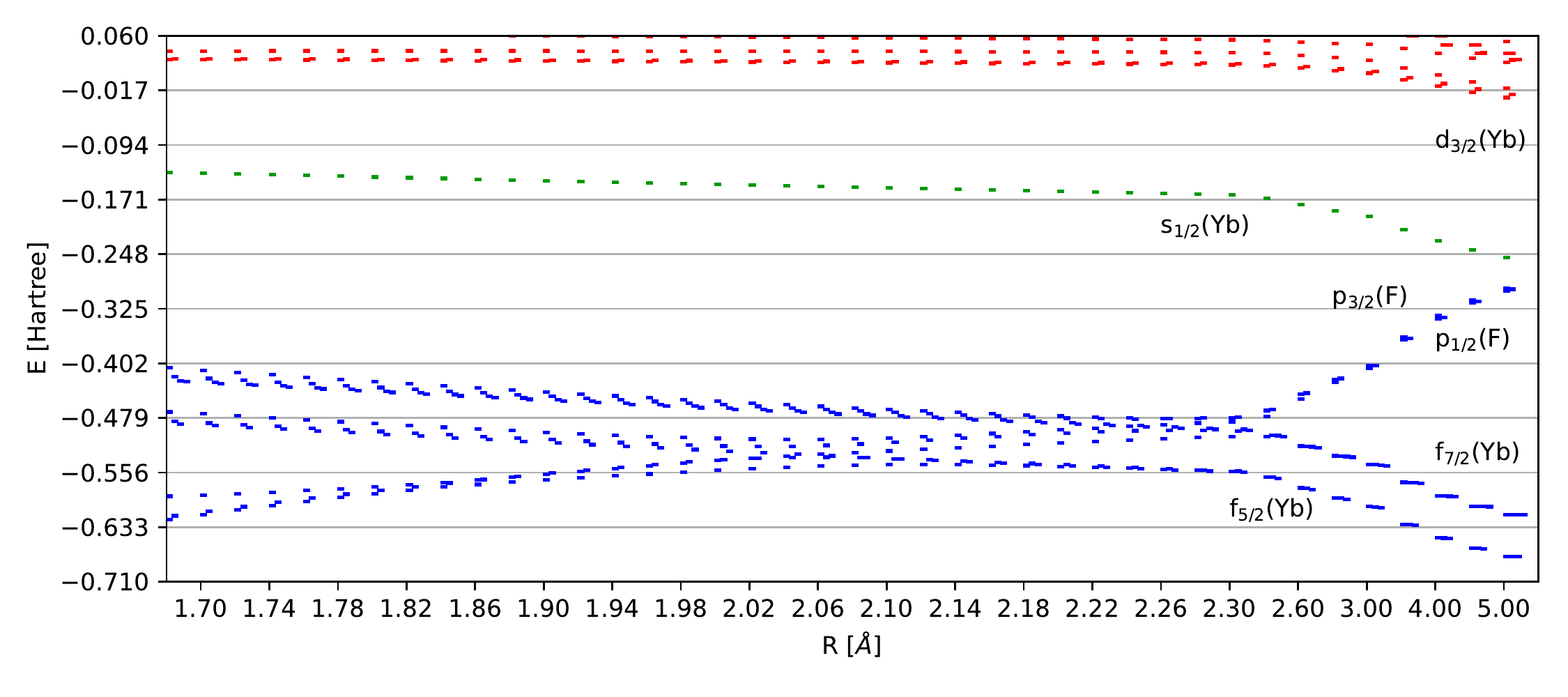}
\caption{Orbital energies of YbF for different radii. Blue green, and red are occupied, partially filled and virtual orbitals, respectively. }
\label{fig:ORB_SCF_v3z}
\end{figure}

\begin{figure}[hbtp]
\centering
\includegraphics[width=0.9\textwidth]{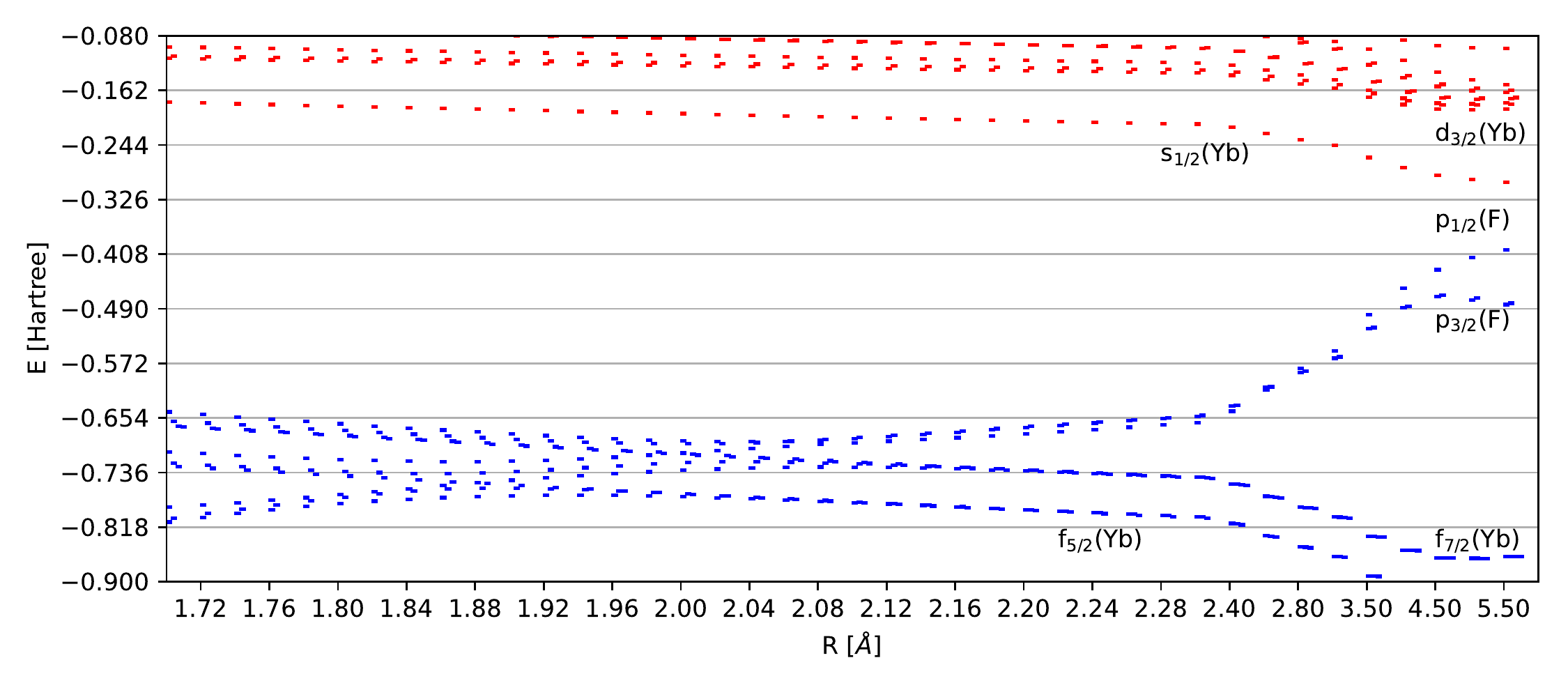}
\caption{Orbital energies of YbF$^+$ for different radii. Blue and red are occupied and virtual orbitals, respectively.}
\label{fig:ORB_CATION4_NEW}
\end{figure}

\clearpage

\subsection{Atomic orbitals}

In this section we compare atomic orbitals that were either optimized for the f$^{13}$ or f$^{14}$ configurations. 

\begin{longtable}{ m{3.0cm} m{5.5cm} m{5.5cm}}
\caption{Orbitals obtained from an AOC-SCF computations for the Yb atom either with a fully occupied (f$^{14}$) or single hole (f$^{13}$) f-shell, the total energy in the first case is -14062.55463420 Hartree, for the second one it is -14062.50155764 Hartree. For each orbital we list the orbital energy according to a Koopmans definition.\cite{Thyssen2001, Cox1975}}
\label{tab:orbitals_atom}
\\
\hline
designation & f$^{14}$ & f$^{13}$ \\
 \hline
\endfirsthead
\multicolumn{3}{c}%
{\tablename\ \thetable\ -- \textit{Continued from previous page}} \\
\hline
designation & f$^{14}$ & f$^{13}$ \\
 \hline
\endhead
\hline \multicolumn{3}{c}{\textit{Continued on next page}} \\
\endfoot
\hline
\endlastfoot
1/15 \newline $5s_{1/2}$(Yb)
& -2.658500494 \newline \includegraphics[width=4.9cm]{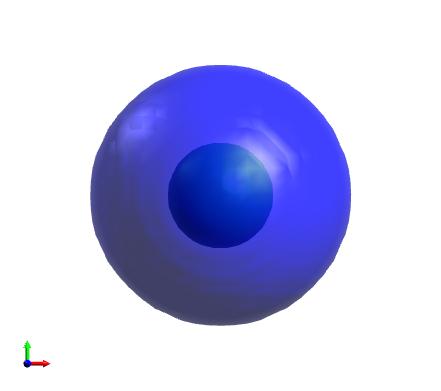}
&  -2.896356031 \newline \includegraphics[width=4.9cm]{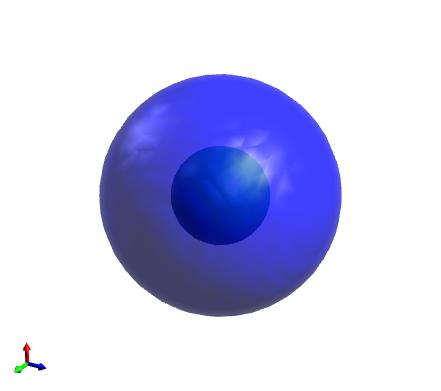} \\
\hline
2/10 \newline $5p_{1/2}$(Yb) &
-1.646207984  \newline \includegraphics[width=4.9cm]{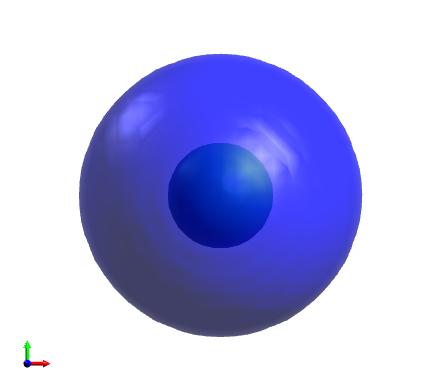} &
-1.856152194 \newline \includegraphics[width=4.9cm]{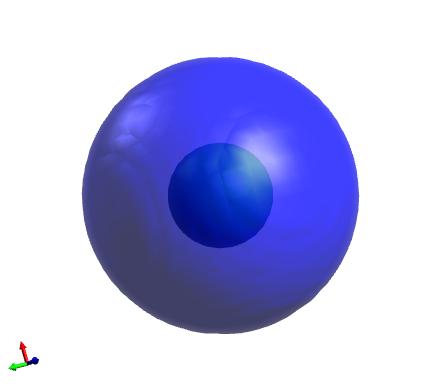} \\
\\
\hline
2/11 \newline $5p_{3/2}$(Yb) &
-1.410261382 \newline \includegraphics[width=4.9cm]{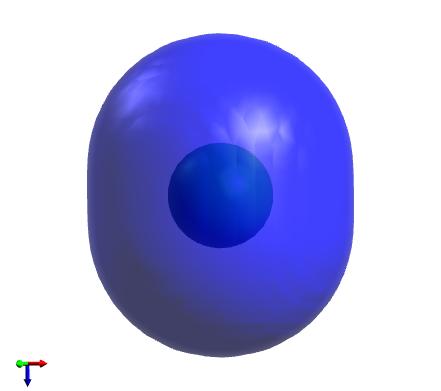} & 
-1.604023002 \newline 
 \includegraphics[width=4.9cm]{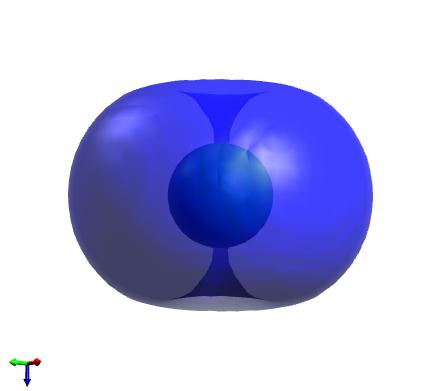}
\\
\hline
2/12 \newline $5p_{3/2}$(Yb)  
& -1.410261382 \newline \includegraphics[width=4.9cm]{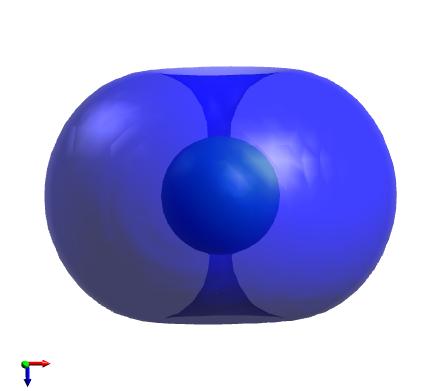} 
& -1.604023002 \newline \includegraphics[width=4.9cm]{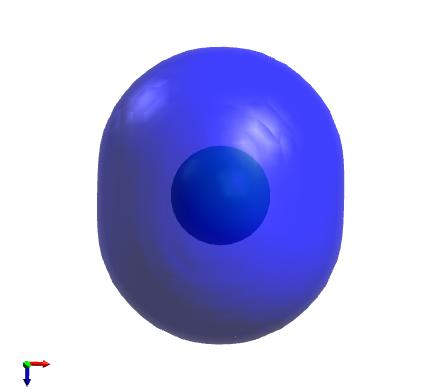} \\
\hline
2/13 \newline $4f_{5/2, 7/2}$(Yb) &
-0.767646169 \newline \includegraphics[width=4.9cm]{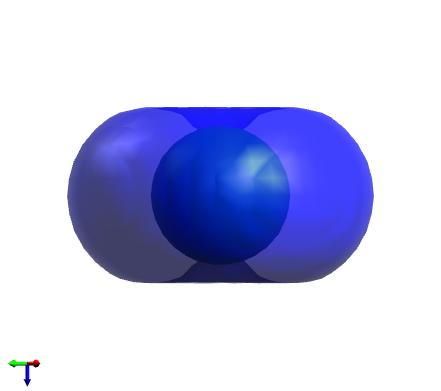} & -1.068583089 \newline \includegraphics[width=4.9cm]{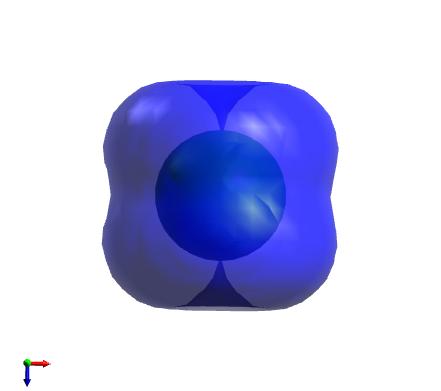} \\
\hline
2/14 \newline $4f_{5/2, 7/2}$(Yb) & 
-0.767646168 \newline \includegraphics[width=4.9cm]{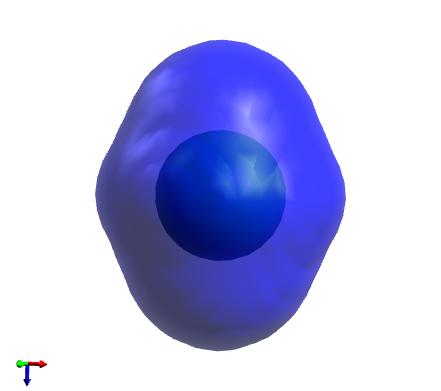} &
-1.068583087 \newline  \includegraphics[width=4.9cm]{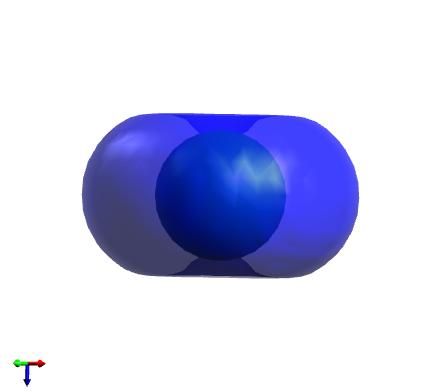} \\
\hline
2/15 \newline $4f_{5/2, 7/2}$(Yb) & 
-0.767646168 \newline \includegraphics[width=4.9cm]{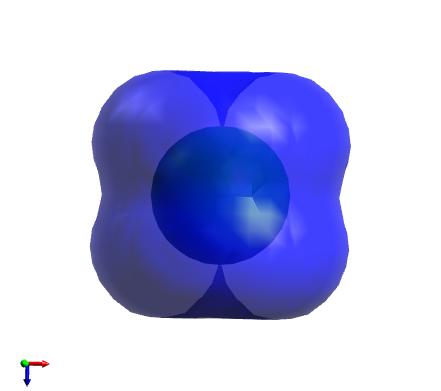} &
-1.068583087 \newline \includegraphics[width=4.9cm]{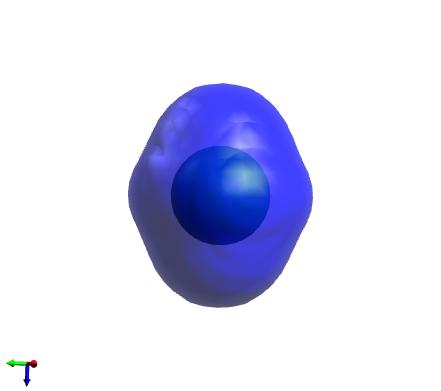} \\
\hline
2/16 \newline $4f_{5/2, 7/2}$(Yb) & 
-0.708324589 \newline \includegraphics[width=4.9cm]{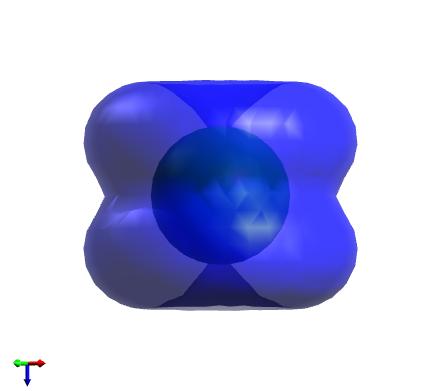} &
-1.010576973 \newline \includegraphics[width=4.9cm]{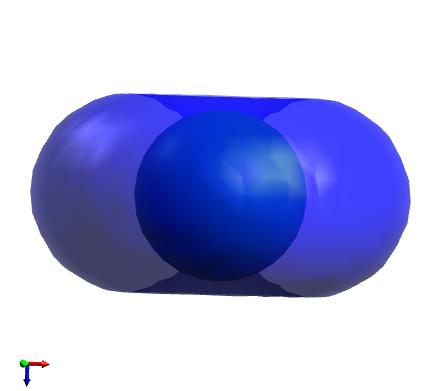} \\
\hline
2/17 \newline $4f_{5/2, 7/2}$(Yb)  & 
 -0.708324585 \newline \includegraphics[width=4.9cm]{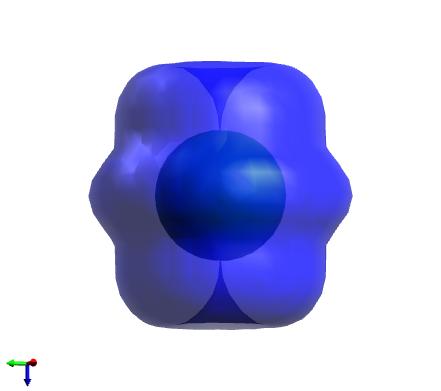} &
 -1.010576972 \newline \includegraphics[width=4.9cm]{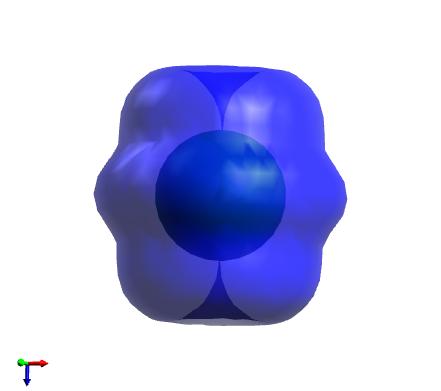} \\
\hline
2/18 \newline $4f_{5/2, 7/2}$(Yb) & 
 -0.708324585 \newline \includegraphics[width=4.9cm]{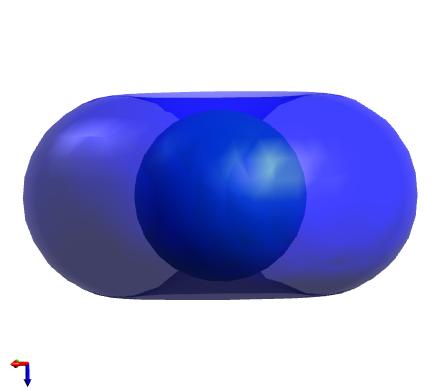} &
 -1.010576972 \newline  \includegraphics[width=4.9cm]{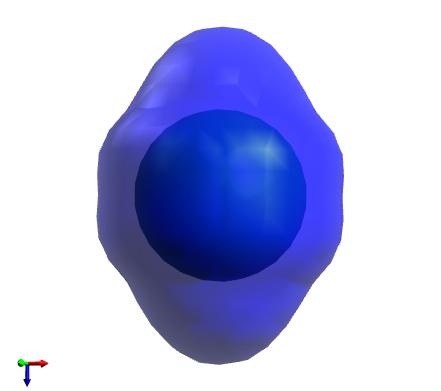} \\
\hline
2/18 \newline $4f_{5/2, 7/2}$(Yb) & 
 -0.708324583 \newline \includegraphics[width=4.9cm]{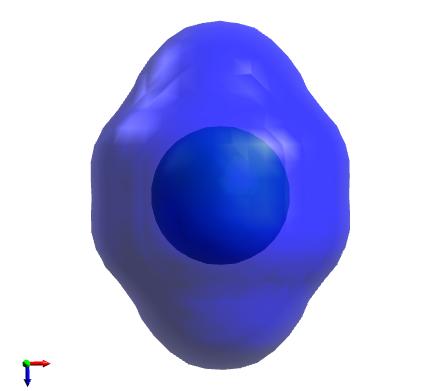} &
 -1.010576968 \newline \includegraphics[width=4.9cm]{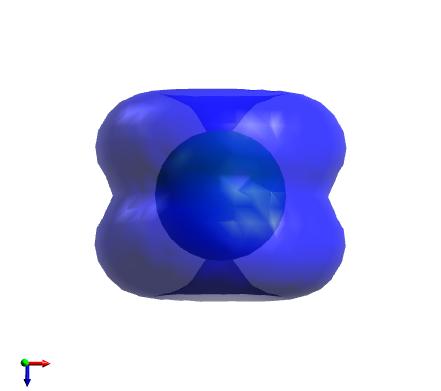} \\
\hline
1/16 \newline $5s_{1/2}$(Yb) & 
 -0.310402947 \newline \includegraphics[width=4.9cm]{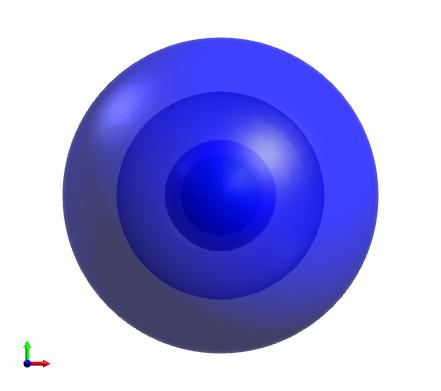} &
 -0.364220506 \newline \includegraphics[width=4.9cm]{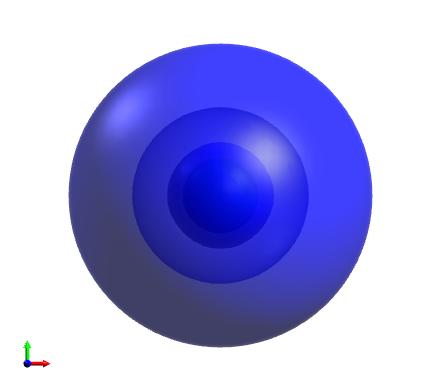} \\
\hline
\end{longtable}

\clearpage

\subsection{Diatomic orbitals - equilibrium distance}

This section contains orbitals of the YbF diatomic for a internuclear distance close to the ground state equilibrium one. 

\begin{longtable}{ m{2.5cm} m{5.5cm} m{5.5cm}}
\caption{Orbitals obtained from an AOC-SCF computations either with a fully occupied (f$^{14}$) or single hole (f$^{13}$) f-shell for YbF with a distance of 2~\AA{}.For each orbital we list the orbital energy according to a Koopman's definition.\cite{Thyssen2001, Cox1975} The contributing atomic orbitals contributing have been given in the first column.}
\label{tab:orbitals_short}
\\
\hline
 designation & f$^{14}$ &  f$^{13}$ \\
 \hline
\endfirsthead
\multicolumn{3}{c}%
{\tablename\ \thetable\ -- \textit{Continued from previous page}} \\
\hline
 designation & f$^{14}$ & f$^{13}$ \\
 \hline
\endhead
\hline \multicolumn{3}{c}{\textit{Continued on next page}} \\
\endfoot
\hline
\endlastfoot
25 \newline $5s_{1/2}$(Yb) &
-2.475129022 \newline \includegraphics[width=4.5cm]{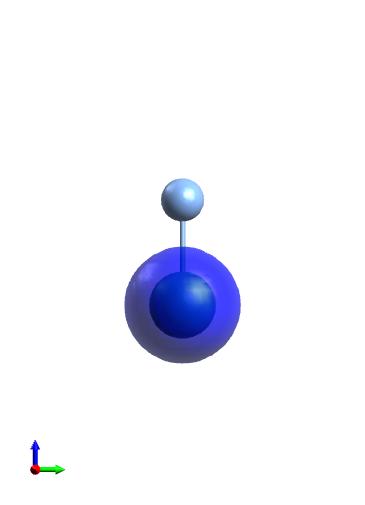} & 
-2.606565796 \newline \includegraphics[width=4cm]{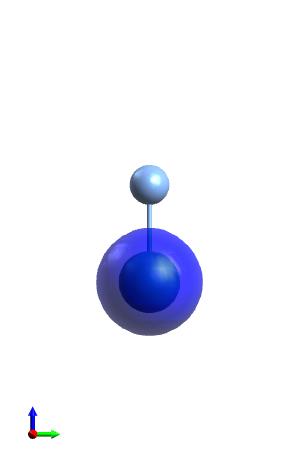}
\\
\hline
26 \newline $5p_{1/2}$(Yb) \newline $2s_{1/2}$(F) & 
-1.511889433 \includegraphics[width=4cm]{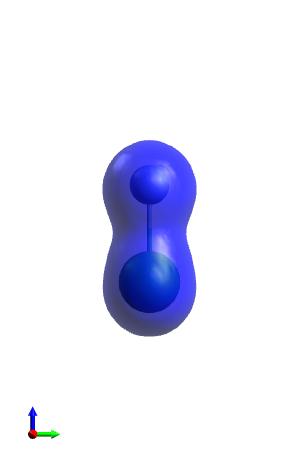} & 
-1.611193447 \includegraphics[width=4cm]{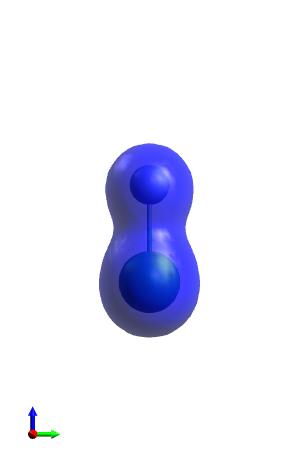}
\\
\hline
27 \newline $5p_{1/2}$(Yb) \newline $2s_{1/2}$(F) & 
-1.435750686 \includegraphics[width=4cm]{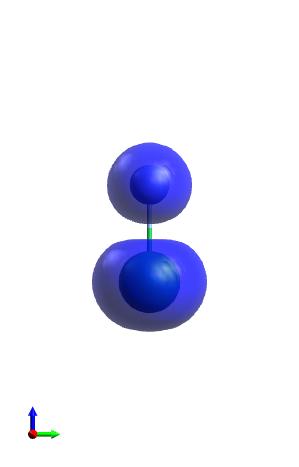} & 
-1.539604687 \includegraphics[width=4cm]{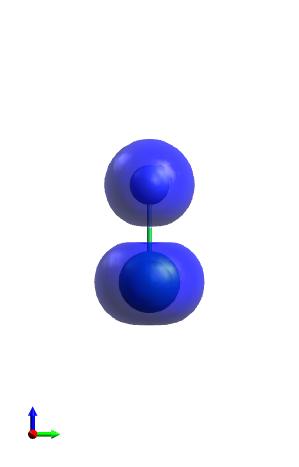}
\\
\hline
28 \newline $5p_{3/2}$(Yb)  & 
-1.229656975 \includegraphics[width=4cm]{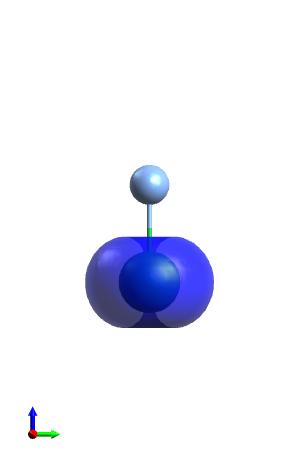}  &
-1.323836039 \includegraphics[width=4cm]{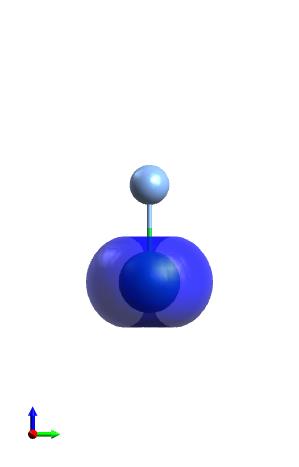}
\\
\hline
29 \newline $5p_{3/2}$(Yb)  & 
-1.210630046 \includegraphics[width=4cm]{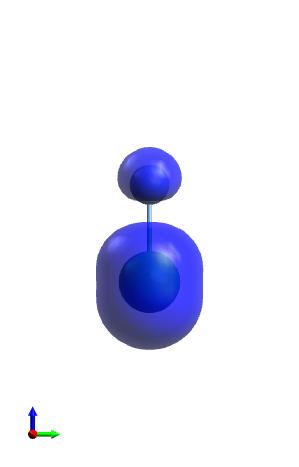} &  
-1.311522305 \includegraphics[width=4cm]{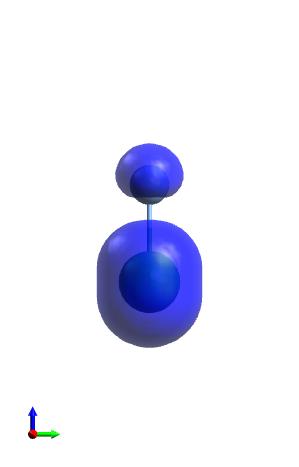}
\\
\hline
30 \newline $4f_{5/2, 7/2}$(Yb) \newline $2p_{1/2, 3/2}$(F) & 
-0.594742161 \includegraphics[width=4cm]{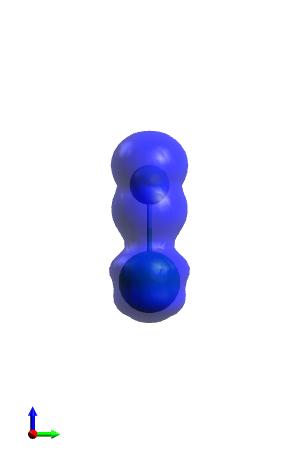} & 
-0.774346397 \includegraphics[width=4cm]{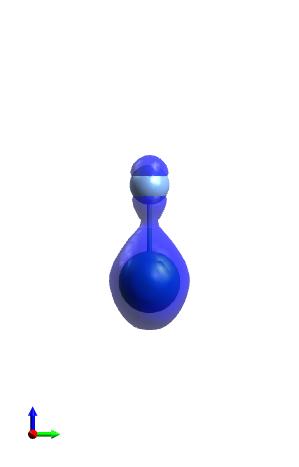}
\\
\hline
31 \newline $4f_{5/2, 7/2}$(Yb) \newline $2p_{1/2, 3/2}$(F) & 
-0.584034308 \includegraphics[width=4cm]{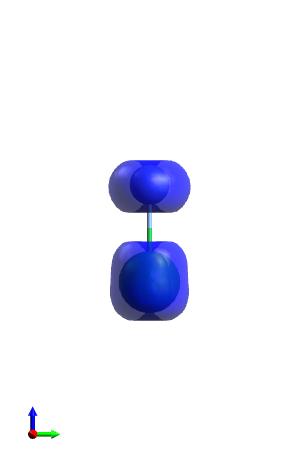} &
-0.773626615 \includegraphics[width=4cm]{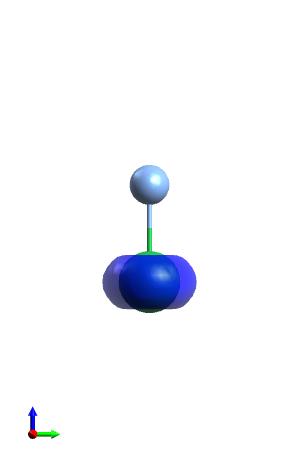}
\\
\hline
32 \newline $4f_{5/2, 7/2}$(Yb) \newline $2p_{1/2, 3/2}$(F) &
-0.582311174 \includegraphics[width=4cm]{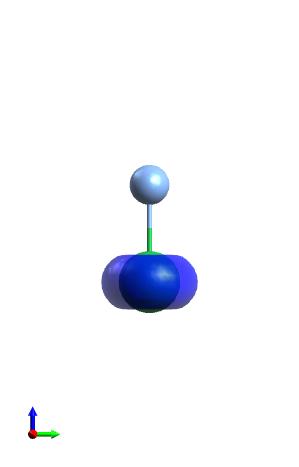} & 
-0.772523010 \includegraphics[width=4cm]{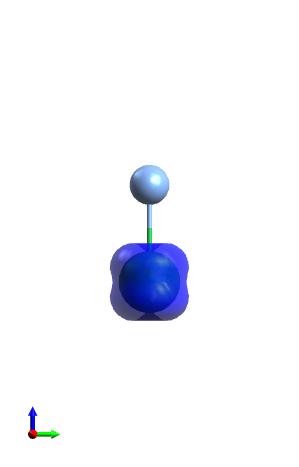}
\\
\hline
33 \newline $4f_{5/2, 7/2}$(Yb) \newline $2p_{1/2, 3/2}$(F) & 
-0.574792083 \includegraphics[width=4cm]{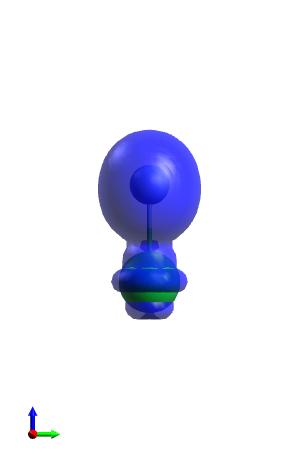} & 
-0.719689709 \includegraphics[width=4cm]{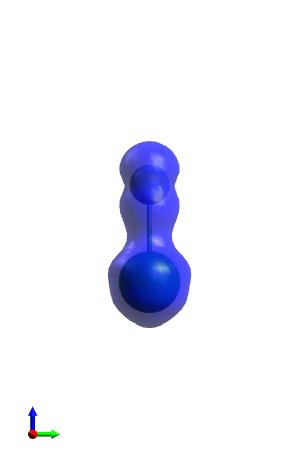}
\\
\hline
34 \newline $4f_{5/2, 7/2}$(Yb) \newline $2p_{1/2, 3/2}$(F) & 
-0.561976545  \includegraphics[width=4cm]{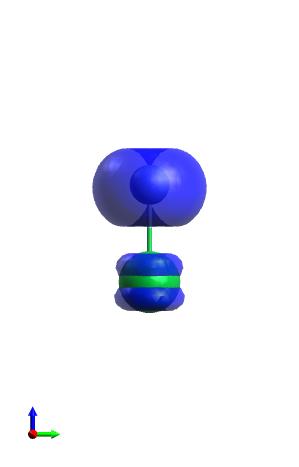} & 
-0.715455417 \includegraphics[width=4cm]{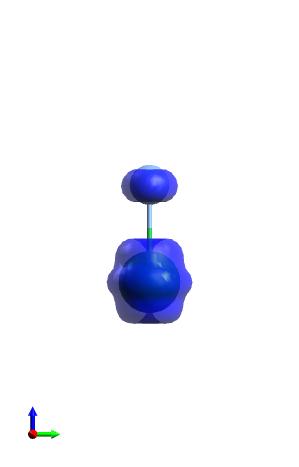}
\\
\hline
\hline
35 \newline $4f_{5/2, 7/2}$(Yb) \newline $2p_{1/2, 3/2}$(F) &
-0.55211333 \includegraphics[width=4cm]{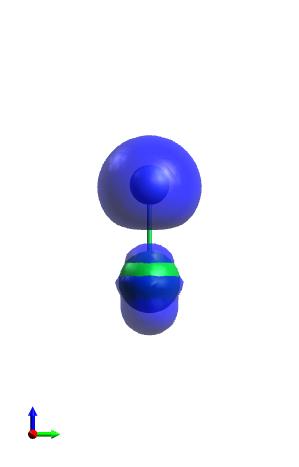} & 
-0.715038003 \includegraphics[width=4cm]{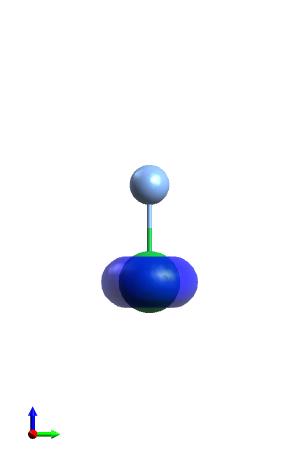}
\\
\hline
\hline
36 \newline $4f_{5/2, 7/2}$(Yb) \newline $2p_{1/2, 3/2}$(F) &
-0.523207455 \includegraphics[width=4cm]{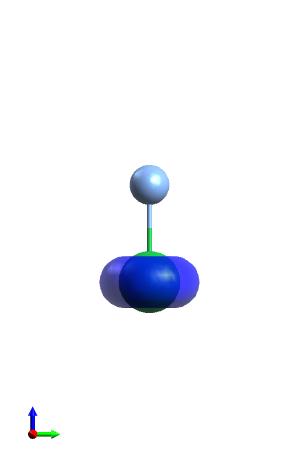} & 
-0.713718461 \includegraphics[width=4cm]{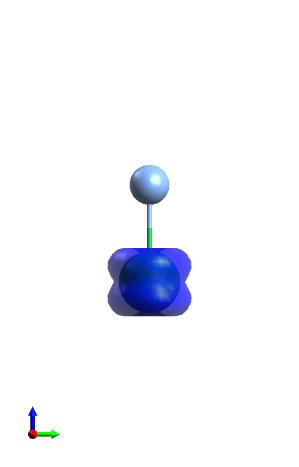}
\\
\hline
37 \newline $4f_{5/2, 7/2}$(Yb) \newline $2p_{1/2, 3/2}$(F) &
-0.521323830 \includegraphics[width=4cm]{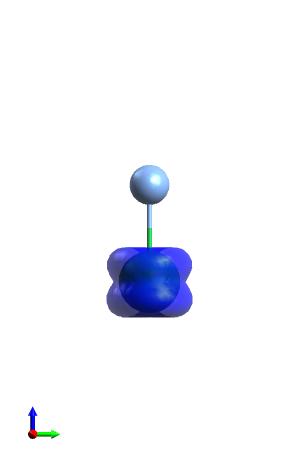} & 
-0.653507491 \includegraphics[width=4cm]{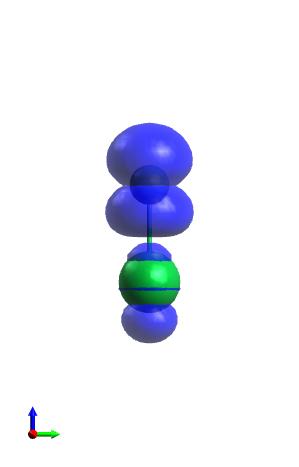}
\\
\hline
38 \newline $4f_{5/2, 7/2}$(Yb) \newline $2p_{1/2, 3/2}$(F) &
-0.515413462 \includegraphics[width=4cm]{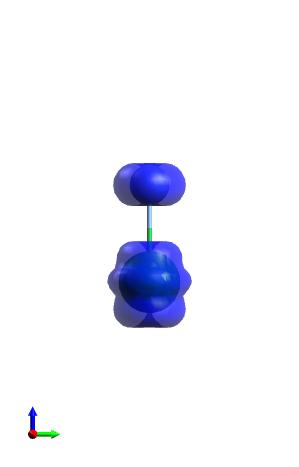} & 
-0.644769626 \includegraphics[width=4cm]{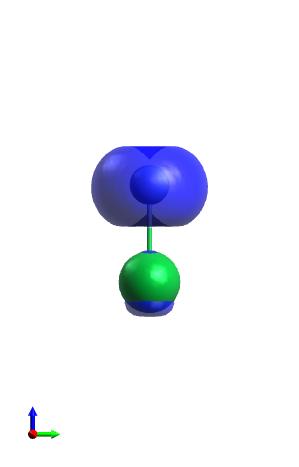}
\\
\hline
39 \newline $4f_{5/2, 7/2}$(Yb) \newline $2p_{1/2, 3/2}$(F) &
-0.509924741 \includegraphics[width=4cm]{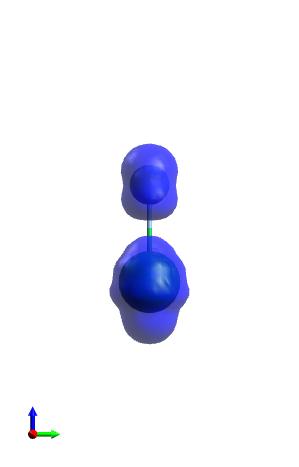} & 
-0.639112727 \includegraphics[width=4cm]{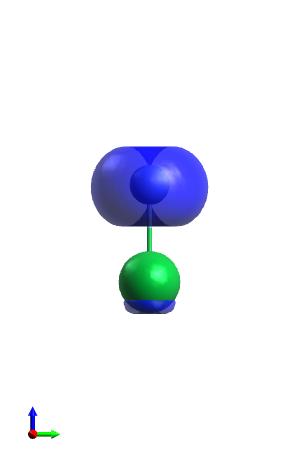}
\\
\hline
40 \newline $4f_{5/2, 7/2}$(Yb) \newline $2p_{1/2, 3/2}$(F) &
-0.129066590 \includegraphics[width=4cm]{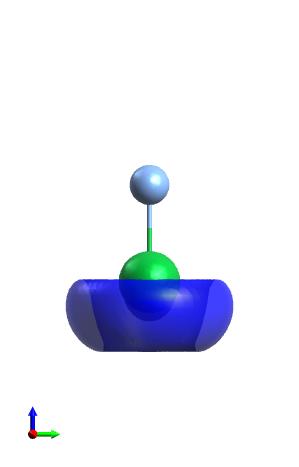} & 
-0.241206568 \includegraphics[width=4cm]{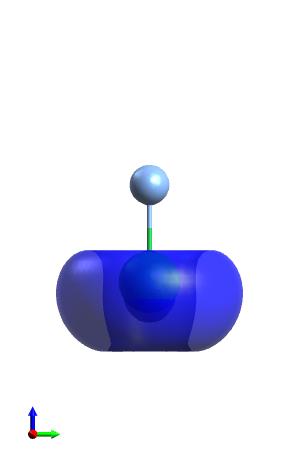}
\\
\hline
\end{longtable}

\clearpage

\subsection{Diatomic orbitals - long range}

This section contains orbitals of the YbF diatomic for a internuclear distance significantly larger than the minima of most potential energy curves. 

\begin{longtable}{ m{2.5cm} m{5.5cm}  m{5.5cm}}
\caption{Orbitals obtained from an AOC-SCF computations either with a fully occupied (f$^{14}$) or single hole (f$^{13}$) f-shell for YbF with a distance of 3~\AA{}. The contributing atomic orbitals contributing have been given in the first column.}
\label{tab:orbitals_long}
\\
\hline
 designation & f$^{14}$ & f$^{13}$ \\
 \hline
\endfirsthead
\multicolumn{3}{c}%
{\tablename\ \thetable\ -- \textit{Continued from previous page}} \\
\hline
 designation & f$^{14}$ (3.0) & f$^{13}$ (3.0) \\
 \hline
\endhead
\hline \multicolumn{3}{c}{\textit{Continued on next page}} \\
\endfoot
\hline
\endlastfoot
25 \newline $5s_{1/2}$(Yb) & \includegraphics[width=4cm]{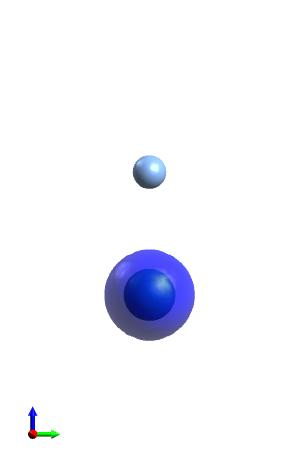} &
\includegraphics[width=4cm]{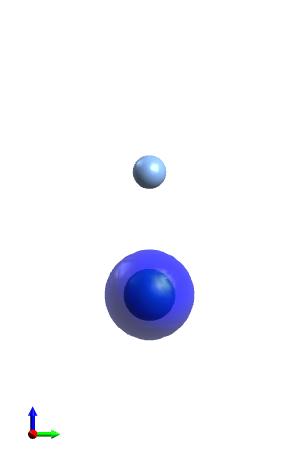}
\\
\hline
26 \newline $5p_{1/2}$(Yb) \newline $2s_{1/2}$(F) & 
\includegraphics[width=4cm]{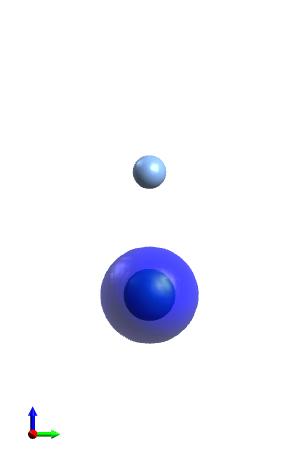} &
\includegraphics[width=4cm]{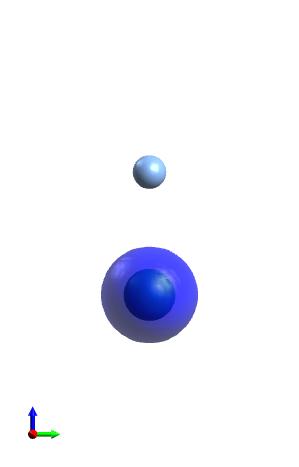}
\\
\hline
27 \newline $5p_{1/2}$(Yb) \newline $2s_{1/2}$(F) & 
\includegraphics[width=4cm]{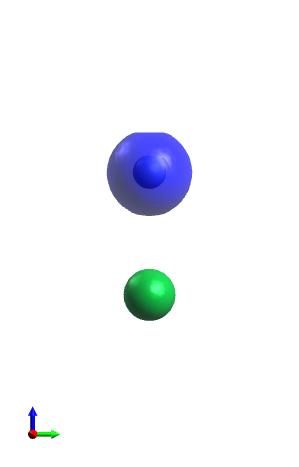} &
\includegraphics[width=4cm]{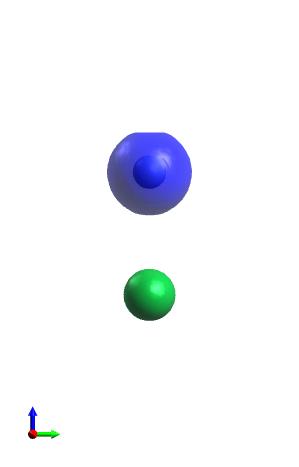}
\\
\hline
28 \newline $5p_{3/2}$(Yb)  &  
\includegraphics[width=4cm]{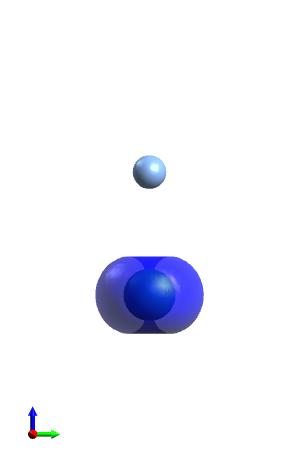} &
\includegraphics[width=4cm]{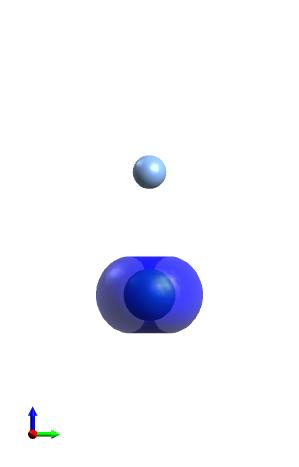}
\\
\hline
29 \newline $5p_{3/2}$(Yb)  & 
\includegraphics[width=4cm]{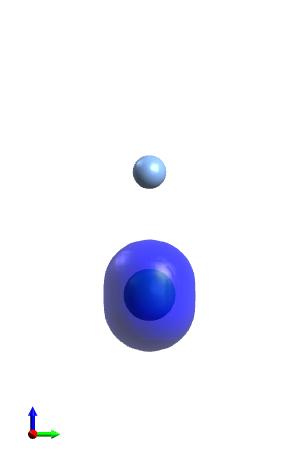} &
\includegraphics[width=4cm]{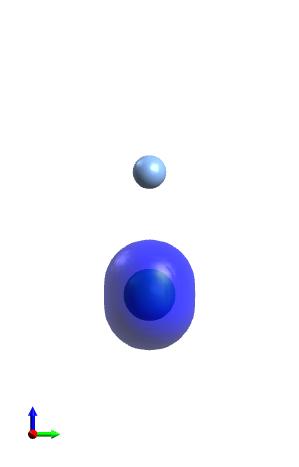}
\\
\hline
30 \newline $4f_{5/2, 7/2}$(Yb) \newline $2p_{1/2, 3/2}$(F) & 
\includegraphics[width=4cm]{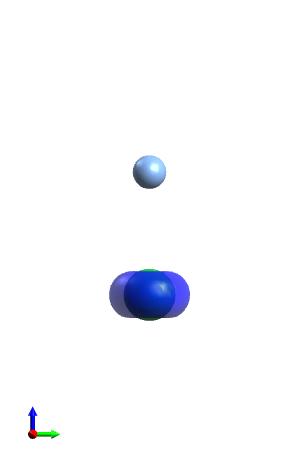} &
\includegraphics[width=4cm]{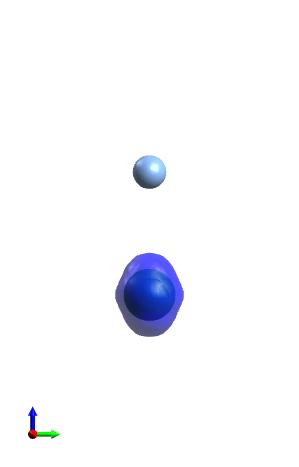}
\\
\hline
31 \newline $4f_{5/2, 7/2}$(Yb) \newline $2p_{1/2, 3/2}$(F) & 
\includegraphics[width=4cm]{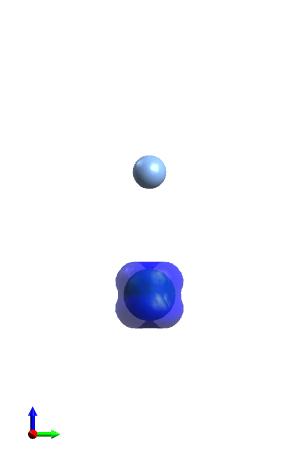} &
\includegraphics[width=4cm]{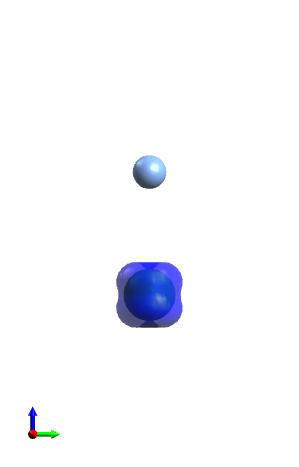}
\\
\hline
32 \newline $4f_{5/2, 7/2}$(Yb) \newline $2p_{1/2, 3/2}$(F) &
\includegraphics[width=4cm]{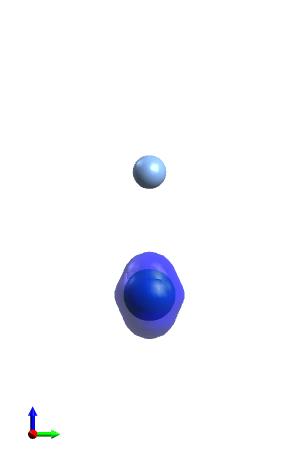} &
\includegraphics[width=4cm]{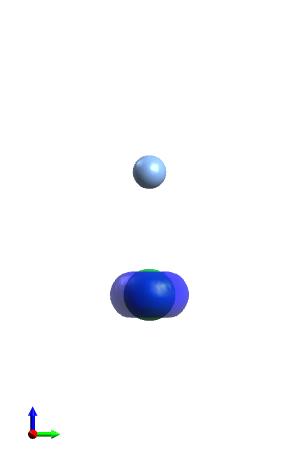}
\\
\hline
33 \newline $4f_{5/2, 7/2}$(Yb) \newline $2p_{1/2, 3/2}$(F) & 
\includegraphics[width=4cm]{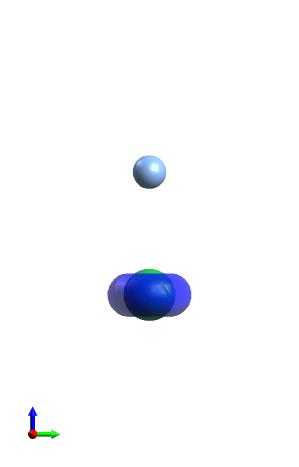} &
\includegraphics[width=4cm]{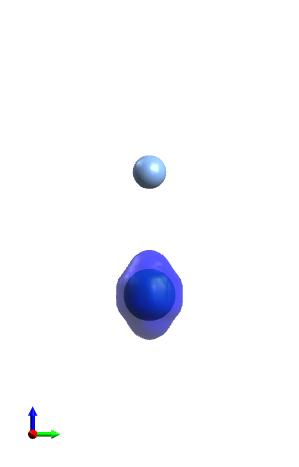}
\\
\hline
34 \newline $4f_{5/2, 7/2}$(Yb) \newline $2p_{1/2, 3/2}$(F) & 
\includegraphics[width=4cm]{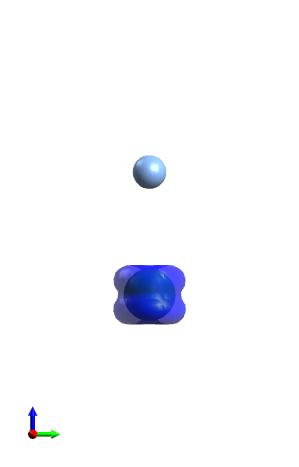} &
\includegraphics[width=4cm]{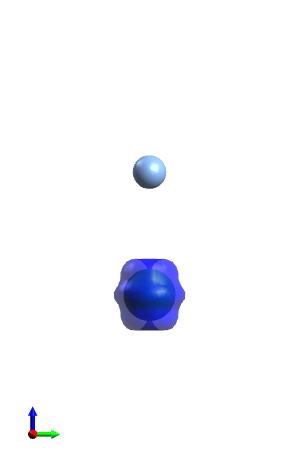}
\\
\hline
\hline
35 \newline $4f_{5/2, 7/2}$(Yb) \newline $2p_{1/2, 3/2}$(F) &
\includegraphics[width=4cm]{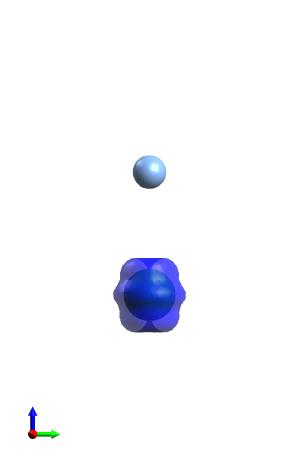} &
\includegraphics[width=4cm]{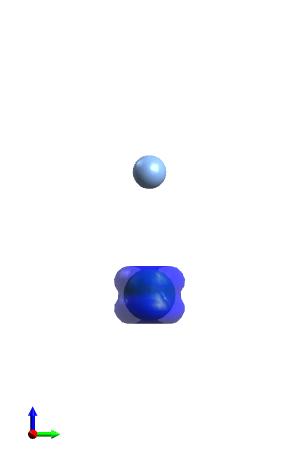}
\\
\hline
\hline
36 \newline $4f_{5/2, 7/2}$(Yb) \newline $2p_{1/2, 3/2}$(F) &
\includegraphics[width=4cm]{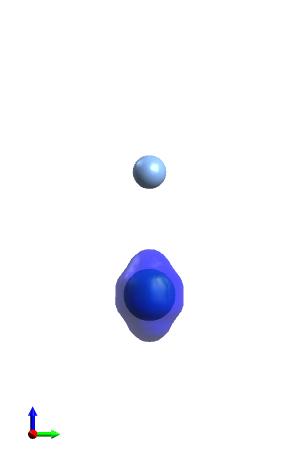} &
\includegraphics[width=4cm]{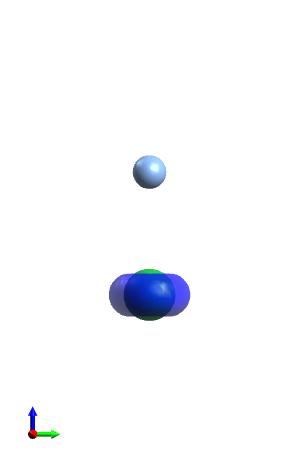}
\\
\hline
37 \newline $4f_{5/2, 7/2}$(Yb) \newline $2p_{1/2, 3/2}$(F) &
\includegraphics[width=4cm]{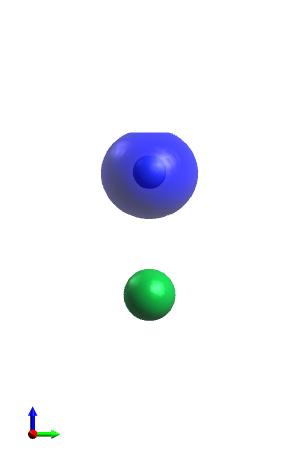} &
\includegraphics[width=4cm]{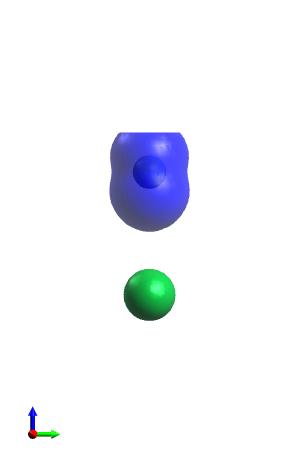}
\\
\hline
38 \newline $4f_{5/2, 7/2}$(Yb) \newline $2p_{1/2, 3/2}$(F) &
\includegraphics[width=4cm]{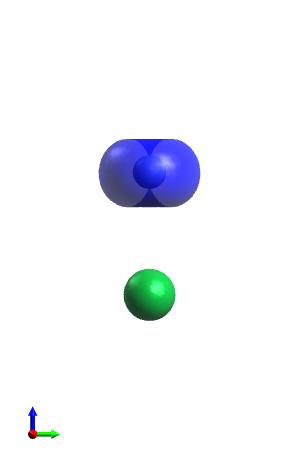} &
\includegraphics[width=4cm]{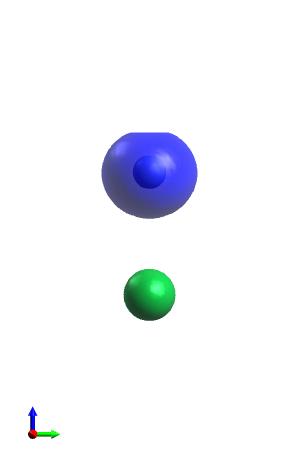}
\\
\hline
39 \newline $4f_{5/2, 7/2}$(Yb) \newline $2p_{1/2, 3/2}$(F) &
\includegraphics[width=4cm]{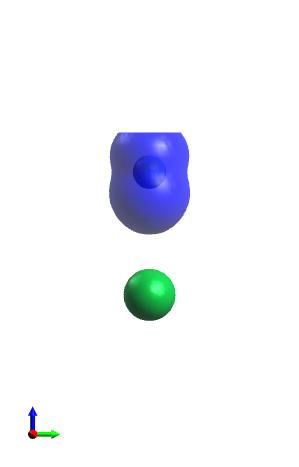} &
\includegraphics[width=4cm]{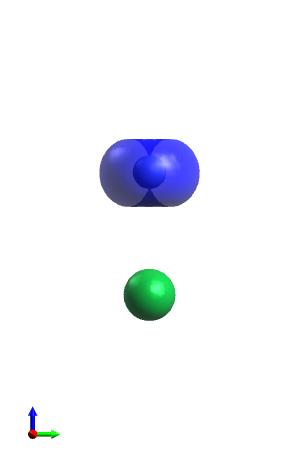}
\\
\hline
40 \newline $4f_{5/2, 7/2}$(Yb) \newline $2p_{1/2, 3/2}$(F) &
\includegraphics[width=4cm]{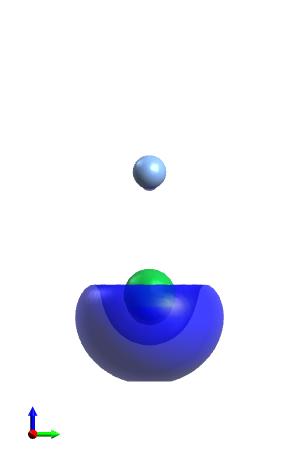} &
\includegraphics[width=4cm]{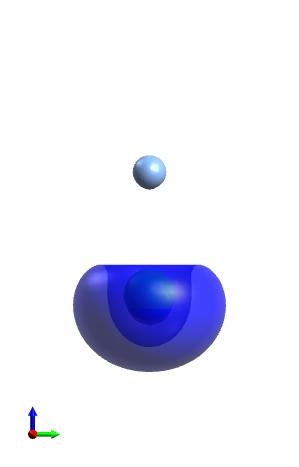}
\\
\hline
\end{longtable}

\clearpage

\section{Kramers restricted configuration interaction (KRCI)}
\label{sec:KRCI}

\subsection{Hamiltonian}

Due to limitations in the implementation it is not possible to run the more accurate ${^2}$DC$^{M}$\cite{Sikkema2009} in combination with Kramers-restricted configuration interaction. 
Therefore, we have run the computations with the X2C-AMFI\cite{Ilias2007,prog:amfi} Hamiltonian. In order to estimate the influence of this approximation we compare the results with the 4-component Dirac-Coulomb-Breit Hamiltonian (DCB) from the literature~\cite{Eliav1995}. in figure in tables~\ref{tab:Yb_cation_KRCI_lv} and \ref{tab:Yb_cation_hole_all} for the Yb cation. 
\begin{table}[hbtp]
\scriptsize
\caption{Transition energies and transition dipole moments for the Yb cation. Reference values have been obtained from the NIST database\cite{NIST_ASD}, the computed values were obtained for different basis sets with Kramers-restricted configuration interaction. }
\label{tab:Yb_cation_KRCI_lv}
\begin{tabular}{ c | l | r | r | r | r | r | r | r | r | r | r | r | r | r | r}
&&\multicolumn{2}{c |}{ NIST\cite{NistDiatomic} }&\multicolumn{2}{c |}{ 2z (x2c) }&\multicolumn{2}{c |}{3z (x2c) }&\multicolumn{2}{c |}{4z (x2c)} &\multicolumn{2}{c |}{ 2z (DCB\cite{Eliav1995}) }&\multicolumn{2}{c |}{3z (DCB\cite{Eliav1995}) }&\multicolumn{2}{c}{4z (DCB\cite{Eliav1995})}
\\
\hline
state&conf&E&S&E&TDM&E&TDM&E&TDM&E&TDM&E&TDM&E&TDM
\\
\hline
$^2$S$_{1/2}$       & 4f$^{14}$6s &     0 &      &     0 &      &     0 &     &      0 & 	  &     0 &    &     0 &     & 0    & 
\\
$^2$D$_{3/2}$       & 4f$^{14}$5d & 22961 &      & 23322 & 0.0  & 22802 & 0.0  & 23606 & 0.0  & 23389 & 0.0 & 22871 & 0.0 & 23674 & 0.0
\\
$^2$D$_{5/2}$       & 4f$^{14}$5d & 24333 &      & 23882 & 0.0 & 23321 & 0.0  & 24117 & 0.0  & 23945 & 0.0 & 23386 & 0.0 & 24182 & 0.0
\\
$^2$P$^\circ_{1/2}$ & 4f$^{14}$6p & 27062 & 6.1  & 25210 & 5.3  & 24533 & 5.5  & 25331 & 5.3. & 25241 & 5.3 & 24564 & 5.5 & 25362 & 5.3
\\
$^2$P$^\circ_{3/2}$ & 4f$^{14}$6p & 30392 & 11.4 & 28104 & 16.4 & 27385 & 17.3 & 28153 & 16.6 & 28139 & 	16.4 & 27422 & 17.3 & 28189 & 16.6
\end{tabular}
\end{table}
\begin{table}[hbtp]
\caption{Transition energies between states with an open f-shell obtained with configuration interaction. Transition energies relative to the $^2$F$^\circ_{7/2}$ state are listed. }
\label{tab:Yb_cation_hole_all}
\begin{tabular}{ c | l | r | r | r | r | r | r | r | r | r | r}
&&&\multicolumn{3}{c |}{ x2c }&\multicolumn{3}{c |}{ DCB\cite{Eliav1995}}
\\
\hline
state&conf&NIST\cite{NistDiatomic} & 2z & 3z & 4z & 2z & 3z & 4z
\\
\hline
$^2$F$^\circ_{7/2}$ &4f$^{13}\sigma_{6s}^2$              &     0 &     0 &     0 &     0 &     0 &    0 & 0
\\
$^3\left[3/2\right]^\circ_{5/2}$ &4f$^{13}$5d 6s  &  5340 &  4260 &  5538 &  4618 &  4349 &  5630 &  4711
\\
$^3\left[3/2\right]^\circ_{3/2}$ &4f$^{13}$5d 6s  &  7339 &  6387 &  7822 &  7123 &  6478 &  7917 &  7218
\\
$^3\left[11/2\right]^\circ_{9/2} $&4f$^{13}$5d 6s &  8806 &  8214 &  9325 &  8314 &  8304 &  9419 &  8408
\\
$^3\left[11/2\right]^\circ_{11/2}$&4f$^{13}$5d 6s &  9144 &  8320 &  9431 &  8447 &  8409 &  9524 &  8540
\\
\end{tabular}
\end{table}
The effect of the basis set is much larger and since no significant impact can be expected we used the X2C-AMFI Hamiltonian to obtain the data. 

\clearpage

\subsection{Closed f-shell - potential energy curves}

\begin{figure}[hbtp]
\centering
\includegraphics[width=\textwidth]{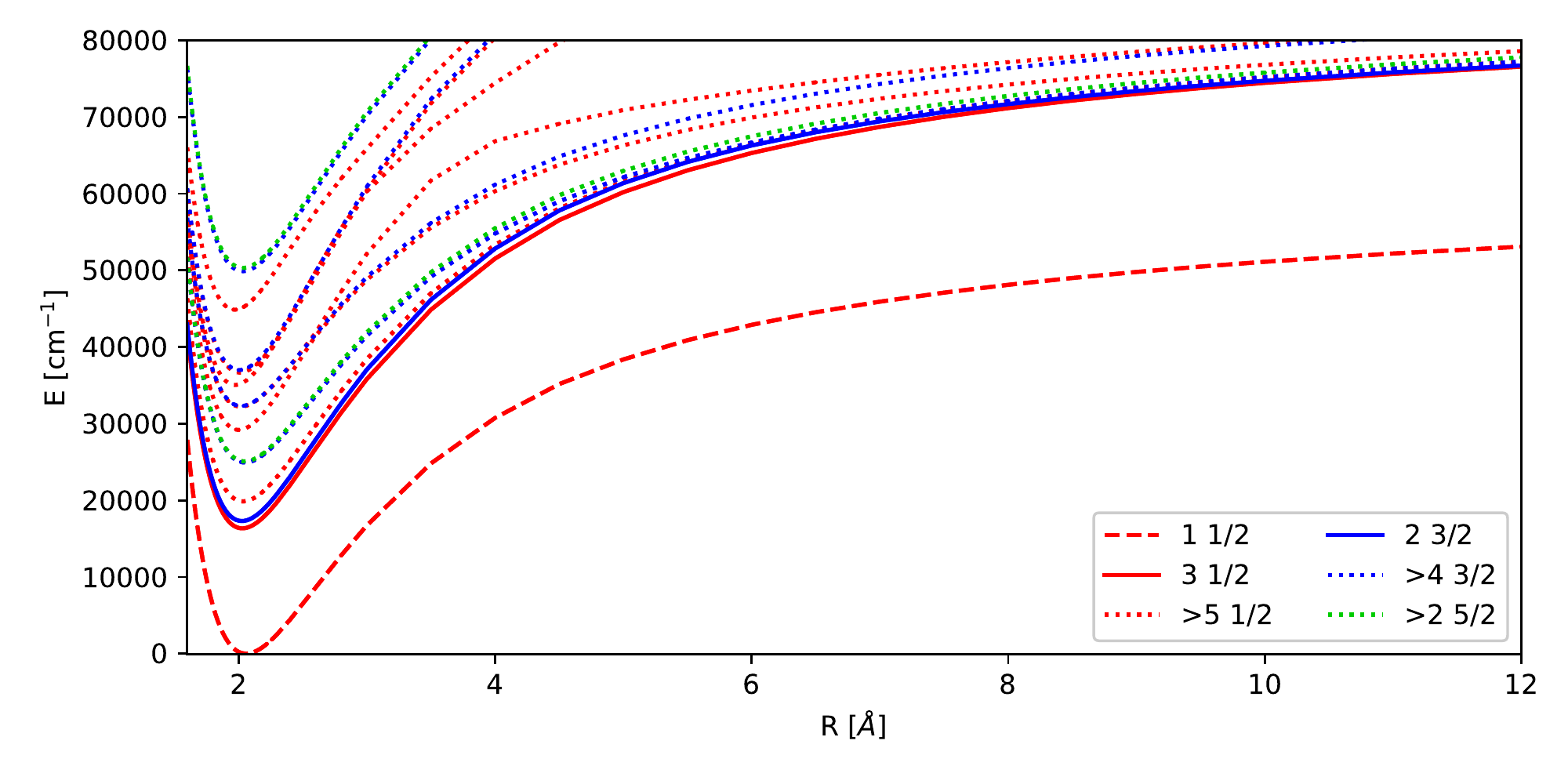}
\caption{KRCI PECs for states with f$^{14}$ for the quadruple zeta basis sets. }
\label{fig:PES_CI_v4z_C}
\end{figure}

\begin{table}[hbtp]
\caption{Spectroscopic constants for the different electronic states ($\Omega=1/2, 3/2, 5/2$) with f$^{14}$  obtained by KRCI for the double and triple zeta basis sets. 
Vibrational constant ($\omega_e$), anharmonicty constant ($\omega_e\chi_e$), and  transition energy (T$_e$), are given in \wn{}, the equilibrium bond distance (r$_e$) in \AA{}.}
\label{tab:KRCI_f14_v2z_v3z}
\begin{tabular}{ r | r | r | r | r | r | r | r | r | r   }
 $\Omega$ &  state  &    \multicolumn{4}{c |}{ v2z } & \multicolumn{4}{c }{v3z} \\
 \hline
  & &  r$_e$    &     $\omega_e$   &   $\omega_e\chi_e$ & T$_e$  &  r$_e$    &     $\omega_e$   &   $\omega_e\chi_e$ & T$_e$\\
 \hline
   1/2    &    1 &      2.0609 &         493 &        2.29  &            0   & 2.0594 &         493 &        2.27  &            0 \\
   1/2    &    2 &      2.0316 &         520 &        2.30  &        17319   & 2.0299 &         521 &        2.27  &        16663 \\
   1/2    &    3 &      2.0397 &         513 &        2.49  &        21252   & 2.0382 &         513 &        2.33  &        20270 \\
   1/2    &    4 &      2.1095 &         444 &        2.25  &        35868   & 2.0079 &         547 &        2.16  &        30993 \\
   1/2    &    5 &      2.0440 &         503 &        1.88  &        37977   & 2.0531 &         481 &        2.51  &        33700 \\
   1/2    &    6 &      2.0080 &         574 &        2.83  &        42795   & 1.9871 &         579 &        2.31  &        37776 \\
   1/2    &    7 &      2.0608 &         509 &        2.01  &        52474   & 2.0149 &         537 &        1.98  &        42915 \\
   3/2    &    1 &      2.0281 &         524 &        2.32  &        18425   & 2.0266 &         525 &        2.28  &        17667 \\
   3/2    &    2 &      2.0779 &         472 &        2.31  &        26817   & 2.0592 &         486 &        2.34  &        25450 \\
   3/2    &    3 &      2.1120 &         448 &        2.27  &        36017   & 2.0545 &         483 &        2.50  &        33792 \\
   3/2    &    4 &      2.0558 &         509 &        2.03  &        54498   & 2.0114 &         538 &        1.98  &        43720 \\
   5/2    &    1 &      2.0761 &         474 &        2.31  &        27016   & 2.0575 &         488 &        2.33  &        25572 \\
\end{tabular}
\end{table}

\begin{table}[hbtp]
\caption{Spectroscopic constants for the different electronic states ($\Omega=1/2, 3/2, 5/2$) with f$^{14}$  obtained by KRCI for the quadruple zeta and extrapolation to the complete basis set. 
Vibrational constant ($\omega_e$), anharmonicty constant ($\omega_e\chi_e$), and  transition energy (T$_e$), are given in \wn{}, the equilibrium bond distance (r$_e$) in \AA{}.}
\label{tab:KRCI_f14_v4z_cbs}
\begin{tabular}{ r | r | r | r | r | r | r | r | r | r   }
 $\Omega$ &  state  &    \multicolumn{4}{c |}{ v4z } & \multicolumn{4}{c }{CBS} \\
 \hline
  & &  r$_e$    &     $\omega_e$   &   $\omega_e\chi_e$ & T$_e$  &  r$_e$    &     $\omega_e$   &   $\omega_e\chi_e$ & T$_e$\\
 \hline
   1/2    &    1 &      2.0587 &         492 &        2.30  &            0    &  2.0829 &         465 &        2.40  &            0\\
   1/2    &    2 &      2.0290 &         521 &        2.30  &        16338    &  2.0504 &         496 &        2.38  &        16189\\
   1/2    &    3 &      2.0351 &         515 &        2.37  &        19857    &  2.0552 &         490 &        2.49  &        19631\\
   1/2    &    4 &      2.0012 &         554 &        2.23  &        29183    &  2.0154 &         537 &        2.31  &        28043\\
   1/2    &    5 &      2.0169 &         509 &        2.97  &        32206    &  2.0131 &         505 &        3.44  &        31208\\
   1/2    &    6 &      1.9727 &         587 &        2.35  &        35042    &  1.9792 &         570 &        2.41  &        33312\\
   1/2    &    7 &      2.0103 &         542 &        1.87  &        36658    &  2.0269 &         525 &        1.79  &        32245\\
   3/2    &    1 &      2.0260 &         524 &        2.31  &        17296    &  2.0473 &         499 &        2.37  &        17123\\
   3/2    &    2 &      2.0496 &         492 &        2.40  &        24927    &  2.0669 &         470 &        2.50  &        24583\\
   3/2    &    3 &      2.0193 &         509 &        2.86  &        32297    &  2.0160 &         503 &        3.22  &        31295\\
   3/2    &    4 &      2.0058 &         546 &        1.98  &        36990    &  2.0215 &         530 &        1.96  &        32245\\
   5/2    &    1 &      2.0474 &         495 &        2.39  &        25067    &  2.0639 &         474 &        2.48  &        24744\\
\end{tabular}
\end{table}

\clearpage

\subsection{Closed f-shell - dipole moments}

\begin{figure}[hbtp]
\centering
\includegraphics[width=\textwidth]{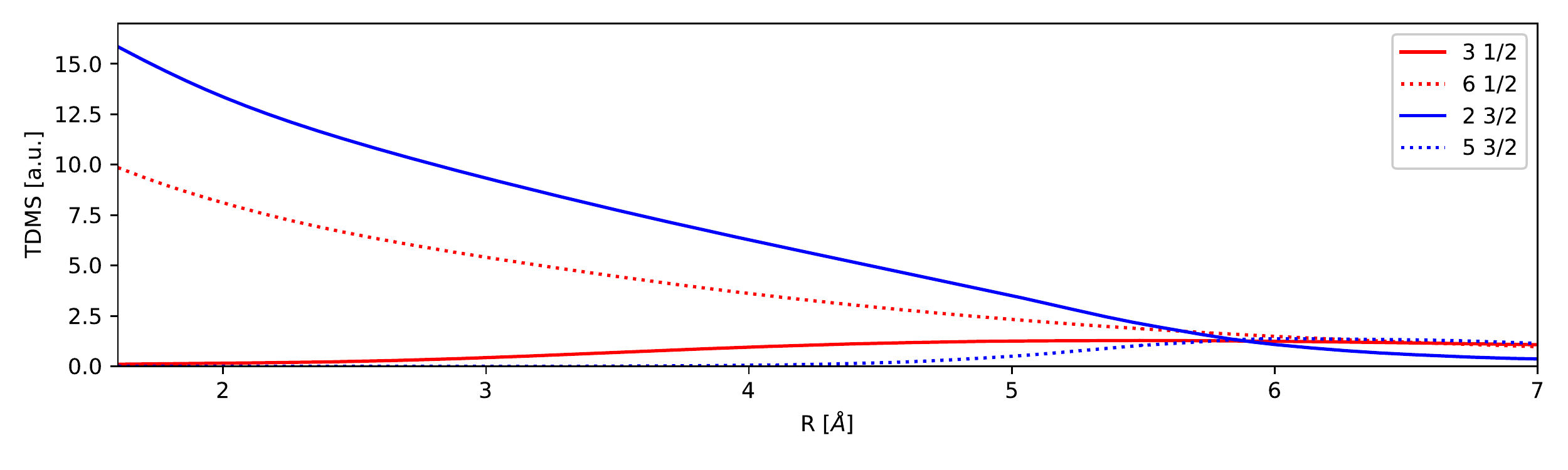}
\caption{KRCI TDMs for states with f$^{14}$ for the double zeta basis sets. }
\label{fig:TDMs_CI_v2z_C}
\end{figure}
\begin{figure}[hbtp]
\centering
\includegraphics[width=\textwidth]{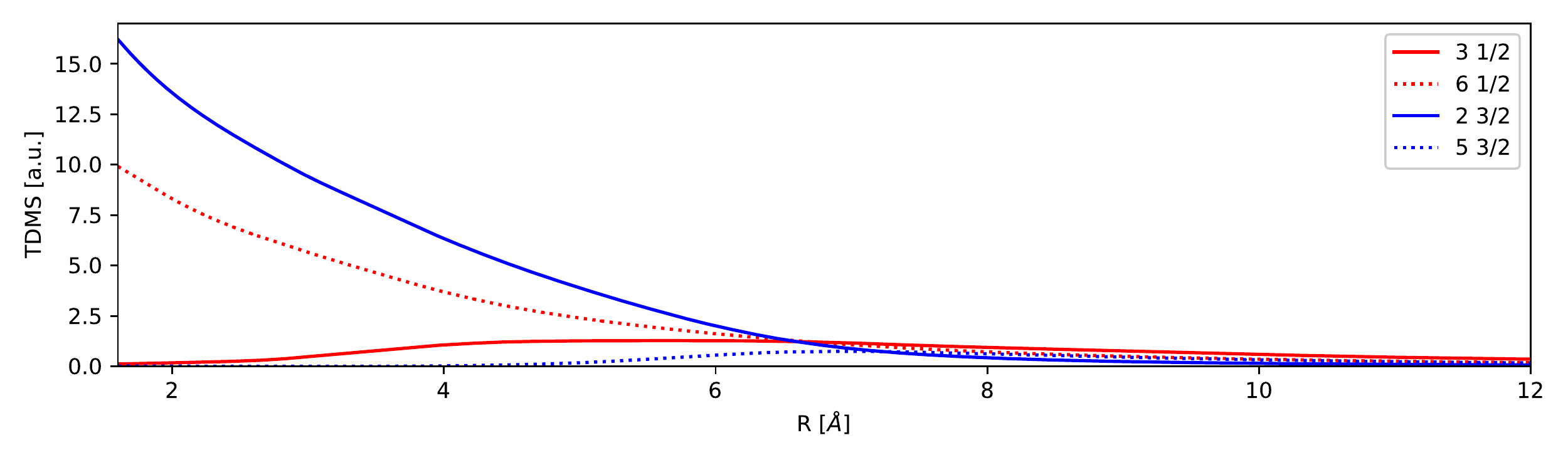}
\caption{KRCI TDMs for states with f$^{14}$ for the triple zeta basis sets. }
\label{fig:TDMs_CI_v3z_C}
\end{figure}
\begin{figure}[hbtp]
\centering
\includegraphics[width=\textwidth]{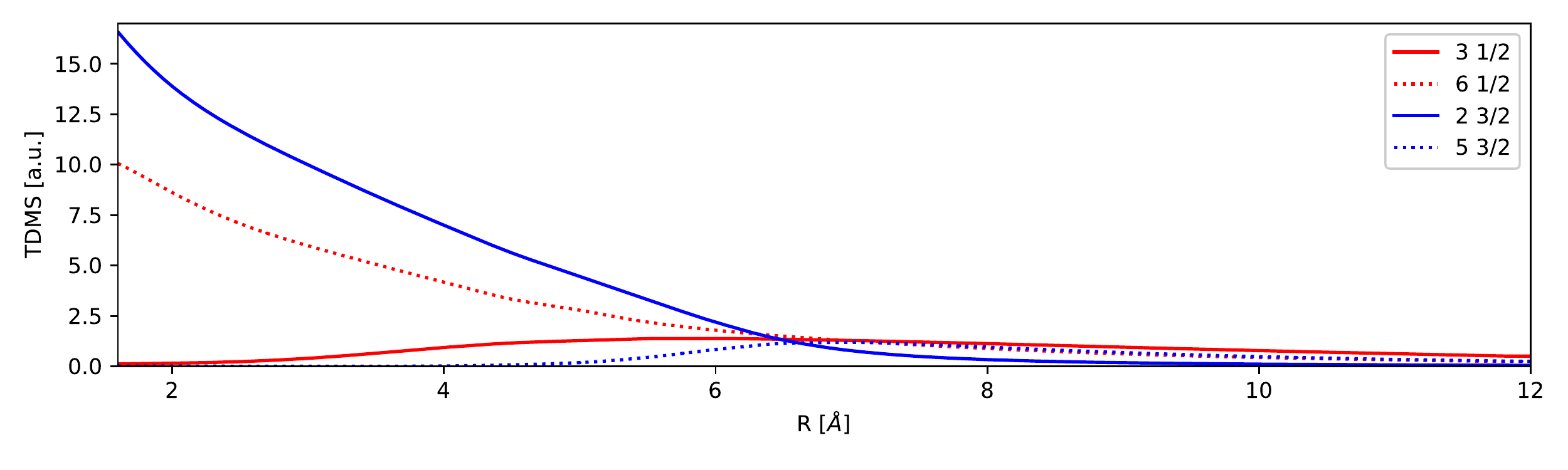}
\caption{KRCI TDMs for states with f$^{14}$ for the quadruple zeta basis sets. }
\label{fig:TDMs_CI_v4z_C}
\end{figure}

\begin{figure}[hbtp]
\centering
\includegraphics[width=\textwidth]{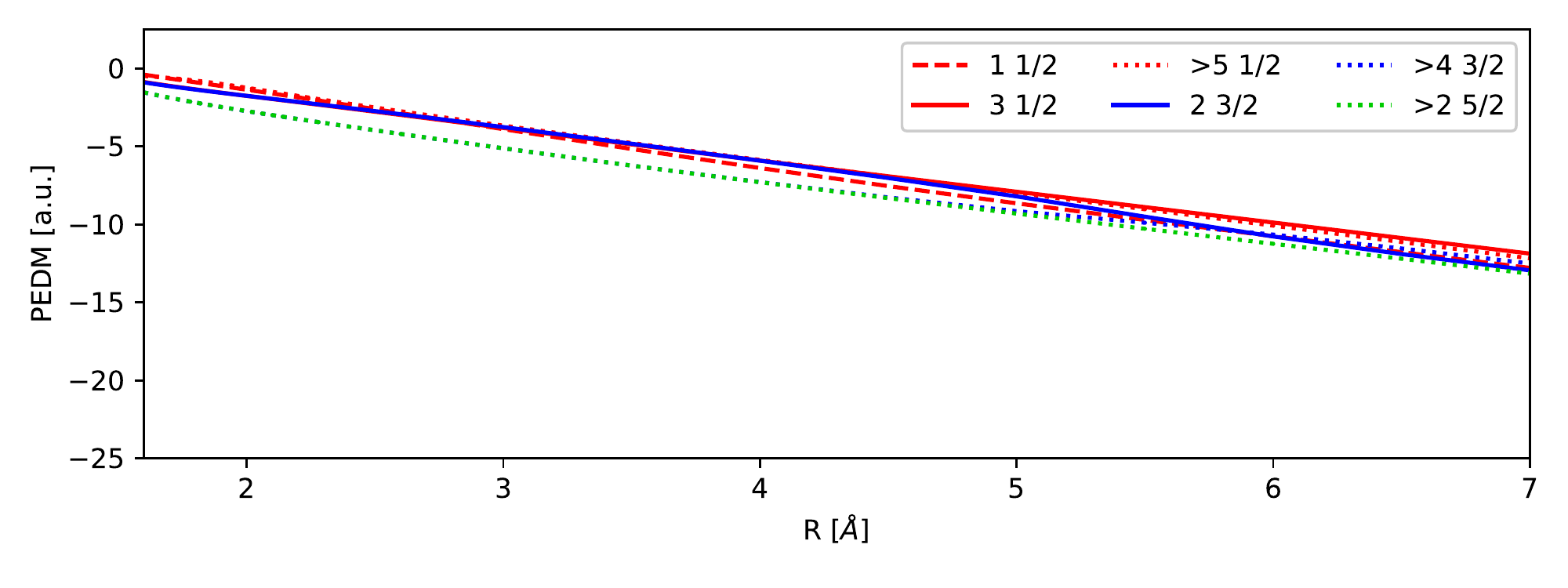}
\caption{KRCI PEDMs for states with f$^{14}$ for the double zeta basis sets. }
\label{fig:PEDM_CI2_v2z}
\end{figure}
\begin{figure}[hbtp]
\centering
\includegraphics[width=\textwidth]{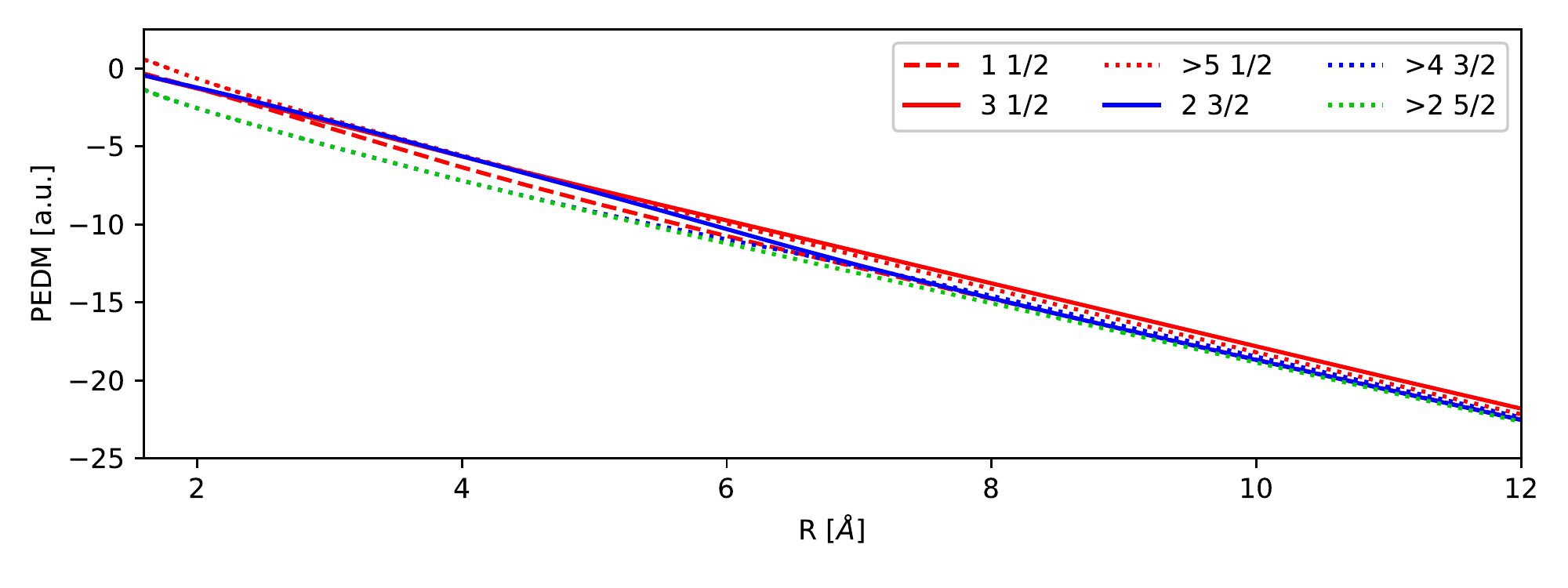}
\caption{KRCI PEDMs for states with f$^{14}$ for the triple zeta basis sets. }
\label{fig:PEDM_CI2_v3z}
\end{figure}    
\begin{figure}[hbtp]
\centering
\includegraphics[width=\textwidth]{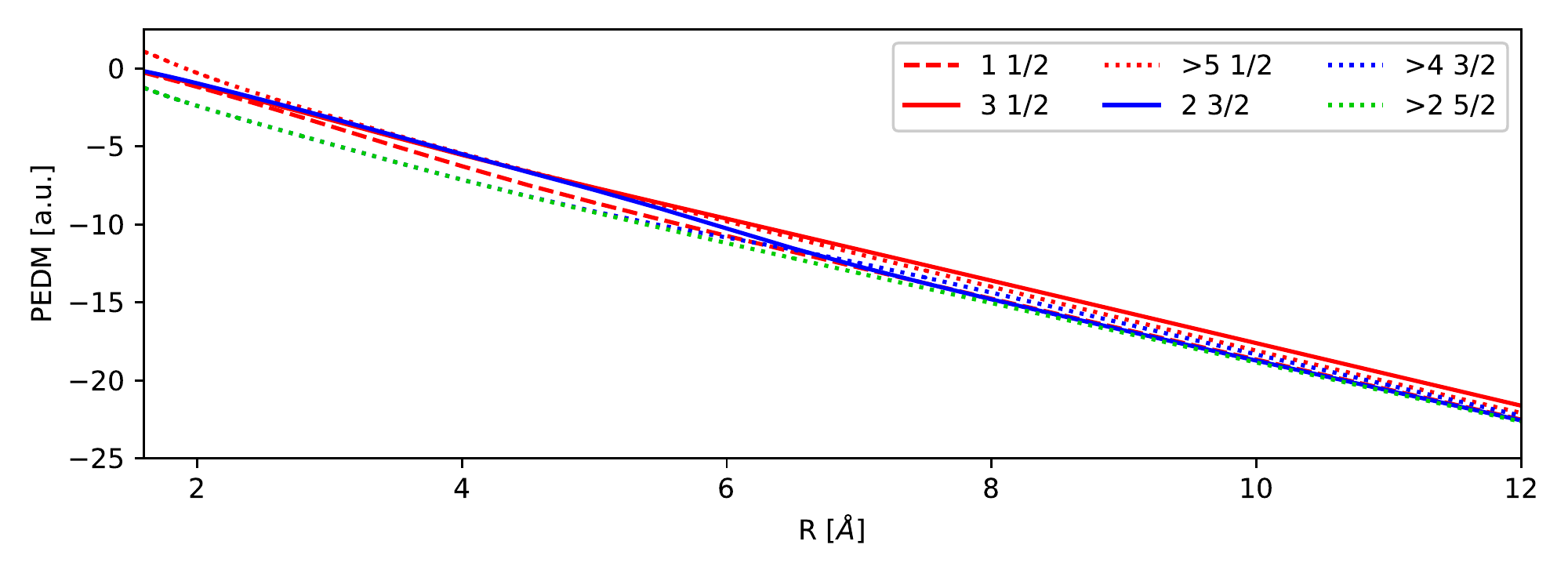}
\caption{KRCI PEDMs for states with f$^{14}$ for the quadruple zeta basis sets. }
\label{fig:PEDM_CI2_v4z}
\end{figure}

\clearpage

\subsection{Open f-shell - potential energy curves }

\begin{figure}[hbtp]
\centering
\includegraphics[width=\textwidth]{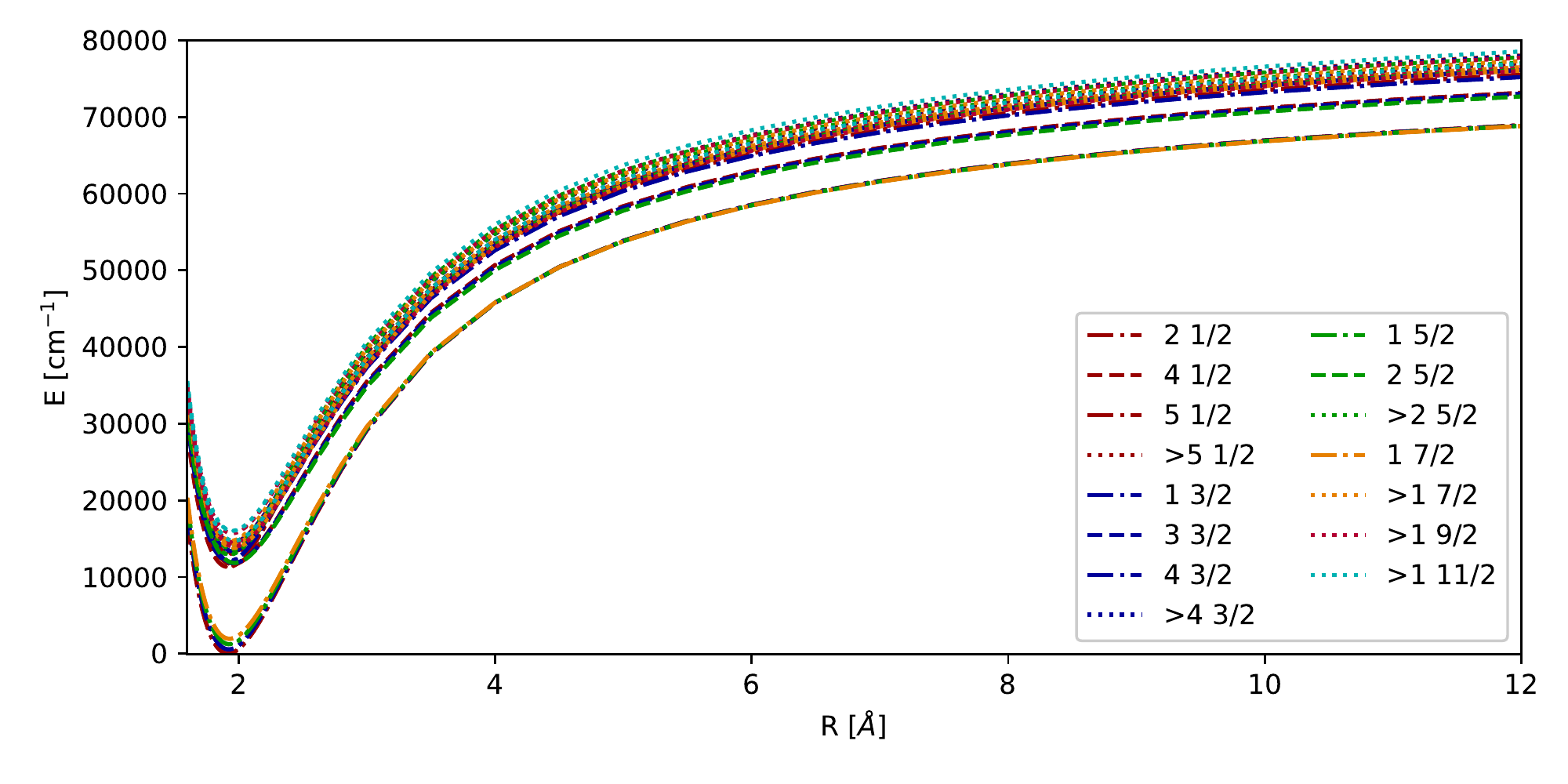}
\caption{KRCI PECs for states with f$^{13}$ for the triple zeta basis sets. }
\label{fig:PES_CI_v4z_I}
\end{figure}

\begin{table}[hbtp]
\caption{Spectroscopic constants for the different electronic states ($\Omega=1/2, 3/2, 5/2$) with f$^{13}$  obtained by KRCI for the double and triple zeta basis sets. 
Vibrational constant ($\omega_e$), anharmonicty constant ($\omega_e\chi_e$), and  transition energy (T$_e$), are given in \wn{}, the equilibrium bond distance (r$_e$) in \AA{}.}
\label{tab:KRCI_f13_v2z_v3z}
\begin{tabular}{ r | r | r | r | r | r | r | r | r | r   }
 $\Omega$ &  state  &    \multicolumn{4}{c |}{ v2z } & \multicolumn{4}{c }{v3z} \\
 \hline
  & &  r$_e$    &     $\omega_e$   &   $\omega_e\chi_e$ & T$_e$  &  r$_e$    &     $\omega_e$   &   $\omega_e\chi_e$ & T$_e$\\
 \hline
   1/2    &    1 &      1.9254 &         627 &        2.53  &            0 &          1.9188 &         631 &        2.73  &            0 \\ 
   1/2    &    2 &      1.9762 &         510 &        0.23  &        10834 &          1.9540 &         517 &        1.24  &        11083 \\ 
   1/2    &    3 &      1.9321 &         702 &        7.89  &        11243 &          1.9314 &         684 &        4.10  &        11352 \\ 
   1/2    &    4 &      1.9568 &         558 &        2.79  &        12833 &          1.9404 &         584 &        3.07  &        12811 \\ 
   1/2    &    5 &      1.9555 &         575 &        2.64  &        13071 &          1.9469 &         582 &        2.73  &        12946 \\ 
   1/2    &    6 &      1.9689 &         578 &        2.38  &        13663 &          1.9631 &         569 &        2.12  &        13947 \\ 
   1/2    &    7 &      1.9645 &         578 &        2.25  &        13884 &          1.9566 &         580 &        2.61  &        14069 \\ 
   3/2    &    1 &      1.9305 &         624 &        2.54  &          557 &          1.9239 &         628 &        2.76  &          545 \\ 
   3/2    &    2 &      1.9797 &         556 &        2.43  &        10849 &          1.9677 &         561 &        2.48  &        11212 \\ 
   3/2    &    3 &      1.9331 &         597 &        5.44  &        12028 &          1.9262 &         625 &        3.62  &        12045 \\ 
   3/2    &    4 &      1.9706 &         587 &        0.62  &        12938 &          1.9527 &         576 &        2.65  &        13073 \\ 
   3/2    &    5 &      1.9639 &         575 &        1.88  &        13087 &          1.9519 &         569 &        2.22  &        13194 \\ 
   3/2    &    6 &      1.9621 &         576 &        2.73  &        13632 &          1.9596 &         576 &        2.24  &        13797 \\ 
   3/2    &    7 &      1.9592 &         591 &        2.88  &        13966 &          1.9568 &         594 &        2.69  &        14083 \\ 
   5/2    &    1 &      1.9341 &         619 &        2.54  &         1272 &          1.9277 &         623 &        2.77  &         1227 \\ 
   5/2    &    2 &      1.9837 &         554 &        2.44  &        10843 &          1.9736 &         559 &        2.49  &        11295 \\ 
   5/2    &    3 &      1.9420 &         548 &        5.37  &        12945 &          1.9293 &         580 &        6.14  &        12923 \\ 
   5/2    &    4 &      1.9568 &         625 &        3.55  &        13211 &          1.9563 &         622 &        0.70  &        13264 \\ 
   5/2    &    5 &      1.9664 &         593 &        0.60  &        13543 &          1.9553 &         580 &        1.95  &        13670 \\ 
   5/2    &    6 &      1.9558 &         578 &        2.67  &        14218 &          1.9474 &         588 &        2.81  &        14225 \\ 
   7/2    &    1 &      1.9355 &         613 &        2.50  &         2017 &         1.9290 &         617 &        2.75  &         1931 \\ 
   7/2    &    2 &      1.9753 &         566 &        2.46  &        13192 &          1.9665 &         570 &        2.45  &        13424 \\ 
   7/2    &    3 &      1.9712 &         569 &        2.47  &        13565 &          1.9613 &         573 &        2.48  &        13819 \\ 
   7/2    &    4 &      1.9626 &         569 &        2.41  &        14333 &          1.9458 &         585 &        2.94  &        14475 \\ 
   9/2    &    1 &      1.9776 &         563 &        2.48  &        13226 &          1.9707 &         568 &        2.47  &        13576 \\ 
   9/2    &    2 &      1.9755 &         566 &        2.48  &        13544 &          1.9665 &         571 &        2.50  &        13933 \\ 
   9/2    &    3 &      1.9709 &         571 &        2.51  &        14319 &          1.9592 &         577 &        2.54  &        14880 \\ 
   11/2    &    1 &      1.9769 &         564 &        2.48  &        13466 &          1.9691 &         569 &        2.49  &        14001 \\ 
   11/2    &    2 &      1.9744 &         568 &        2.49  &        14177 &          1.9628 &         576 &        2.52  &        14904 \\
\end{tabular}
\end{table}

\begin{table}[hbtp]
\caption{Spectroscopic constants for the different electronic states ($\Omega=1/2, 3/2, 5/2$) with f$^{13}$  obtained by KRCI for the quadruple zeta  and extrapolation to the complete basis sets. 
Vibrational constant ($\omega_e$), anharmonicty constant ($\omega_e\chi_e$), and  transition energy (T$_e$), are given in \wn{}, the equilibrium bond distance (r$_e$) in \AA{}.}
\label{tab:KRCI_f13_v4z_cbs}
\begin{tabular}{ r | r | r | r | r | r | r | r | r | r   }
 $\Omega$ &  state  &    \multicolumn{4}{c |}{ v4z } & \multicolumn{4}{c }{CBS} \\
 \hline
  & &  r$_e$    &     $\omega_e$   &   $\omega_e\chi_e$ & T$_e$  &  r$_e$    &     $\omega_e$   &   $\omega_e\chi_e$ & T$_e$\\
 \hline
 \hline
   1/2    &    1 &      1.9183 &         632 &        2.60  &            0 &          1.9200 &         631 &        2.51  &            0 \\ 
   1/2    &    2 &      1.9191 &         583 &        8.23  &        11329 &          1.9038 &         655 &       13.57  &        11428 \\ 
   1/2    &    3 &      1.9497 &         641 &        0.17  &        11672 &          1.9603 &         620 &        0.08  &        11876 \\ 
   1/2    &    4 &      1.9366 &         593 &        2.73  &        13044 &          1.9360 &         598 &        2.50  &        13206 \\ 
   1/2    &    5 &      1.9426 &         588 &        2.65  &        13240 &          1.9417 &         592 &        2.56  &        13443 \\ 
   1/2    &    6 &      1.9566 &         576 &        2.40  &        14295 &          1.9539 &         582 &        2.71  &        14531 \\ 
   1/2    &    7 &      1.9503 &         580 &        2.75  &        14504 &          1.9479 &         578 &        2.71  &        14808 \\ 
   3/2    &    1 &      1.9236 &         628 &        2.61  &          543 &          1.9253 &         628 &        2.50  &          540 \\ 
   3/2    &    2 &      1.9589 &         530 &        0.82  &        11810 &          1.9505 &         516 &        0.80  &        12202 \\ 
   3/2    &    3 &      1.9306 &         667 &        4.87  &        12069 &          1.9331 &         711 &        8.45  &        12062 \\ 
   3/2    &    4 &      1.9483 &         582 &        2.55  &        13356 &          1.9473 &         585 &        2.47  &        13549 \\ 
   3/2    &    5 &      1.9440 &         578 &        2.90  &        13495 &          1.9408 &         584 &        3.30  &        13699 \\ 
   3/2    &    6 &      1.9545 &         575 &        2.67  &        14251 &          1.9528 &         574 &        2.97  &        14567 \\ 
   3/2    &    7 &      1.9575 &         591 &        1.96  &        14429 &          1.9601 &         589 &        1.50  &        14667 \\ 
   5/2    &    1 &      1.9276 &         623 &        2.61  &         1227 &          1.9296 &         622 &        2.46  &         1223 \\ 
   5/2    &    2 &      1.9690 &         563 &        2.45  &        11828 &          1.9677 &         564 &        2.42  &        12196 \\ 
   5/2    &    3 &      1.9282 &         614 &        3.10  &        12923 &          1.9302 &         635 &        1.41  &        12919 \\ 
   5/2    &    4 &      1.9579 &         592 &        1.08  &        13556 &          1.9627 &         571 &        0.93  &        13747 \\ 
   5/2    &    5 &      1.9482 &         582 &        2.57  &        14030 &          1.9451 &         584 &        3.07  &        14278 \\ 
   5/2    &    6 &      1.9449 &         593 &        2.66  &        14442 &          1.9453 &         596 &        2.53  &        14590 \\ 
   7/2    &    1 &     1.9293 &         617 &        2.59  &         1935 &          1.9315 &         616 &        2.43  &         1933 \\ 
   7/2    &    2 &     1.9627 &         573 &        2.43  &        13764 &          1.9620 &         574 &        2.42  &        13994 \\ 
   7/2    &    3 &     1.9552 &         576 &        2.39  &        14284 &          1.9528 &         577 &        2.31  &        14607 \\ 
   7/2    &    4 &     1.9424 &         595 &        2.81  &        14729 &          1.9422 &         601 &        2.69  &        14905 \\ 
   9/2    &    1 &      1.9669 &         570 &        2.49  &        13962 &      1.9663 &         571 &        2.49  &        14224 \\ 
   9/2    &    2 &      1.9613 &         575 &        2.48  &        14516 &      1.9596 &         576 &        2.45  &        14923 \\ 
   9/2    &    3 &      1.9494 &         584 &        2.59  &        15691 &      1.9444 &         588 &        2.62  &        16262 \\ 
   11/2    &    1 &      1.9645 &         572 &        2.49  &        14706 &      1.9633 &         574 &        2.48  &        15202 \\ 
   11/2    &    2 &      1.9525 &         584 &        2.52  &        16023 &      1.9472 &         588 &        2.51  &        16817 \\ 
\end{tabular}
\end{table}

\clearpage

\subsection{Open f-shell - dipole moments }

\begin{figure}[hbtp]
\centering
\includegraphics[width=\textwidth]{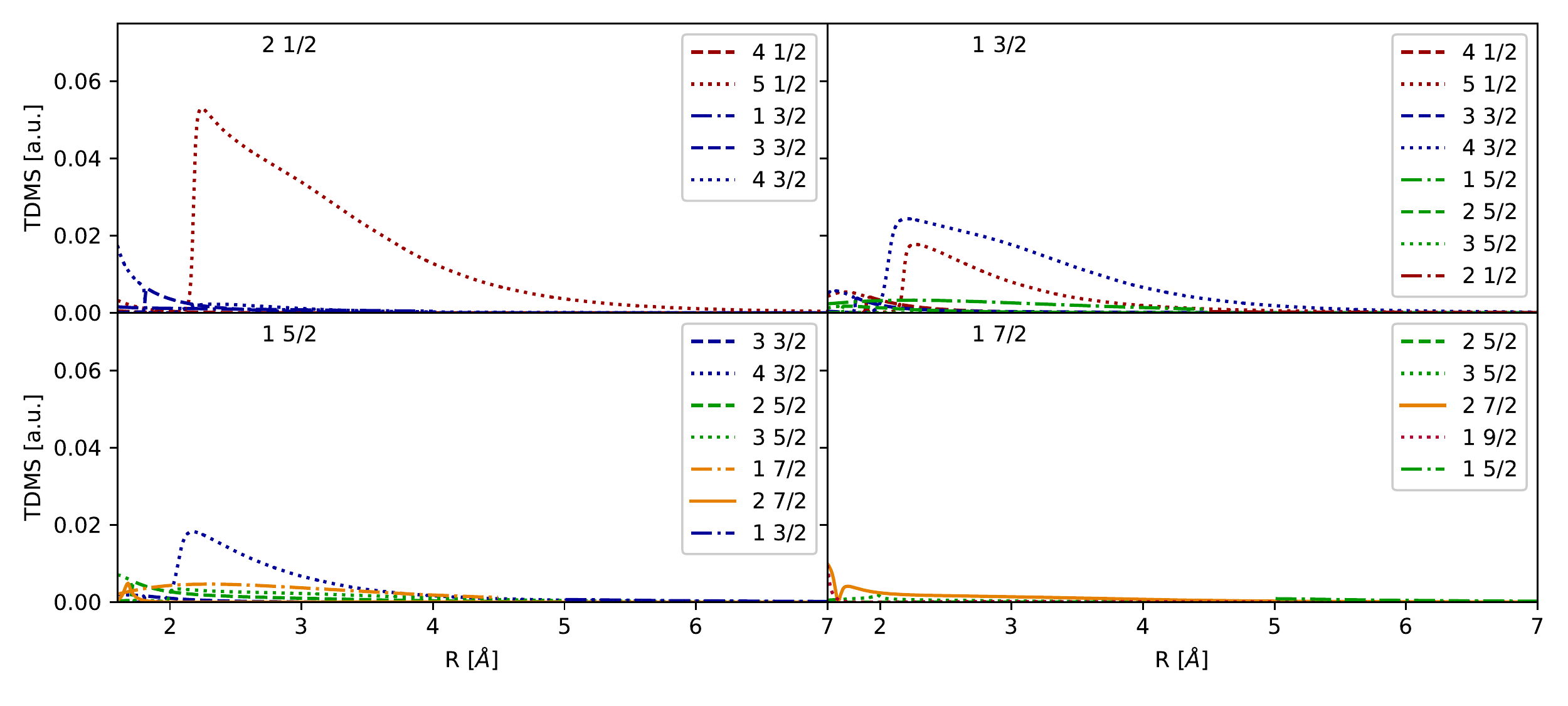}
\caption{KRCI TDMs for states with f$^{14}$ for the double zeta basis sets. }
\label{fig:TDMs_CI_v2z_H}
\end{figure}
\begin{figure}[hbtp]
\centering
\includegraphics[width=\textwidth]{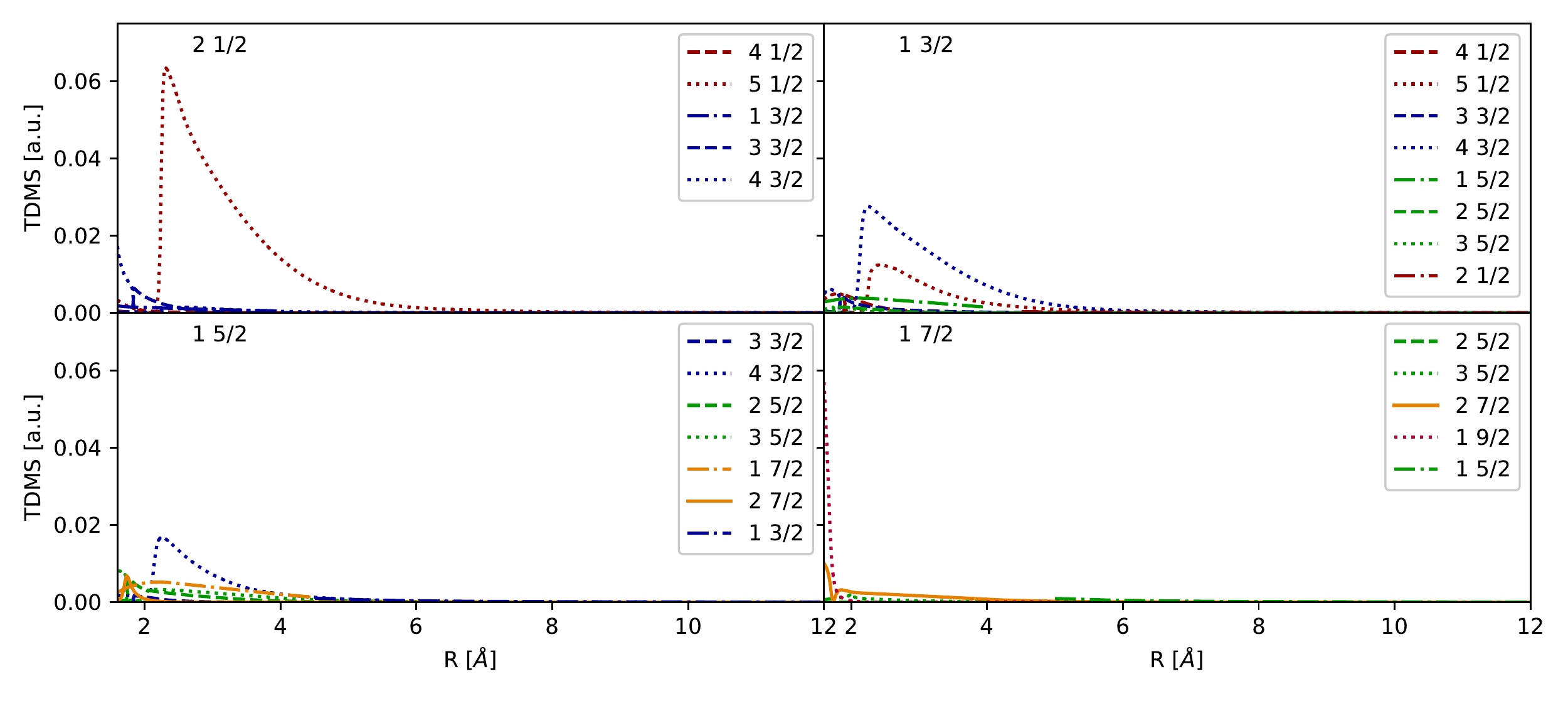}
\caption{KRCI TDMs for states with f$^{14}$ for the triple zeta basis sets. }
\label{fig:TDMs_CI_v3z_H}
\end{figure}
\begin{figure}[hbtp]
\centering
\includegraphics[width=\textwidth]{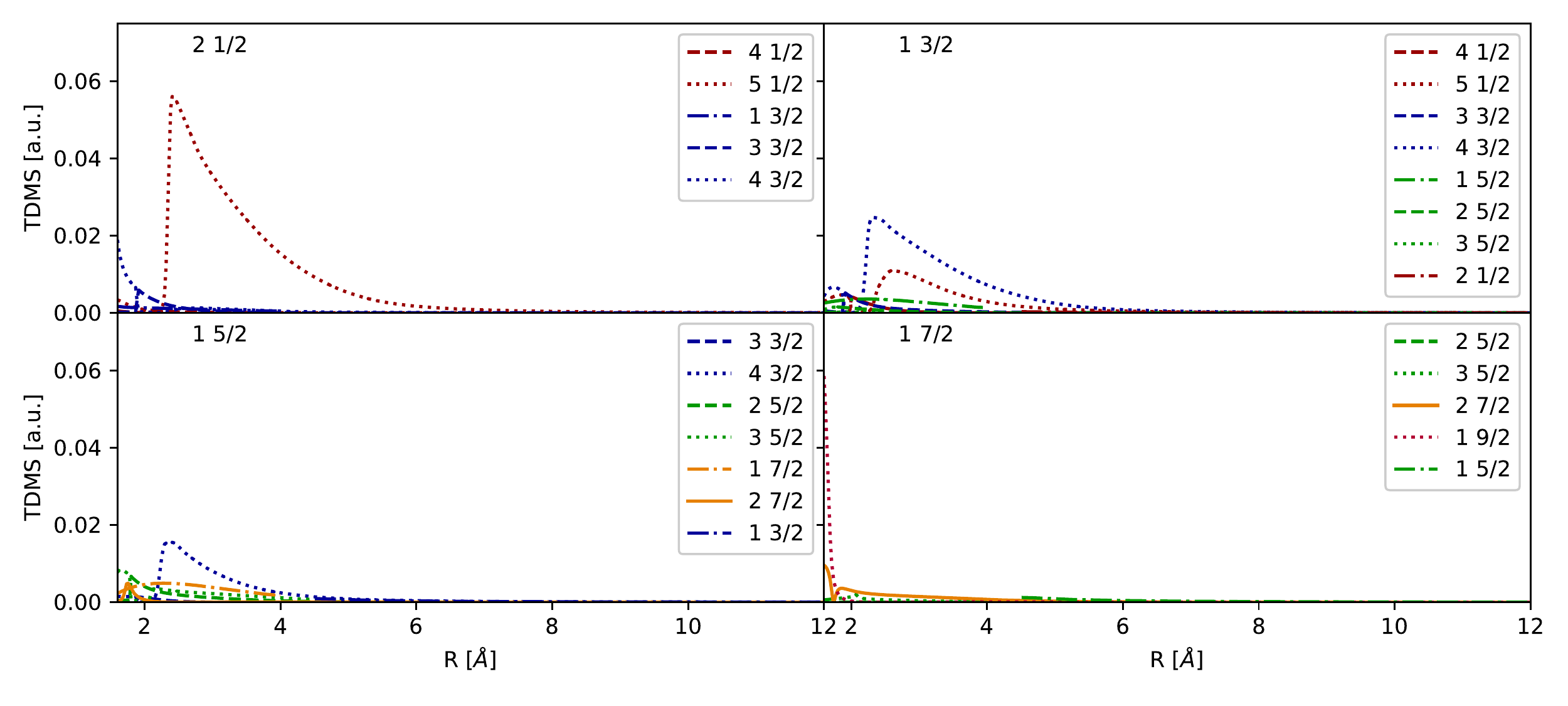}
\caption{KRCI TDMs for states with f$^{14}$ for the quadruple zeta basis sets. }
\label{fig:TDMs_CI_v4z_H}
\end{figure}

\begin{figure}[hbtp]
\centering
\includegraphics[width=\textwidth]{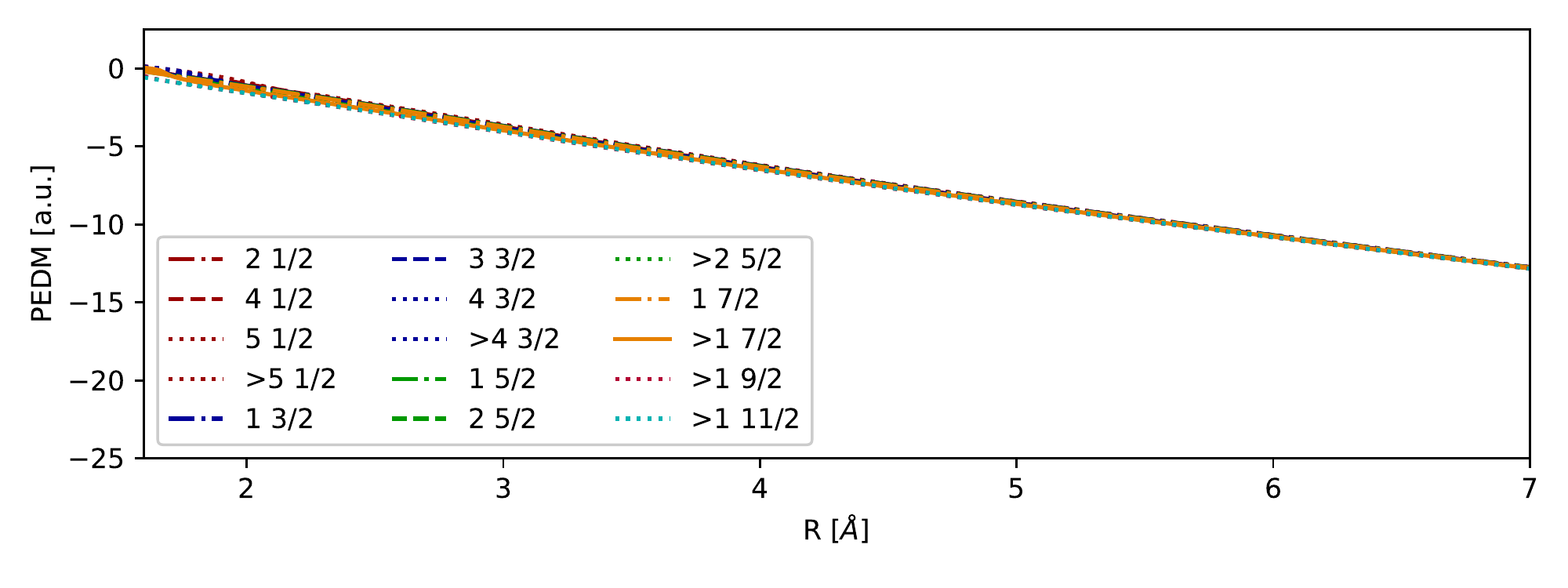}
\caption{KRCI PEDMs for states with f$^{13}$ for the double zeta basis sets. }
\label{fig:PEDM_CI_v2z_I}
\end{figure}
\begin{figure}[hbtp]
\centering
\includegraphics[width=\textwidth]{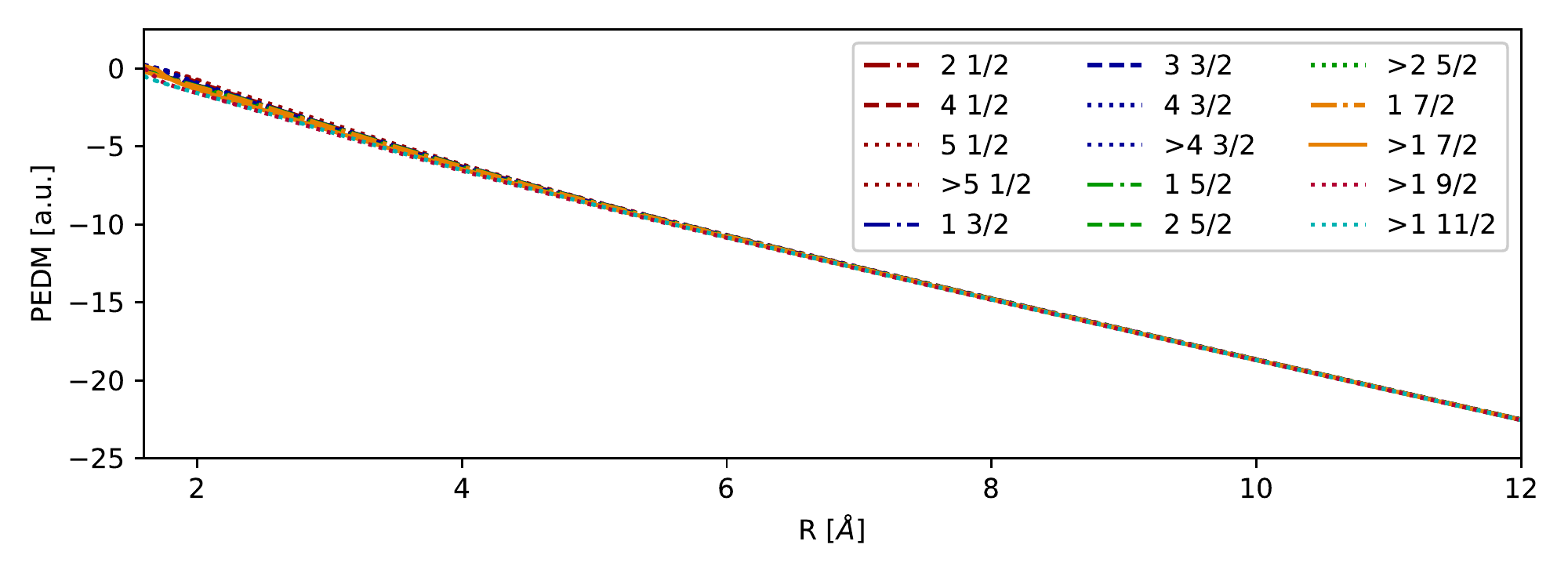}
\caption{KRCI PEDMs for states with f$^{13}$ for the triple zeta basis sets. }
\label{fig:PEDM_CI_v3z_I}
\end{figure}
\begin{figure}[hbtp]
\centering
\includegraphics[width=\textwidth]{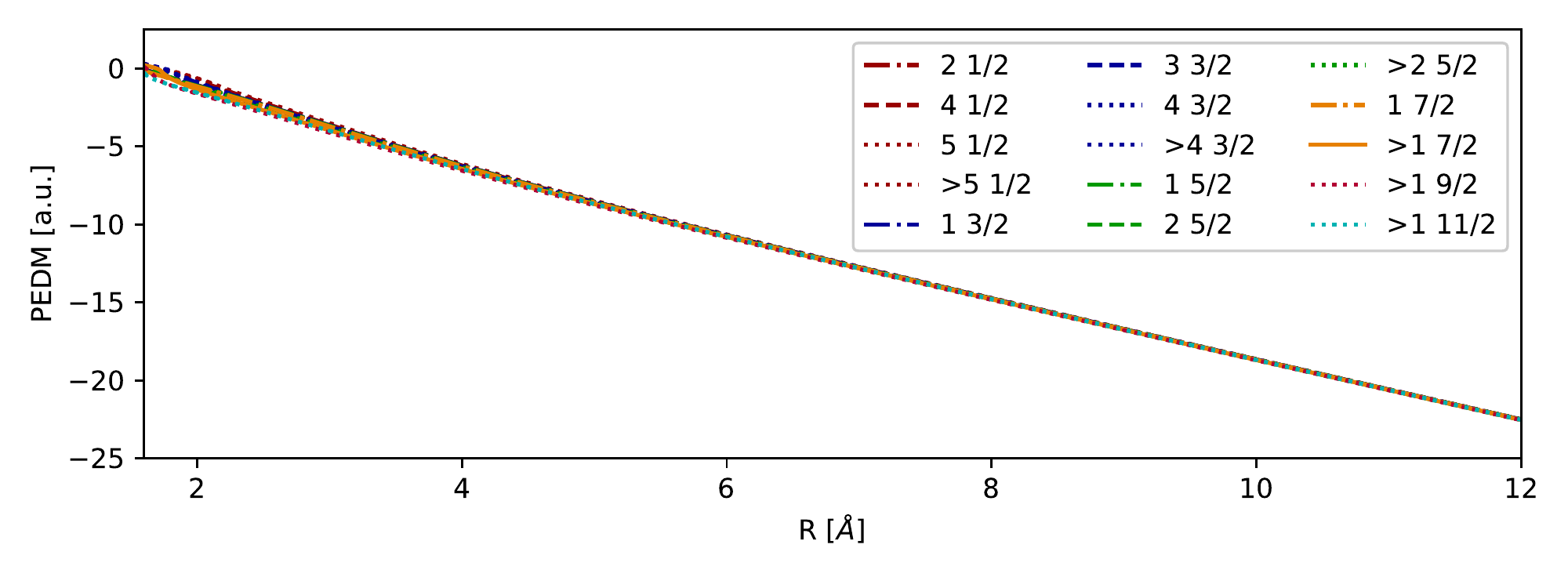}
\caption{KRCI PEDMs for states with f$^{13}$ for the triple zeta basis sets. }
\label{fig:PEDM_CI_v4z_I}
\end{figure}

\clearpage

\subsection{Combined potentials, shifting}

\begin{figure}[hbtp]
\centering
\includegraphics[width=0.99\textwidth]{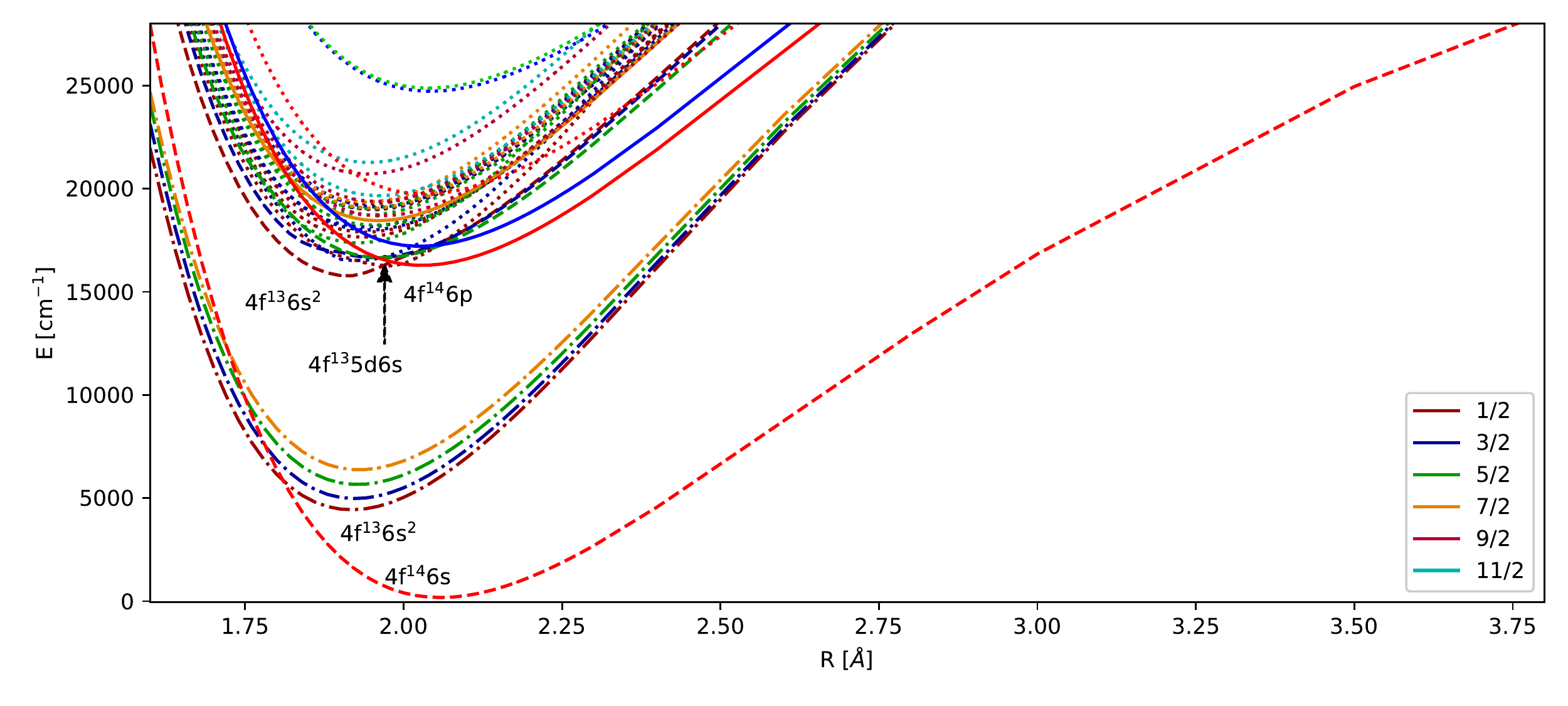}
\includegraphics[width=0.99\textwidth]{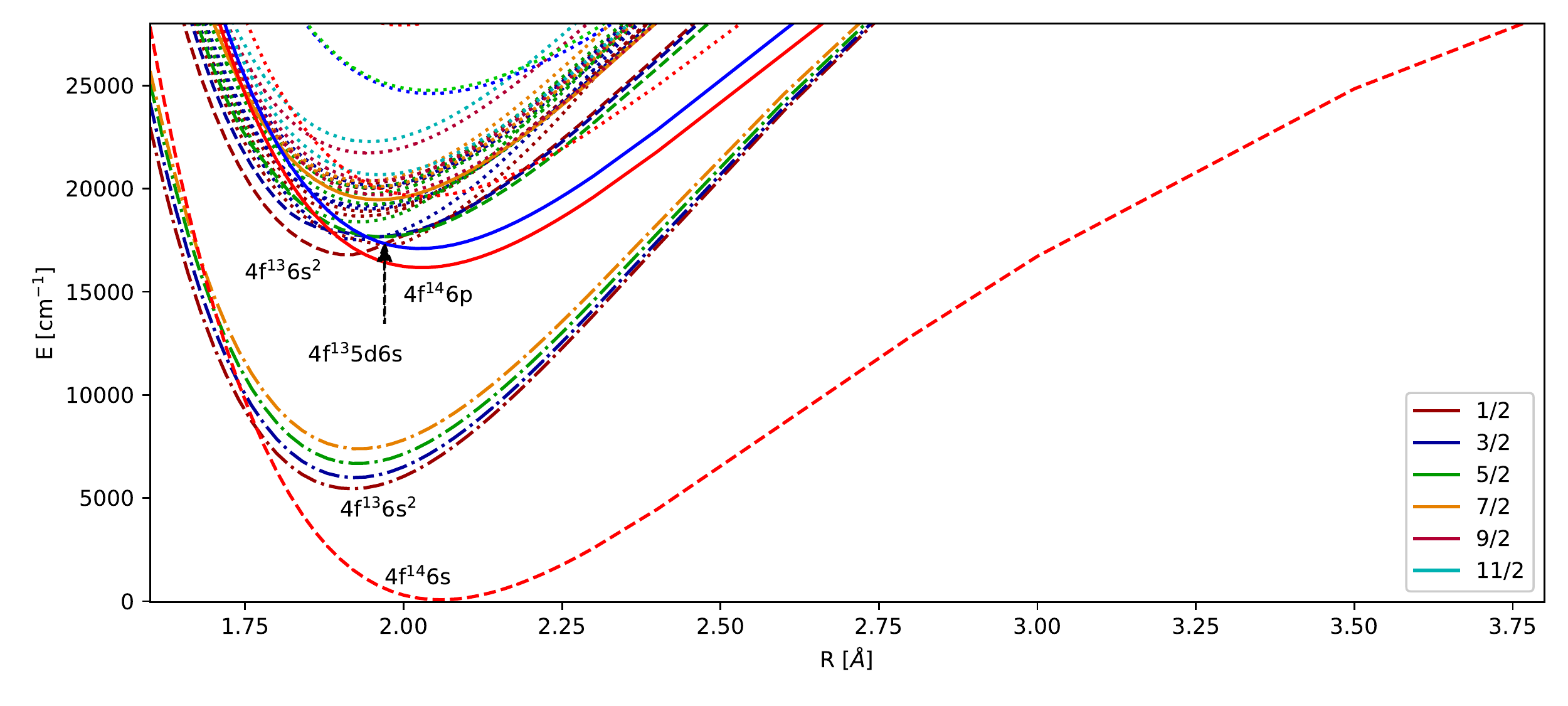}
\caption{Combination of the sets of KRCI potentials. The lowest $\Omega=1/2$ are denoted by the assigned dominant configuration. For the upper the shift was applied at 15~\AA{} for the lower picture at 10\AA{}.}
\label{fig:PES_KRCI_shift}
\end{figure}

The combination of the potential energy by shifting according to the outermost point is shown in figure~\ref{fig:PES_KRCI_shift}. While the minimum of the potential changes from about 45000 to 53000~\wn{}, there are only small changes in the relative position of the minima of different electronic states. 

\begin{figure}[hbtp]
\centering
\includegraphics[width=\textwidth]{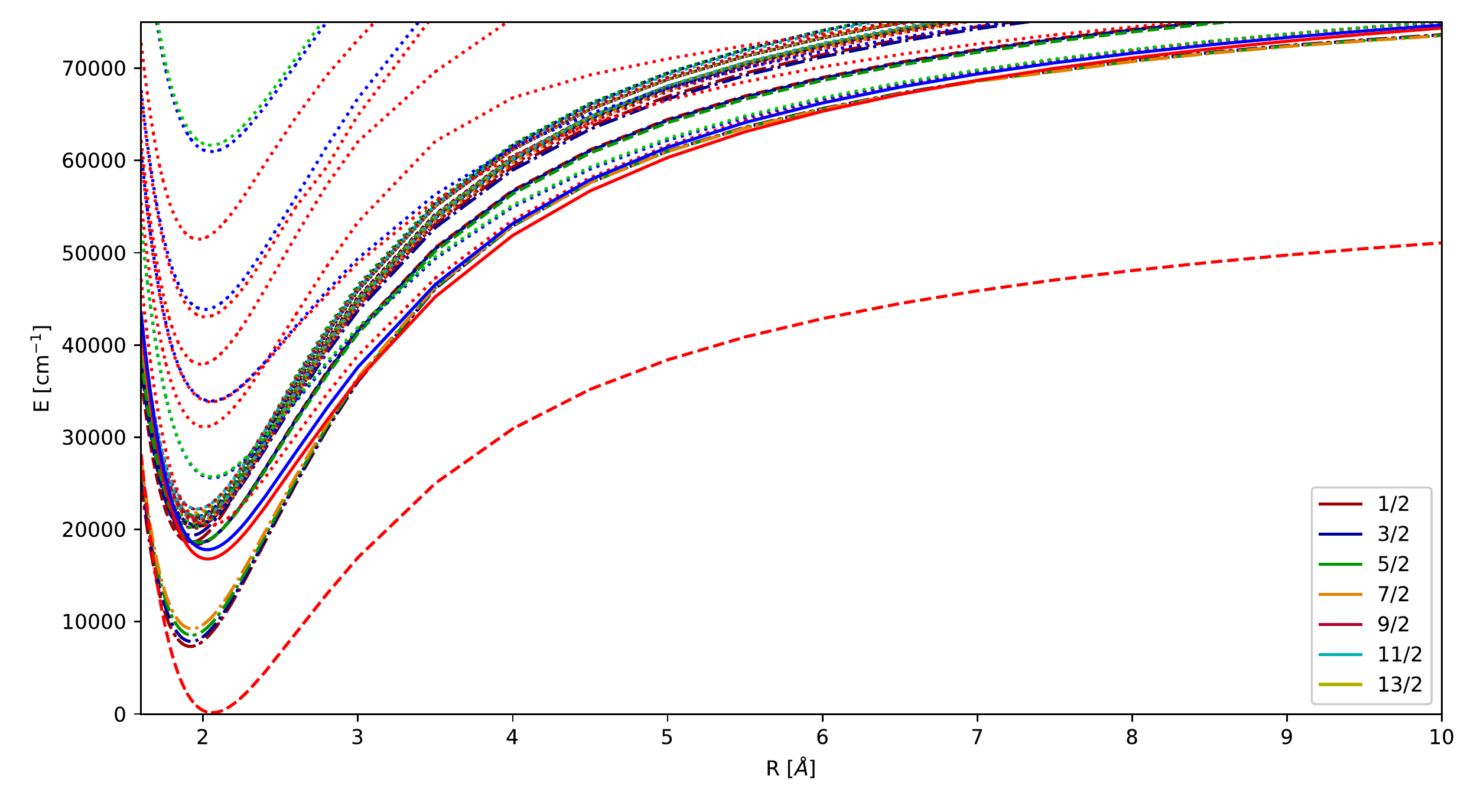}
\caption{KRCI PECs for states with open and closed f-shell for the triple zeta basis sets. }
\label{fig:PES_CI_v3z}
\end{figure}

\begin{figure}[hbtp]
\centering
\includegraphics[width=\textwidth]{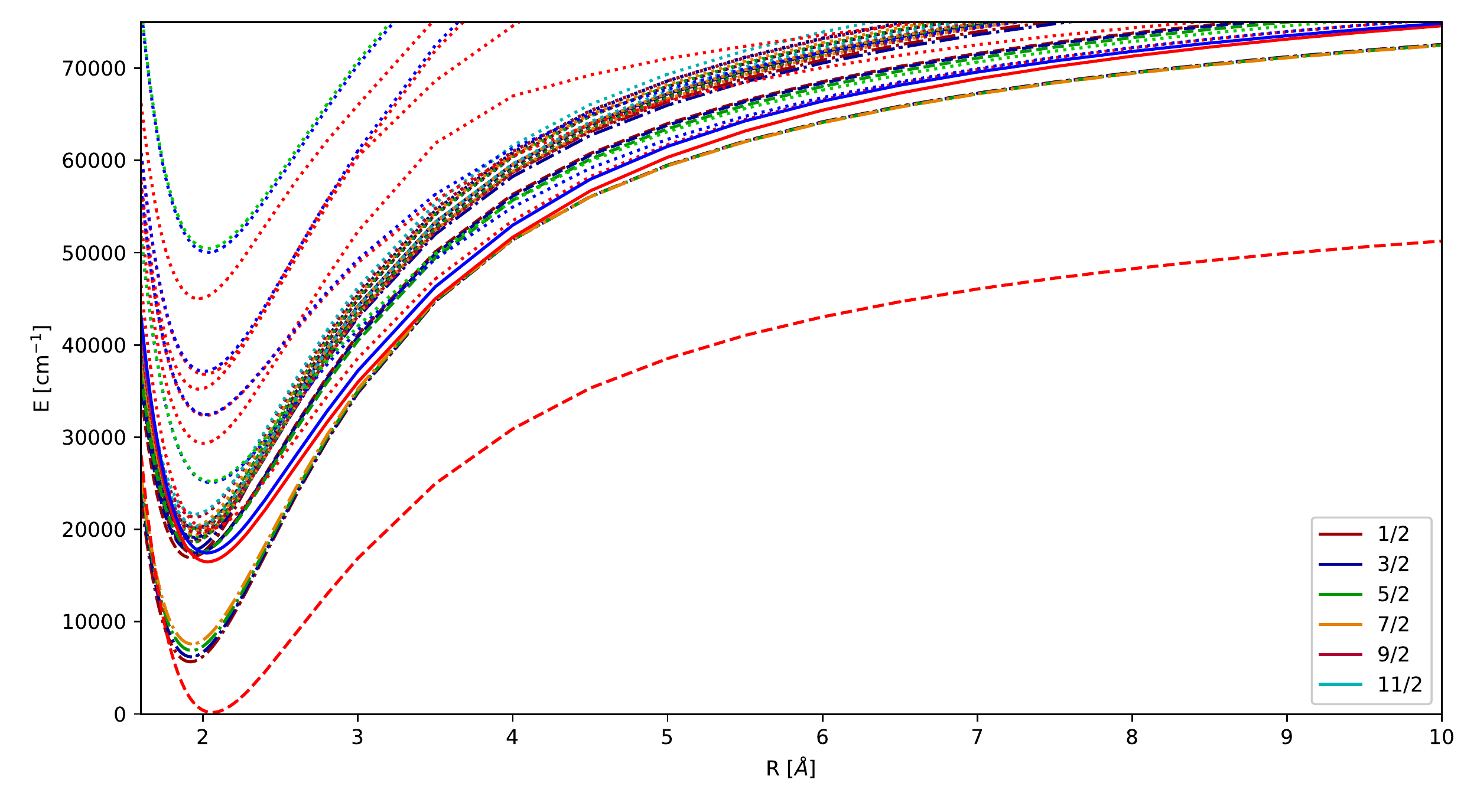}
\caption{KRCI PECs for states with open and closed f-shell for the quadruple zeta basis sets. }
\label{fig:PES_CI_v4z}
\end{figure}

\begin{figure}[hbtp]
\centering
\includegraphics[width=\textwidth]{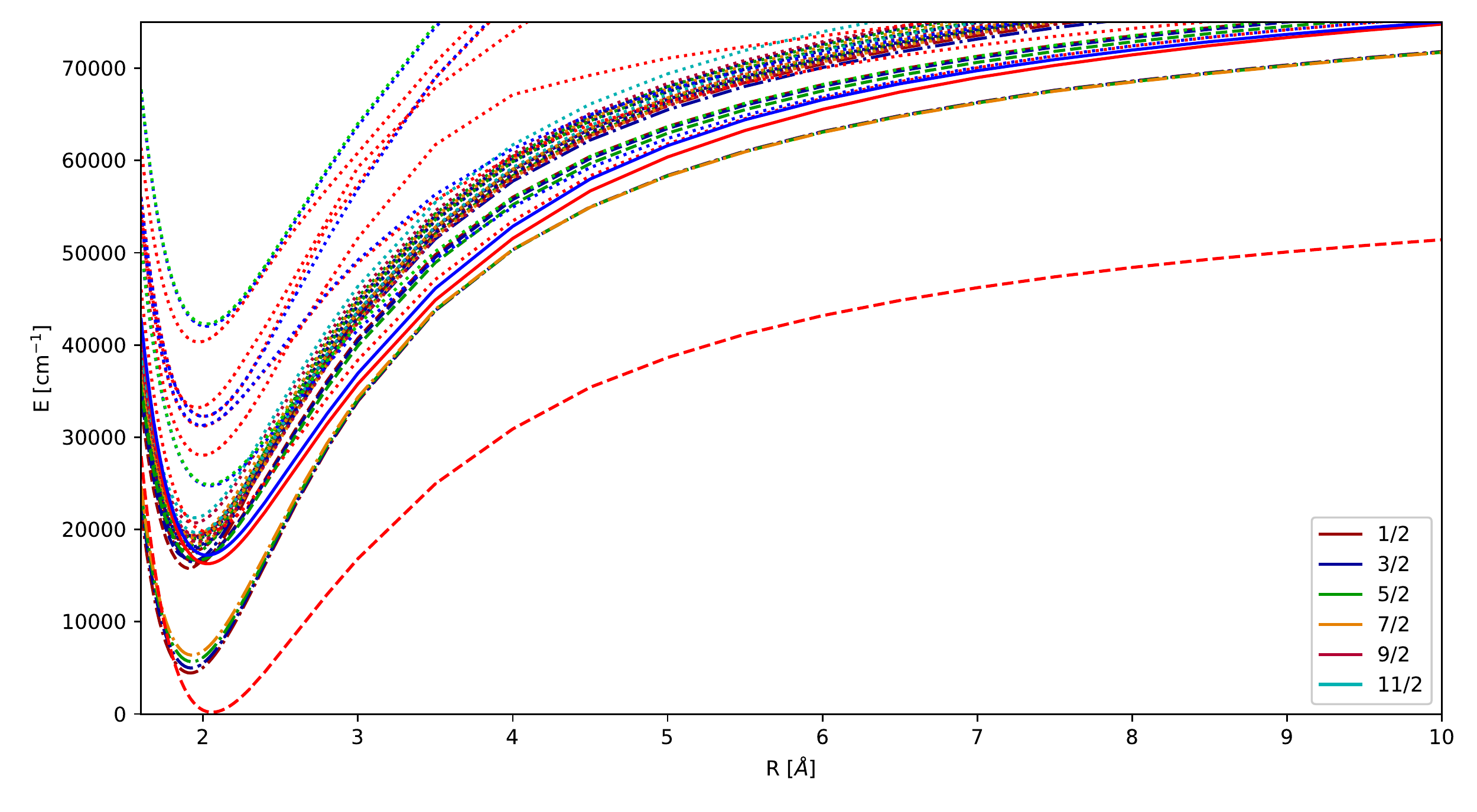}
\caption{KRCI PECs for states with open and closed f-shell for extrapolation to the basis set limit.}
\label{fig:PES_CI_CBS}
\end{figure}

Figures \ref{fig:PES_CI_v3z} to \ref{fig:PES_CI_CBS} show the combined potentials energy curves for different basis set sizes. 

\clearpage

\subsection{Collection - transition dipole moments}

\begin{table}[hbtp]
\caption{KRCI transition dipole moment table for the equilibrium distances of ground state. }
\label{tab:dipole_moment_krci}
\begin{tabular}{ r | r | r | r | r  }
 $\Omega_1$ & state$_1$ & $\Omega_2$ & state$_2$ &TDMs [a.u.]
 \\ 
 \hline
 $\frac{1}{2}$&1&$\frac{1}{2}$&3&0.163
 \\ 
 $\frac{1}{2}$&1&$\frac{1}{2}$&6&8.410
 \\ 
 $\frac{1}{2}$&1&$\frac{3}{2}$&2&13.571
 \\ 
 $\frac{1}{2}$&1&$\frac{3}{2}$&5&0.012
 \\ 
\hline
 $\frac{1}{2}$&2&$\frac{1}{2}$&4&0.000
 \\ 
 $\frac{1}{2}$&2&$\frac{1}{2}$&5&0.001
 \\ 
 $\frac{1}{2}$&2&$\frac{3}{2}$&1&0.001
 \\ 
 $\frac{1}{2}$&2&$\frac{3}{2}$&3&0.006
 \\ 
\hline
 $\frac{3}{2}$&1&$\frac{1}{2}$&4&0.000
 \\ 
 $\frac{3}{2}$&1&$\frac{1}{2}$&5&0.004
 \\ 
 $\frac{3}{2}$&1&$\frac{3}{2}$&3&0.005
 \\ 
 $\frac{3}{2}$&1&$\frac{3}{2}$&4&0.000
 \\ 
 $\frac{3}{2}$&1&$\frac{5}{2}$&1&0.003
 \\ 
 $\frac{3}{2}$&1&$\frac{5}{2}$&2&0.001
 \\ 
\hline
 $\frac{5}{2}$&1&$\frac{3}{2}$&3&0.001
 \\ 
 $\frac{5}{2}$&1&$\frac{3}{2}$&4&0.000
 \\ 
 $\frac{5}{2}$&1&$\frac{5}{2}$&2&0.005
 \\ 
 $\frac{5}{2}$&1&$\frac{5}{2}$&3&0.000
 \\ 
 $\frac{5}{2}$&1&$\frac{7}{2}$&1&0.004
 \\ 
 $\frac{5}{2}$&1&$\frac{7}{2}$&2&0.001
 \\ 
 \hline
 $\frac{7}{2}$&1&$\frac{5}{2}$&2&0.000
 \\ 
 $\frac{7}{2}$&1&$\frac{5}{2}$&3&0.001
 \\ 
 $\frac{7}{2}$&1&$\frac{7}{2}$&1&0.003
 \\ 
 $\frac{7}{2}$&1&$\frac{7}{2}$&2&0.001
 \\
\end{tabular}
\end{table}

\clearpage

\section{Intermediate Hamiltonian Fock Space Coupled Cluster (IHFS-CCSD)}
\label{sec:FSCC}

\subsection{Closed f-shell - potential energy curves}

\begin{figure}[hbtp]
\centering
\includegraphics[width=0.99\textwidth]{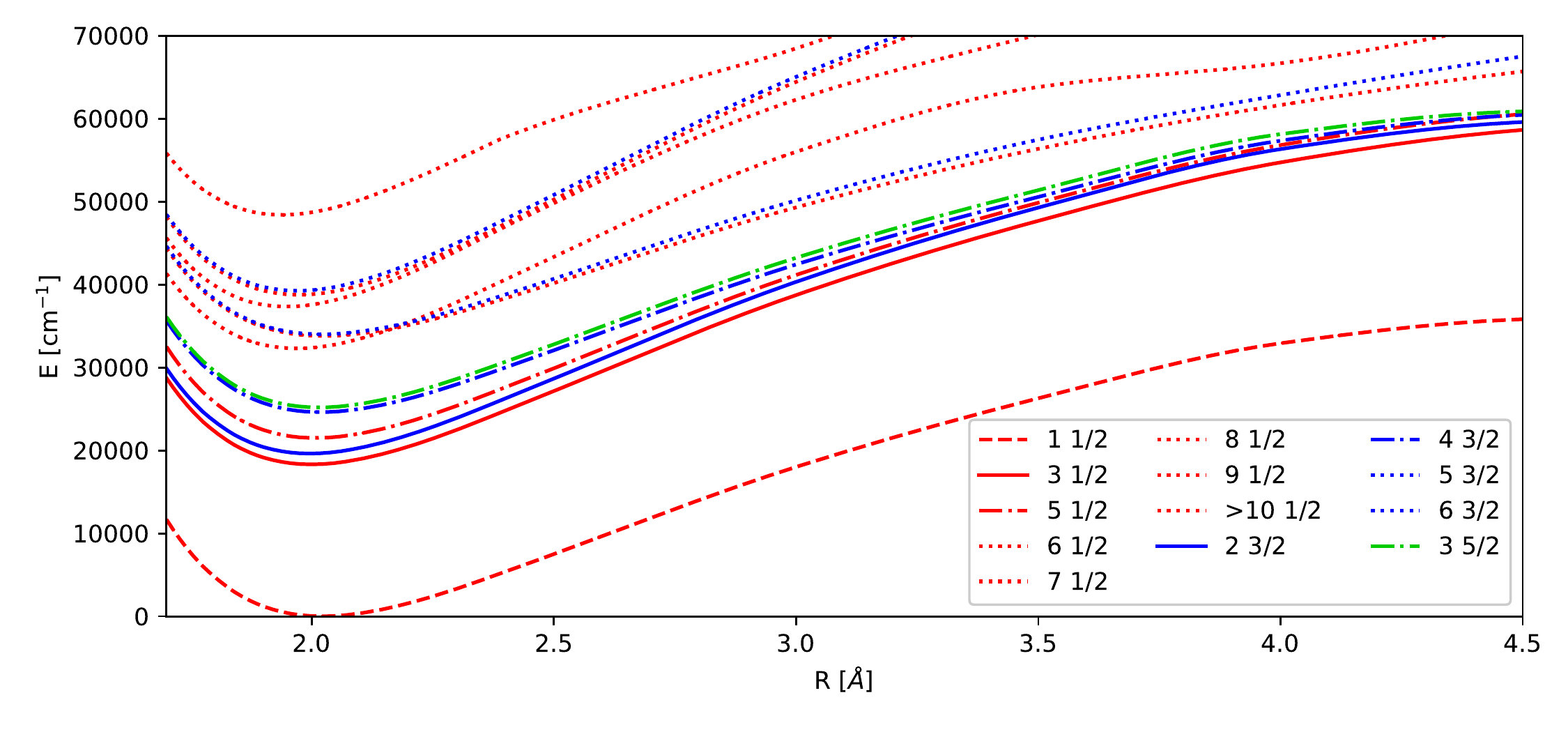}
\caption{IHFS-CCSD PECs for states from the (0h,1p) sector with a cation reference for the quadruple zeta basis sets. }
\label{fig:PES_CATION4}
\end{figure}



\begin{table}[hbtp]
\caption{Spectroscopic constants for the different electronic states ($\Omega=1/2, 3/2, 5/2$) obtained by IHFS-CCSD for the (0h,1p) sector using double and triple zeta basis sets. 
Vibrational constant ($\omega_e$), anharmonicty constant ($\omega_e\chi_e$), and  transition energy (T$_e$), are given in \wn{}, the equilibrium bond distance (r$_e$) in \AA{}.}
\label{tab:FSCC_f14_v2z_v3z}
\begin{tabular}{ r | r | r | r | r | r | r | r | r | r   }
 $\Omega$ &  state  &    \multicolumn{4}{c |}{ v2z } & \multicolumn{4}{c }{v3z} \\
 \hline
  & &  r$_e$    &     $\omega_e$   &   $\omega_e\chi_e$ & T$_e$  &  r$_e$    &     $\omega_e$   &   $\omega_e\chi_e$ & T$_e$\\
 \hline
   1/2    &    1 &      2.0315 &         502 &        2.33  &            0 &      2.0281 &         504 &        2.32  &            0  \\
   1/2    &    2 &      2.0045 &         524 &        2.19  &        18610 &      2.0037 &         526 &        2.22  &        18425 \\
   1/2    &    3 &      2.0133 &         518 &        2.31  &        22113 &      2.0127 &         521 &        2.28  &        21687 \\
   1/2    &    4 &      2.0905 &         443 &        2.21  &        36083 &      1.9680 &         543 &        6.91  &        33550 \\
   1/2    &    5 &      1.9873 &         518 &        3.00  &        38956 &      2.0481 &         524 &        0.02  &        34724 \\
   1/2    &    6 &      1.9994 &         569 &        2.17  &        42974 &      1.9648 &         582 &        2.30  &        39207 \\
   1/2    &    7 &      2.0383 &         508 &        1.98  &        56216 &      1.9964 &         532 &        1.92  &        46115 \\
   1/2    &    8 &      1.8438 &         813 &        3.41  &        66635 &      1.9414 &         591 &        2.37  &        54196 \\
   3/2    &    1 &      2.0010 &         527 &        2.20  &        19980 &      2.0004 &         530 &        2.24  &        19747 \\
   3/2    &    2 &      2.0394 &         484 &        2.17  &        25904 &      2.0268 &         496 &        2.24  &        24977 \\
   3/2    &    3 &      2.0892 &         450 &        2.19  &        36389 &      2.0448 &         470 &        2.39  &        34962 \\
   3/2    &    4 &      2.0319 &         510 &        2.01  &        58253 &      1.9909 &         535 &        1.99  &        46891 \\
   5/2    &    1 &      2.0372 &         486 &        2.18  &        26612 &      2.0243 &         499 &        2.26  &        25563 \\
   ION    &      &      1.9506 &         594 &        2.08  &       48113  &      1.9483 &         597 &        2.15  &      48352   \\
\end{tabular}
\end{table}

\begin{table}[hbtp]
\caption{Spectroscopic constants for the different electronic states ($\Omega=1/2, 3/2, 5/2$) obtained by IHFS-CCSD for the (0h,1p) sector using double and triple zeta  and extrapolation to the complete basis sets. 
Vibrational constant ($\omega_e$), anharmonicty constant ($\omega_e\chi_e$), and  transition energy (T$_e$), are given in \wn{}, the equilibrium bond distance (r$_e$) in \AA{}.}
\label{tab:FSCC_f14_v4z_cbs}
\begin{tabular}{ r | r | r | r | r | r | r | r | r | r   }
 $\Omega$ &  state  &    \multicolumn{4}{c |}{ v4z } & \multicolumn{4}{c }{CBS} \\
 \hline
  & &  r$_e$    &     $\omega_e$   &   $\omega_e\chi_e$ & T$_e$  &  r$_e$    &     $\omega_e$   &   $\omega_e\chi_e$ & T$_e$\\
 \hline
   1/2    &    1 &      2.0220 &         510 &        2.60  &            0 &      2.0176 &         515 &        2.82  &            0 \\
   1/2    &    2 &      1.9988 &         533 &        2.46  &        18323 &      1.9953 &         539 &        2.63  &        18249 \\
   1/2    &    3 &      2.0072 &         527 &        2.54  &        21506 &      2.0032 &         533 &        2.73  &        21375 \\
   1/2    &    4 &      1.9649 &         574 &        4.13  &        32306 &      1.9644 &         581 &        1.82  &        31416 \\
   1/2    &    5 &      2.0298 &         506 &        0.35  &        33748 &      2.0155 &         494 &        0.68  &        33018 \\
   1/2    &    6 &      1.9477 &         599 &        2.29  &        37343 &      1.9359 &         610 &        2.08  &        35954 \\
   1/2    &    7 &      1.9732 &         577 &        2.08  &        38770 &      1.9597 &         608 &        1.95  &        33366\\
   1/2    &    8 &      1.9384 &         592 &        2.46  &        48402 &      1.9363 &         592 &        2.39  &        44179\\
   3/2    &    1 &      1.9955 &         537 &        2.46  &        19629 &      1.9920 &         542 &        2.63  &        19543\\
   3/2    &    2 &      2.0182 &         505 &        2.53  &        24624 &      2.0120 &         512 &        2.73  &        24363\\
   3/2    &    3 &      2.0208 &         482 &        2.69  &        34007 &      2.0031 &         493 &        2.96  &        33271\\
   3/2    &    4 &      1.9703 &         576 &        2.10  &        39254 &      1.9582 &         605 &        1.96  &        33646\\
   5/2    &    1 &      2.0154 &         508 &        2.53  &        25193 &      2.0089 &         515 &        2.77  &        24919\\
   ION    &      &      1.9450 &         604 &        1.98  &      48426 &      1.9437 &         609 &        1.86  &         49901    \\
\end{tabular}
\end{table}

\clearpage

\subsection{Open f-shell - potential energy curves}

\begin{figure}[hbtp]
\centering
\includegraphics[width=0.45\textwidth]{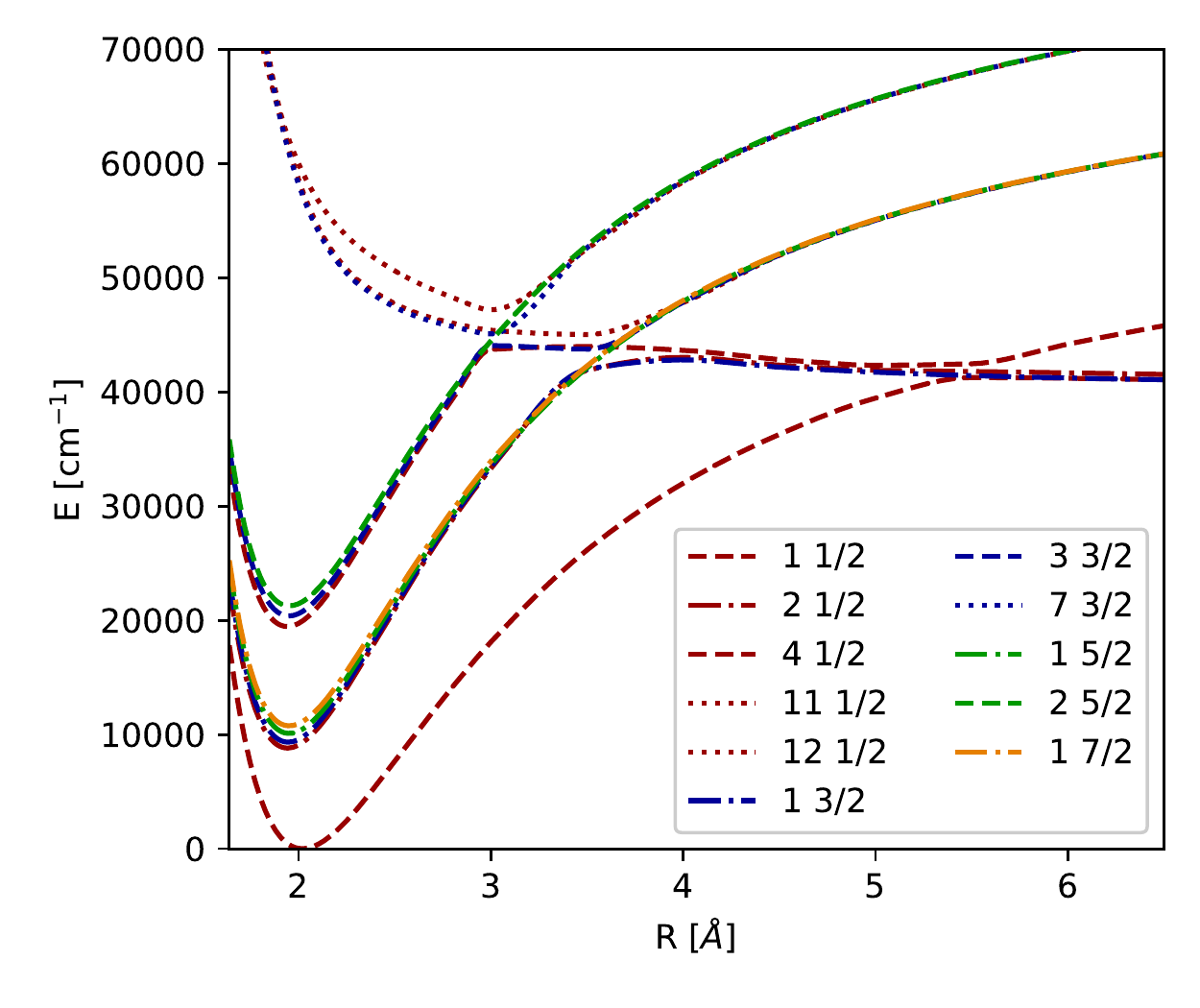}
\includegraphics[width=0.45\textwidth]{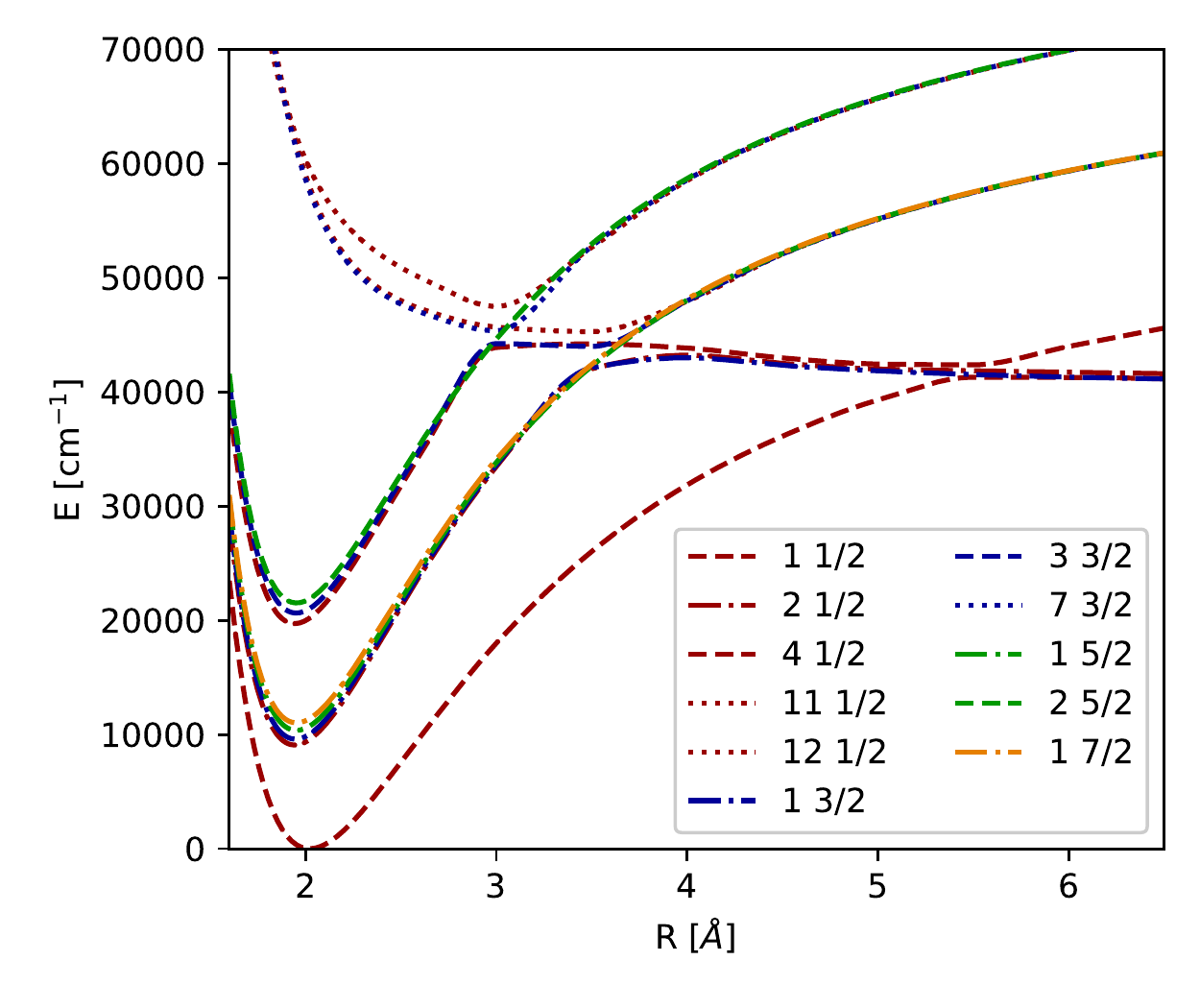}
\caption{IHFS-CCSD PECs for (1h,0p) sector starting from the anion reference for the quadruple zeta basis sets and a shift of 0.2 and 0.1 Hartree.}
\label{fig:PES_ANION4_12}
\end{figure}

\begin{figure}[hbtp]
\centering
\includegraphics[width=0.99\textwidth]{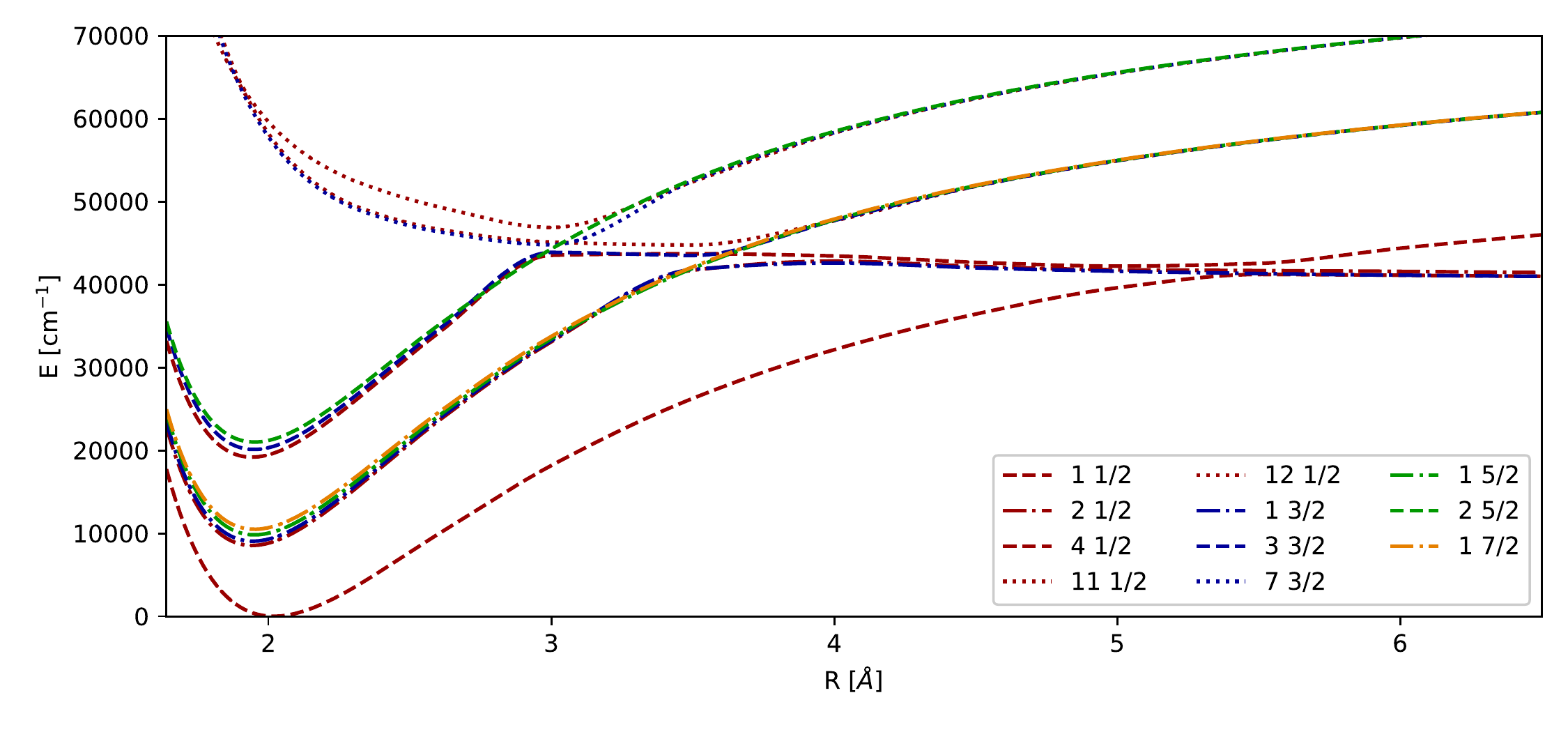}
\caption{XIHFS-CCSD PECs for (1h,0p) sector starting from the anion reference for the quadruple zeta basis sets, extrapolation using the results in figue~\ref{fig:PES_ANION4_12}}
\label{fig:PES_ANION4}
\end{figure}




\begin{table}[hbtp]
\caption{Spectroscopic constants for the different electronic states ($\Omega=1/2, 3/2, 5/2$) obtained by XIHFS-CCSD for the (1h,0p) sector using double and triple zeta basis sets. 
Vibrational constant ($\omega_e$), anharmonicty constant ($\omega_e\chi_e$), and  transition energy (T$_e$), are given in \wn{}, the equilibrium bond distance (r$_e$) in \AA{}.}
\label{tab:FSCC_f13_v2z_v3z}
\begin{tabular}{ r | r | r | r | r | r | r | r | r | r   }
 $\Omega$ &  state  &    \multicolumn{4}{c |}{ v2z } & \multicolumn{4}{c }{v3z} \\
 \hline
  & &  r$_e$    &     $\omega_e$   &   $\omega_e\chi_e$ & T$_e$  &  r$_e$    &     $\omega_e$   &   $\omega_e\chi_e$ & T$_e$\\
 \hline
   1/2    &    1 &      2.0239 &         497 &        2.06  &            0 &     2.0244 &         504 &        2.26  &            0 \\
   1/2    &    2 &      1.9481 &         619 &        3.17  &         4282 &      1.9463 &         598 &        2.81  &         7059 \\
   1/2    &    3 &      1.9410 &         605 &        2.73  &        14808 &      1.9444 &         594 &        2.72  &        17729 \\
   3/2    &    1 &      1.9440 &         602 &        2.73  &         4630 &      1.9479 &         591 &        2.73  &         7521 \\
   3/2    &    2 &      1.9478 &         599 &        2.75  &        15710 &      1.9518 &         588 &        2.72  &        18616 \\
   5/2    &    1 &      1.9488 &         597 &        2.74  &         5407 &      1.9530 &         586 &        2.72  &         8263 \\
   5/2    &    2 &      1.9491 &         592 &        2.73  &        16625 &      1.9533 &         581 &        2.71  &        19483 \\
   7/2    &    1 &      1.9487 &         591 &        2.73  &         6102 &      1.9529 &         581 &        2.71  &         8918 \\
   ION    &      &      2.1018 &         425 &        2.71  &         -8767    &      2.0935 &         430 &        2.66  &     -9418         \\
\end{tabular}
\end{table}

\begin{table}[hbtp]
\caption{Spectroscopic constants for the different electronic states ($\Omega=1/2, 3/2, 5/2$) obtained by XIHFS-CCSD for the (1h,0p) sector using quadruple zeta and extrapolation to the complete basis sets. 
Vibrational constant ($\omega_e$), anharmonicty constant ($\omega_e\chi_e$), and  transition energy (T$_e$), are given in \wn{}, the equilibrium bond distance (r$_e$) in \AA{}.}
\label{tab:FSCC_f13_v4z_cbs}
\begin{tabular}{ r | r | r | r | r | r | r | r | r | r   }
 $\Omega$ &  state  &    \multicolumn{4}{c |}{ v4z } & \multicolumn{4}{c }{CBS} \\
 \hline
  & &  r$_e$    &     $\omega_e$   &   $\omega_e\chi_e$ & T$_e$  &  r$_e$    &     $\omega_e$   &   $\omega_e\chi_e$ & T$_e$\\
 \hline
   1/2    &    1 &      2.0195 &         509 &        2.35  &            0 &      2.0159 &         513 &        2.42  &            0  \\
   1/2    &    2 &      1.9425 &         599 &        2.80  &         8543 &      1.9396 &         599 &        2.79  &         9627  \\
   1/2    &    3 &      1.9417 &         596 &        2.75  &        19195 &      1.9397 &         597 &        2.78  &        20267  \\
   3/2    &    1 &      1.9455 &         593 &        2.76  &         9057 &      1.9438 &         595 &        2.79  &        10180 \\
   3/2    &    2 &      1.9496 &         590 &        2.76  &        20121 &      1.9480 &         591 &        2.80  &        21222 \\
   5/2    &    1 &      1.9508 &         588 &        2.76  &         9825 &      1.9493 &         589 &        2.78  &        10968 \\
   5/2    &    2 &      1.9513 &         583 &        2.75  &        21011 &      1.9499 &         584 &        2.77  &        22127 \\
   7/2    &    1 &      1.9510 &         582 &        2.74  &        10494 &      1.9496 &         583 &        2.77  &        11645 \\
   ION    &      &      2.0851 &         435 &        2.74  &        -9579 &      2.0823 &         437 &        2.79  &         -8197     \\
\end{tabular}
\end{table}

\clearpage

\subsection{Combined potentials}

In this section we combine the potentials from the previous two sections for different basis sets sizes.

\begin{figure}[hbtp]
\centering
\includegraphics[width=0.99\textwidth]{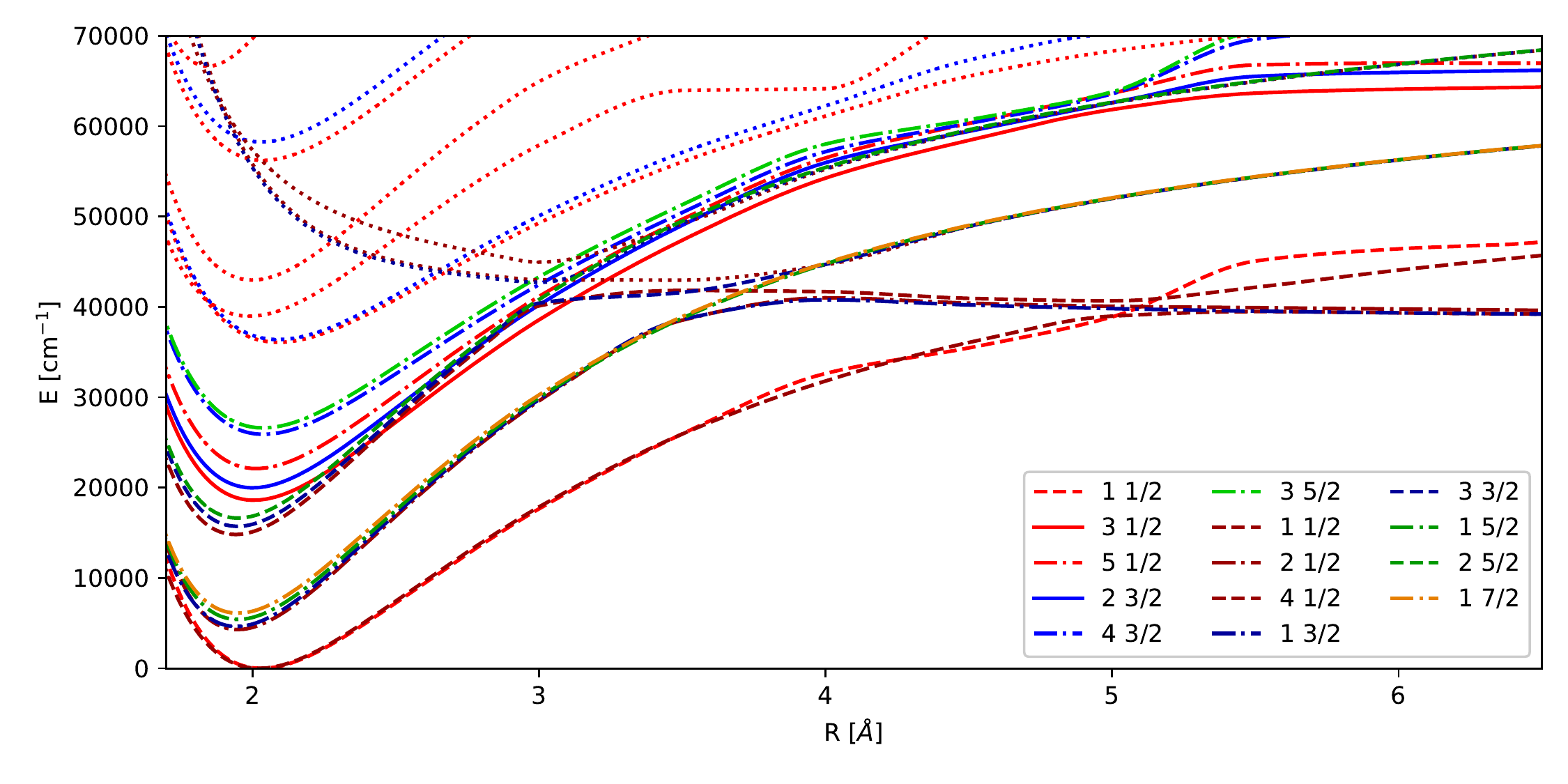}
\caption{IHFS-CCSD PECs generated by the use of two different sectors and reference systems for the double zeta basis sets.}
\label{fig:PES_FSCC_V2Z}
\end{figure}

\begin{figure}[hbtp]
\centering
\includegraphics[width=0.99\textwidth]{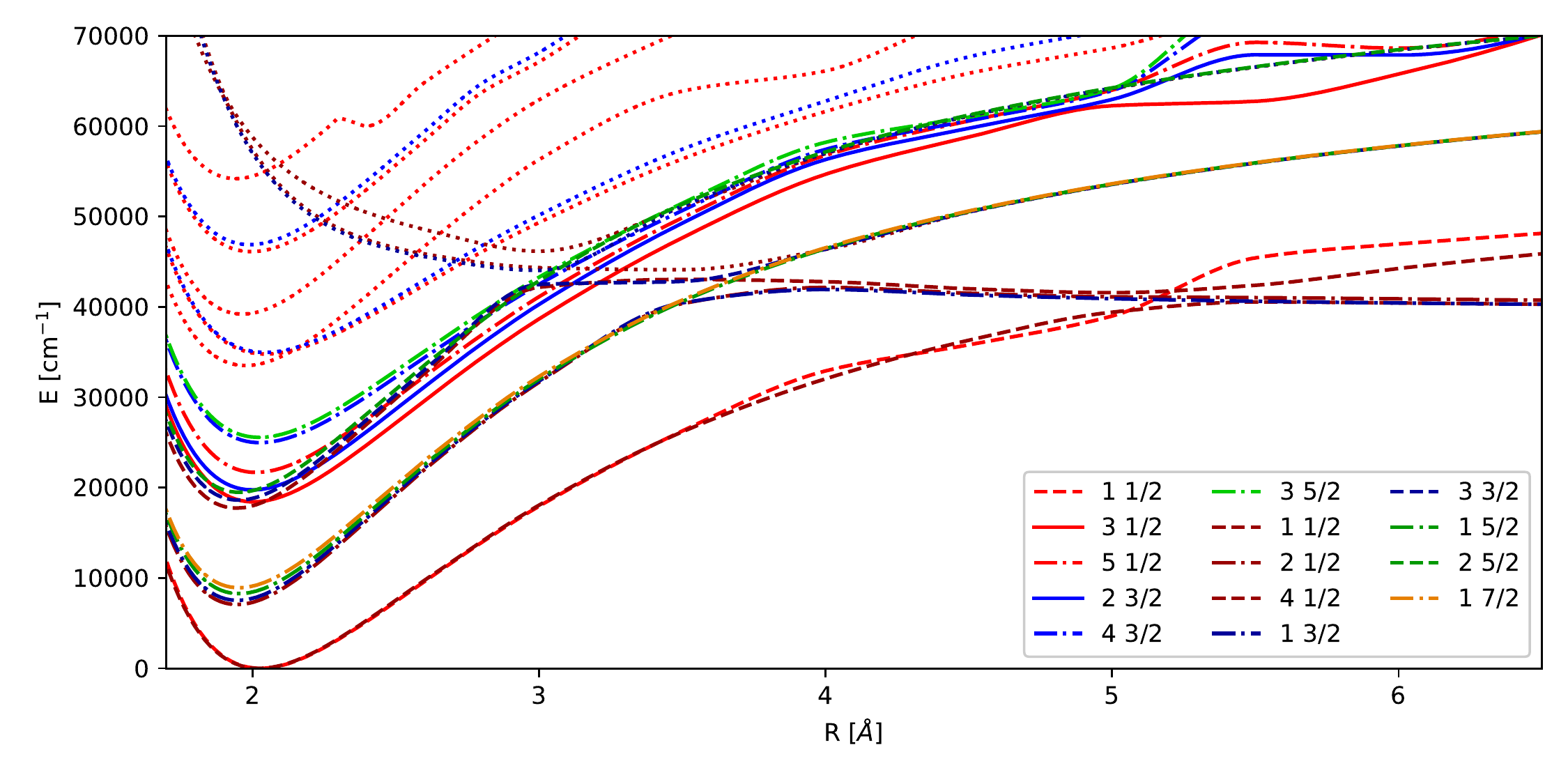}
\caption{IHFS-CCSD PECs generated by the use of two different sectors and reference systems for the triple zeta basis sets.}
\label{fig:PES_FSCC_V3Z}
\end{figure}

\begin{figure}[hbtp]
\centering
\includegraphics[width=0.99\textwidth]{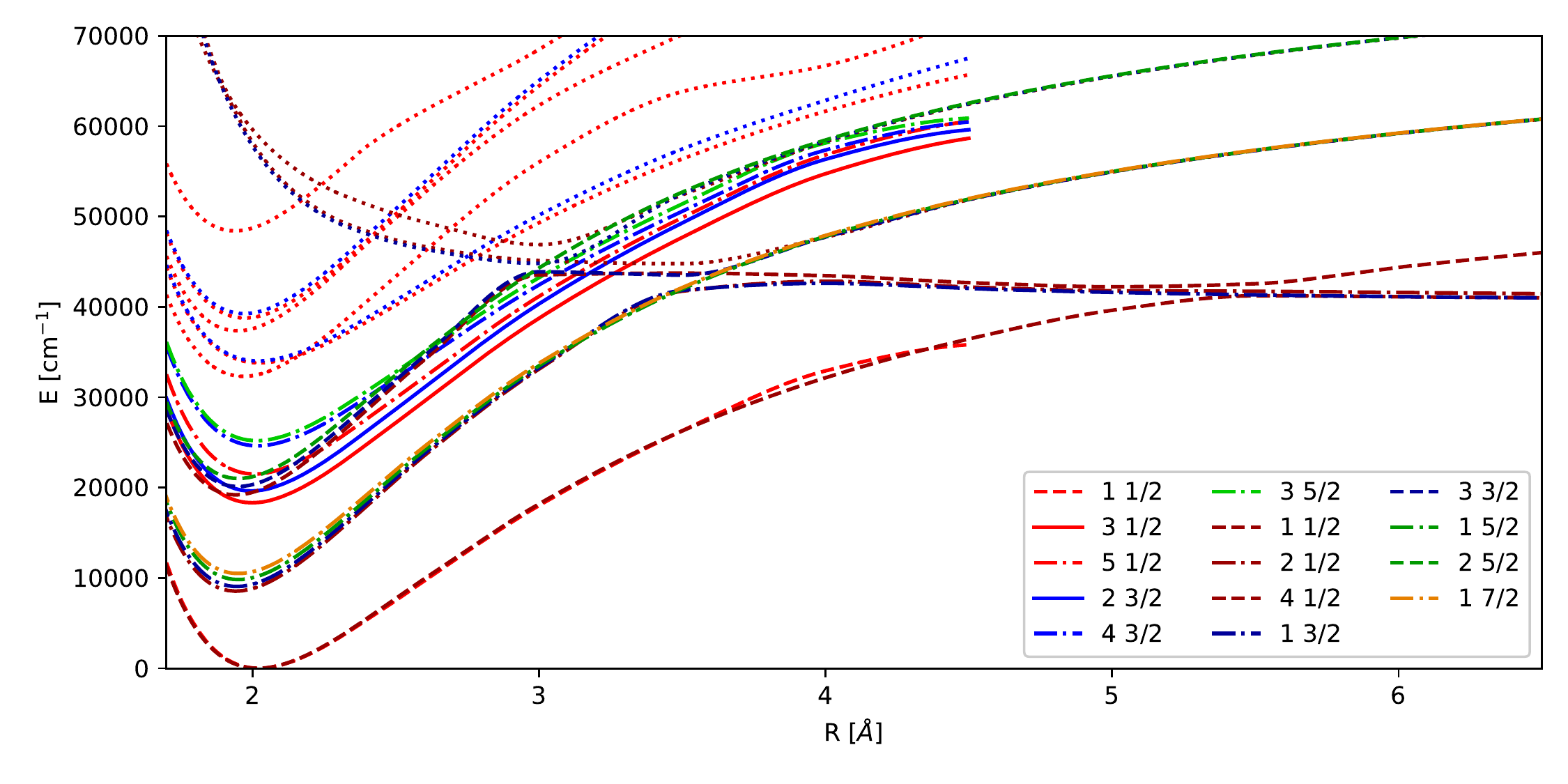}
\caption{IHFS-CCSD PECs generated by the use of two different sectors and reference systems for the quadruple zeta basis sets.}
\label{fig:PES_FSCC_V4Z}
\end{figure}

\begin{figure}[hbtp]
\centering
\includegraphics[width=0.99\textwidth]{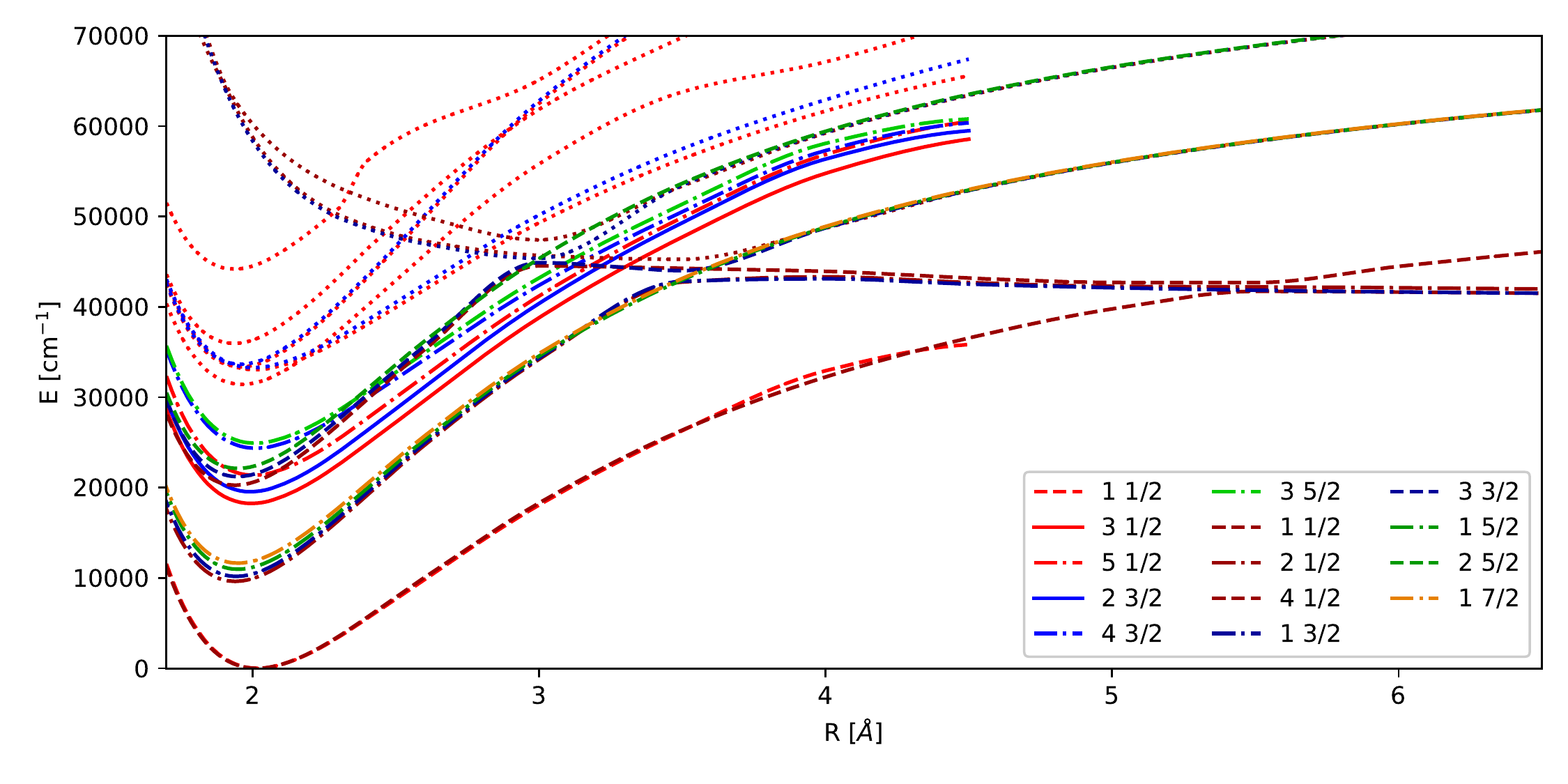}
\caption{IHFS-CCSD PECs generated by the use of two different sectors and reference systems for extrapolation to the basis sets limit.}
\label{fig:PES_FSCC_CBS}
\end{figure}

\clearpage

\subsection{Spectroscopic quantities after mixing}

We extracted the spectroscopic quantities for the adiabatized potentials shown in the main text and listed them here. 

\begin{table}[hbtp]
\small
\caption{Spectroscopic constants for the different electronic states ($\Omega=1/2, 3/2, 5/2$) obtained by IHFS-CCSD after basis set extrapolation, results from two sectors are combined using adiabatization with different coupling constants (C). 
Vibrational constant ($\omega_e$), anharmonicty constant ($\omega_e\chi_e$), and  transition energy (T$_e$), are given in \wn{}, the equilibrium bond distance (r$_e$) in \AA{}.}
\label{tab:FSCC_adiab_100}
\begin{tabular}{ r | r | r | r | r | r | r | r | r | r | r | r | r | r   }
 $\Omega$ &  state  &    \multicolumn{4}{c |}{ C = 10~\wn{} } & \multicolumn{4}{c |}{ C = 100~\wn{} } & \multicolumn{4}{c }{ C = 1000~\wn{} } \\
 \hline
  & &  r$_e$    &     $\omega_e$   &   $\omega_e\chi_e$ & T$_e$  &  r$_e$    &     $\omega_e$   &   $\omega_e\chi_e$ & T$_e$ 
  &     $\omega_e$   &   $\omega_e\chi_e$ & T$_e$ \\
 \hline
   1/2    &    1 &      2.0177 &         515 &        2.80  &            0 &      2.0177 &         515 &        2.80  &            0 &      2.0167 &         515 &        2.80  &            0 \\ 
   1/2    &    2 &      1.9397 &         599 &        2.77  &         9618 &      1.9397 &         599 &        2.77  &         9617 &      1.9418 &         596 &        2.76  &         9522   \\ 
   1/2    &    3 &      1.9953 &         539 &        2.61  &        18249 &      1.9952 &         538 &        2.60  &        18247 &      1.9843 &         533 &        2.35  &        18066 \\ 
   1/2    &    4 &      1.9337 &         609 &        8.95  &        20260 &      1.9349 &         603 &        8.61  &        20258 &      1.9654 &         561 &        3.65  &        20160  \\ 
   1/2    &    5 &      2.0035 &         586 &        0.25  &        21336 &      2.0019 &         586 &        0.38  &        21359 &      1.9860 &         577 &        2.10  &        22209   \\ 
   1/2    &    6 &      1.9644 &         581 &        1.83  &        31417 &      1.9644 &         581 &        1.83  &        31419 &      1.9645 &         580 &        1.96  &        31667  \\ 
   1/2    &    7 &      2.0149 &         470 &        0.14  &        33029 &      2.0149 &         470 &        0.14  &        33031 &      2.0132 &         472 &        0.24  &        33255   \\ 
   1/2    &    8 &      1.9647 &         633 &        2.62  &        33384 &      1.9647 &         633 &        2.62  &        33386 &      1.9656 &         628 &        2.61  &        33633    \\ 
   3/2    &    1 &      1.9438 &         594 &        2.76  &        10171 &      1.9438 &         594 &        2.76  &        10170 &      1.9447 &         593 &        2.76  &        10097   \\ 
   3/2    &    2 &      1.9921 &         542 &        2.61  &        19543 &      1.9919 &         542 &        2.60  &        19540 &      1.9820 &         541 &        2.50  &        19316  \\ 
   3/2    &    3 &      1.9480 &         591 &        2.77  &        21213 &      1.9482 &         591 &        2.80  &        21217 &      1.9621 &         580 &        3.13  &        21524   \\ 
   3/2    &    4 &      2.0121 &         512 &        2.72  &        24363 &      2.0121 &         512 &        2.69  &        24369 &      2.0059 &         528 &        2.33  &        24897 \\ 
   3/2    &    5 &      2.0023 &         473 &        1.95  &        33287 &      2.0023 &         473 &        1.95  &        33290 &      1.9998 &         476 &        2.25  &        33537    \\ 
   3/2    &    6 &      1.9608 &         632 &        3.13  &        33644 &      1.9608 &         632 &        3.13  &        33647 &      1.9630 &         623 &        2.67  &        33954   \\ 
   5/2    &    1 &      1.9493 &         589 &        2.75  &        10959 &      1.9493 &         589 &        2.75  &        10960 &      1.9496 &         588 &        2.76  &        11075  \\ 
   5/2    &    2 &      1.9500 &         584 &        2.75  &        22118 &      1.9500 &         583 &        2.76  &        22117 &      1.9542 &         572 &        3.01  &        22014  \\ 
   5/2    &    3 &      2.0090 &         515 &        2.75  &        24920 &      2.0089 &         516 &        2.72  &        24926 &      2.0024 &         532 &        2.38  &        25513    \\ 
\end{tabular}
\end{table}

\clearpage

\section{Equation-of-motion coupled cluster (EOM-CCSD)}
\label{sec:EOMCC}

\subsection{Yb cation transition energies}

\begin{table}[hbtp]
\caption{Transition energies for the Yb cation. Reference values have been obtained from the Nist database, the computed values were obtained for different basis sets with IP-EOM-CCSD or EA-EOM-CCSD coupled cluster.}
\label{tab:Yb_cation_EOM_long}
\begin{tabular}{ c | l | r | r | r | r | r | r | r  }
&& NIST\cite{NistDiatomic} & \multicolumn{3}{c |}{ EA }&\multicolumn{3}{c }{IP}
\\
\hline
state&conf&E&2z&3z&4z&2z&3z&4z
\\
\hline
$^2$S$_{1/2}$       &4f$^{14}$6s     &  0 & 0 & 0 & 0 & 0 & 0 & 0
\\
\hline
$^2$F$^\circ_{7/2}$ &4f$^{13}$6s$^2$  & 21419 & & & & 12054 & 13524 & 16092
\\
$^2$F$^\circ_{5/2}$  &4f$^{13}$6s$^2$ & 31568 & & & & 22629 & 24139 & 26655
\\
\hline
$^2$D$_{3/2}$       &4f$^{14}$5d     & 22961 & 24209 & 24073 & 24060 & 32956 & 33306 & 33712
\\
$^2$D$_{5/2}$       &4f$^{14}$5d     & 24333 & 25457 & 25351 & 25341 & 33589 & 33865 & 34263
\\
$^2$P$^\circ_{1/2}$  &4f$^{14}$6p    & 27062 & 27780 & 27539 & 27857 & 31553 & 31938 & 32112
\\
$^2$P$^\circ_{3/2}$  &4$^{14}$6p     & 30392 & 31246 & 30954 & 31323 & 34621 & 34975 & 35147
\\
\hline
$^3\left[3/2\right]^\circ_{5/2}$ &4f$^{13}$5d 6s  &26759&&&&&&
\\
$^3\left[3/2\right]^\circ_{3/2}$ &4f$^{13}$5d 6s  &28758&&&&&&
\\
$^3\left[11/2\right]^\circ_{9/2} $&4f$^{13}$5d 6s &30224&&&&&&
\\
$^3\left[11/2\right]^\circ_{11/2}$&4f$^{13}$5d 6s &30563&&&&&&
\\
\end{tabular}
\end{table}

\clearpage

\subsection{Closed f-shell - potential energy curves}

\begin{figure}[hbtp]
\centering
\includegraphics[width=0.99\textwidth]{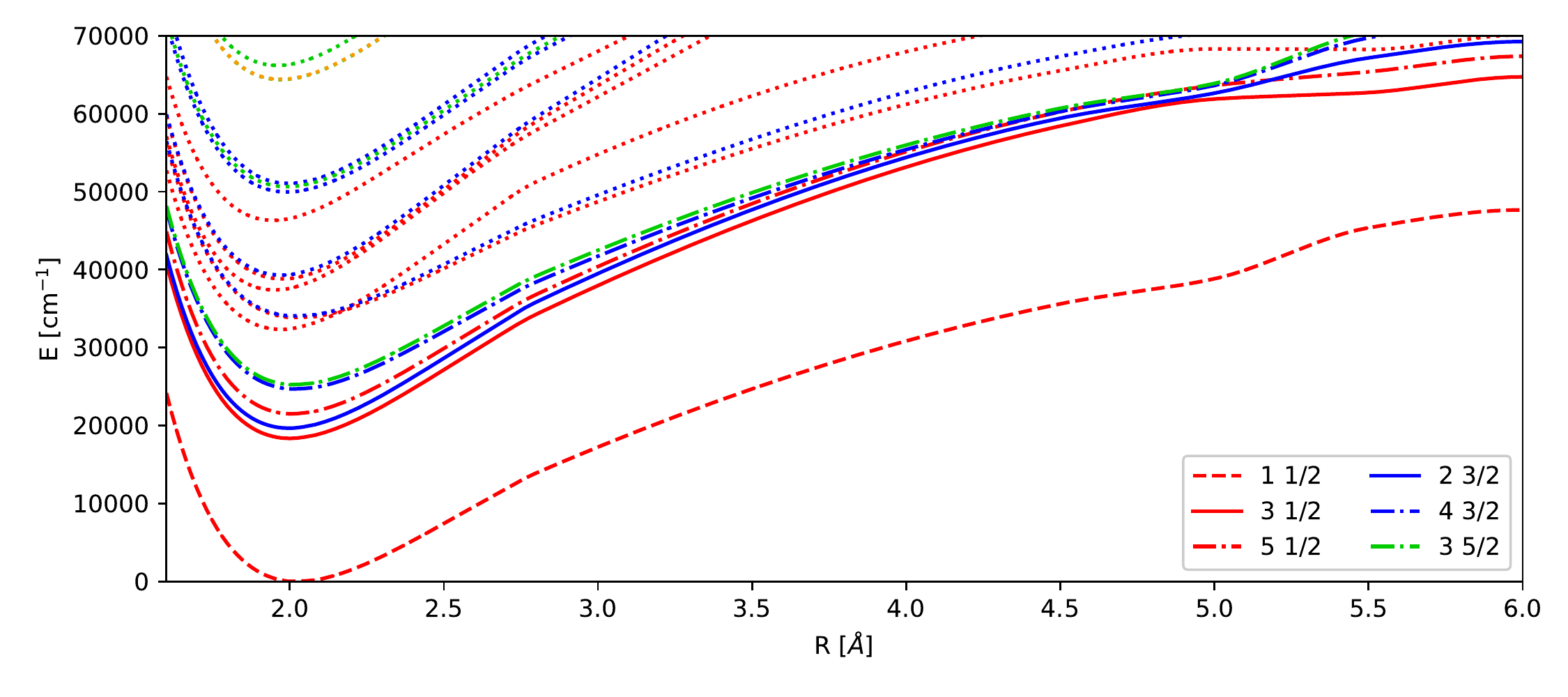}
\caption{EOM-CCSD PECs for states from the (0h,1p) sector with a cation reference for the quadruple zeta basis sets.}
\label{fig:PES_EA_EOM_4z}
\end{figure}

\begin{table}[hbtp]
\small
\caption{Spectroscopic constants for the different electronic states ($\Omega=1/2, 3/2, 5/2$) obtained by EOM-CCSD for the (0h,1p) sector using different basis set sizes. 
Vibrational constant ($\omega_e$), anharmonicty constant ($\omega_e\chi_e$), and  transition energy (T$_e$), are given in \wn{}, the equilibrium bond distance (r$_e$) in \AA{}.}
\label{tab:EA_EOMEA}
\begin{tabular}{ r | r | r | r | r | r | r | r | r | r | r | r | r  | r }
 $\Omega$ &  state  &    \multicolumn{4}{c |}{ v3z } & \multicolumn{4}{c |}{v4z} & \multicolumn{4}{c }{CBS} \\
 \hline
  & &  
  r$_e$    &     $\omega_e$   &   $\omega_e\chi_e$ & T$_e$  &  
  r$_e$    &     $\omega_e$   &   $\omega_e\chi_e$ & T$_e$  &
  r$_e$    &     $\omega_e$   &   $\omega_e\chi_e$ & T$_e$
  \\
 \hline
   1/2    &    1 &      2.0287 &         503 &        2.30  &            0 &      2.0248 &         508 &        2.57  &            0 &       2.0230 &         511 &        2.80  &            0\\ 
   1/2    &    2 &      2.0040 &         525 &        2.22  &        18407 &      2.0012 &         532 &        2.52  &        18384 &      2.0004 &         536 &        2.72  &        18373\\ 
   1/2    &    3 &      2.0128 &         521 &        2.27  &        21631 &      2.0093 &         527 &        2.58  &        21524 &      2.0079 &         532 &        2.78  &        21448 \\ 
   1/2    &    4 &      1.9718 &         513 &        8.14  &        33599 &      1.9646 &         567 &        4.96  &        32407 &      1.9726 &         593 &        0.60  &        31497 \\ 
   1/2    &    5 &      2.0408 &         552 &        0.10  &        34419 &      2.0334 &         510 &        0.28  &        33772 &      2.0294 &         479 &        0.29  &        33299 \\ 
   1/2    &    6 &      1.9652 &         580 &        2.32  &        39226 &      1.9499 &         596 &        2.33  &        37425 &      1.9405 &         607 &        2.20  &        36100 \\ 
   1/2    &    7 &      1.9899 &         550 &        1.95  &        43303 &      1.9753 &         575 &        2.10  &        38850 &      1.9667 &         593 &        2.19  &        35591 \\ 
   1/2    &    8 &      1.9482 &         577 &        2.53  &        51348 &      1.9460 &         582 &        2.56  &        46354 &      1.9456 &         584 &        2.58  &        42725 \\ 
   3/2    &    1 &      2.0006 &         529 &        2.23  &        19738 &      1.9979 &         535 &        2.52  &        19696 &      1.9971 &         540 &        2.72  &        19672 \\ 
   3/2    &    2 &      2.0273 &         496 &        2.25  &        24996 &      2.0210 &         503 &        2.56  &        24691 &      2.0177 &         509 &        2.78  &        24468 \\ 
   3/2    &    3 &      2.0481 &         470 &        2.30  &        34695 &      2.0237 &         480 &        2.71  &        34062 &      2.0068 &         489 &        3.03  &        33563 \\ 
   3/2    &    4 &      1.9862 &         549 &        1.97  &        44290 &      1.9724 &         575 &        2.13  &        39335 &      1.9646 &         593 &        2.14  &        35711 \\ 
   3/2    &    5 &      2.0073 &         525 &        2.10  &        60165 &      1.9900 &         550 &        2.33  &        50005 &      1.9797 &         569 &        2.45  &        42573 \\ 
   3/2    &    6 &      2.0096 &         539 &        2.15  &        62569 &      1.9961 &         563 &        2.39  &        51099 &      1.9881 &         581 &        2.54  &        42720 \\ 
   5/2    &    1 &      2.0248 &         498 &        2.25  &        25584 &      2.0181 &         507 &        2.56  &        25259 &      2.0146 &         513 &        2.80  &        25023 \\ 
   5/2    &    2 &      2.0085 &         523 &        2.10  &        61177 &      1.9905 &         548 &        2.34  &        50686 &      1.9796 &         567 &        2.48  &        43011  \\ 
   5/2    &    3 &      1.9834 &         549 &        2.24  &        75514 &      1.9711 &         569 &        2.39  &        64418 &      1.9642 &         581 &        2.34  &        56321 \\ 
   5/2    &    4 &      1.9526 &         583 &        2.56  &        77193 &      1.9601 &         582 &        2.24  &        66234 &      1.9659 &         583 &        2.18  &        58242 \\ 
   7/2    &    1 &      1.9834 &         549 &        2.24  &        75566 &      1.9712 &         569 &        2.39  &        64446 &      1.9643 &         581 &        2.34  &        56331 \\ 
   7/2    &    2 &      1.9499 &         582 &        2.45  &        77694 &      1.9463 &         590 &        2.48  &        80882 &      1.9006 &         708 &       45.10  &        83343 \\ 
   ION    &    0 &      1.9485 &         596 &        2.16  &     48387    &      1.9471 &         604 &        2.02  &       48471  &      1.9471 &         609 &        1.91  &     48578    \\ 
\end{tabular}
\end{table}

\clearpage

\subsection{Open f-shell - potential energy curves}

\begin{figure}[hbtp]
\centering
\includegraphics[width=0.99\textwidth]{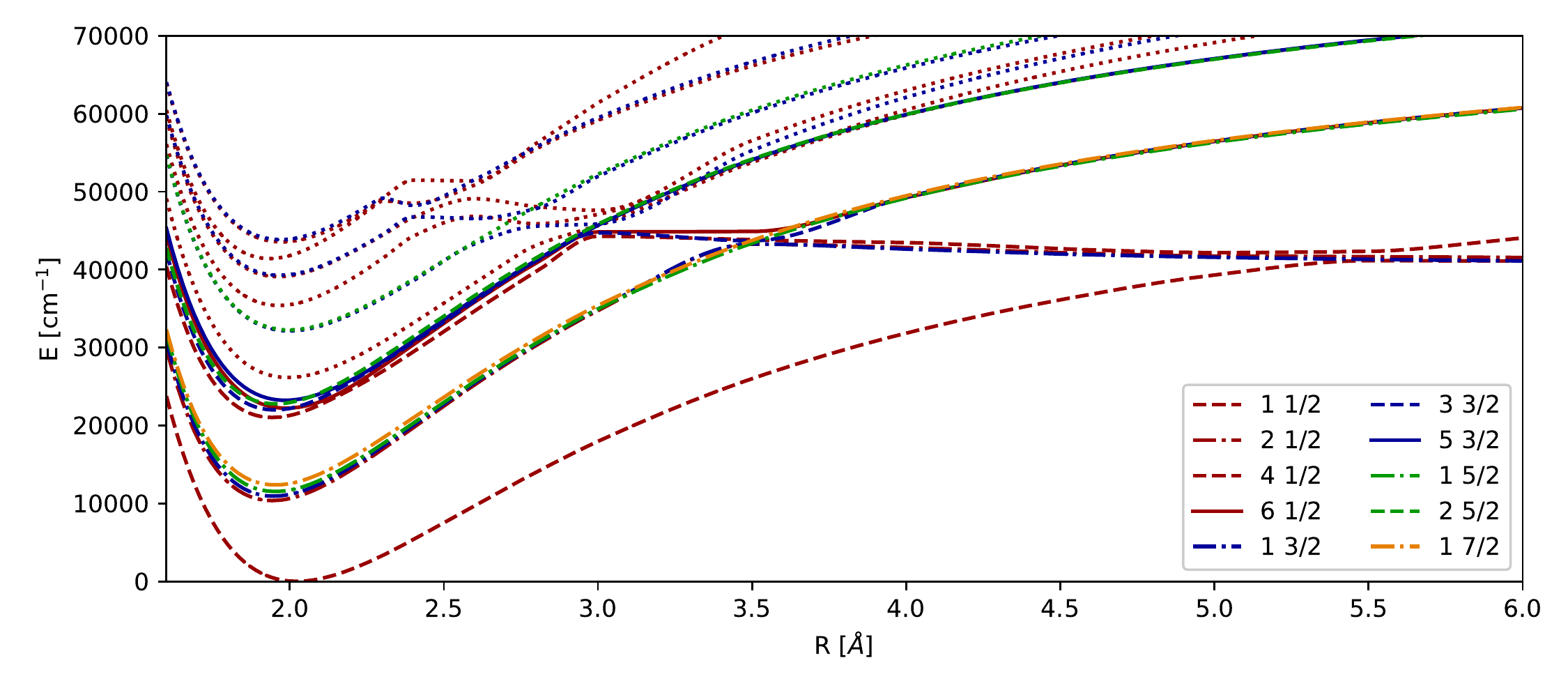}
\caption{EOM-CCSD PECs for states from the (1h, 0p) sector with a anion reference for the quadruple zeta basis sets. }
\label{fig:PES_IP_EOM_4z}
\end{figure}

\begin{table}[hbtp]
\small
\caption{Spectroscopic constants for the different electronic states ($\Omega=1/2, 3/2, 5/2$) obtained by EOM-CCSD for the (1h,0p) sector using different basis set sizes. 
Vibrational constant ($\omega_e$), anharmonicty constant ($\omega_e\chi_e$), and  transition energy (T$_e$), are given in \wn{}, the equilibrium bond distance (r$_e$) in \AA{}.}
\label{tab:IP_EOM_3z}
\begin{tabular}{ r | r | r | r | r | r | r | r | r | r | r | r | r  | r }
 $\Omega$ &  state  &    \multicolumn{4}{c |}{ v3z } & \multicolumn{4}{c |}{v4z} & \multicolumn{4}{c }{CBS} \\
 \hline
  & &  
  r$_e$    &     $\omega_e$   &   $\omega_e\chi_e$ & T$_e$  &  
  r$_e$    &     $\omega_e$   &   $\omega_e\chi_e$ & T$_e$  &
  r$_e$    &     $\omega_e$   &   $\omega_e\chi_e$ & T$_e$
  \\
 \hline
   1/2    &    1 &      2.0247 &         503 &        2.27  &            0 &     2.0233 &         507 &        2.41  &            0 &       2.0250 &         508 &        2.53  &            0 \\ 
   1/2    &    2 &      1.9462 &         597 &        2.78  &         7380 &      1.9431 &         595 &        2.67  &        10361 &     1.9432 &         591 &        2.59  &        12568 \\ 
   1/2    &    3 &      1.9445 &         593 &        2.70  &        18046 &      1.9428 &         588 &        3.23  &        21029 &      1.9432 &         582 &        4.06  &        23241 \\ 
   1/2    &    4 &      1.9922 &         548 &        2.31  &        22306 &      1.9889 &         564 &        1.51  &        22205 &      1.9886 &         573 &        1.18  &        22147 \\ 
   1/2    &    5 &      1.9991 &         543 &        2.36  &        26384 &      1.9941 &         549 &        2.39  &        26151 &      1.9931 &         551 &        2.40  &        25991 \\ 
   1/2    &    6 &      1.9735 &         571 &        2.21  &        36936 &      1.9645 &         584 &        2.21  &        35391 &      1.9606 &         591 &        2.18  &        34280 \\ 
   1/2    &    7 &      1.9973 &         524 &        2.77  &        40983 &      1.9634 &         567 &        2.82  &        39091 &      1.9441 &         595 &        2.76  &        37639 \\ 
   1/2    &    8 &      1.9506 &         609 &        2.41  &        43943 &      1.9346 &         620 &        2.33  &        41412 &      1.9252 &         625 &        2.24  &        39575 \\ 
   3/2    &    1 &      1.9482 &         590 &        2.71  &         7848 &      1.9475 &         590 &        2.66  &        10931 &      1.9494 &         588 &        2.61  &        13211 \\ 
   3/2    &    2 &      1.9521 &         587 &        2.72  &        18941 &      1.9516 &         586 &        2.68  &        21994 &      1.9537 &         584 &        2.64  &        24251 \\ 
   3/2    &    3 &      1.9889 &         551 &        2.31  &        23370 &      1.9855 &         556 &        2.30  &        23227 &      1.9857 &         557 &        2.28  &        23137 \\ 
   3/2    &    4 &      2.0100 &         522 &        2.37  &        32680 &      1.9987 &         533 &        2.47  &        32132 &      1.9936 &         539 &        2.45  &        31734\\ 
   3/2    &    5 &      2.0017 &         523 &        2.70  &        41181 &      1.9659 &         564 &        2.79  &        39262 &      1.9451 &         591 &        2.75  &        37778 \\
   5/2    &    1 &      1.9532 &         585 &        2.72  &         8588 &      1.9530 &         584 &        2.66  &        11529 &      1.9553 &         582 &        2.61  &        13703 \\ 
   5/2    &    2 &      1.9536 &         580 &        2.71  &        19807 &      1.9535 &         579 &        2.65  &        22768 &      1.9559 &         577 &        2.62  &        24957 \\ 
   5/2    &    3 &      2.0085 &         524 &        2.39  &        32839 &      1.9974 &         534 &        2.46  &        32257 &      1.9923 &         540 &        2.45  &        31834 \\ 
   7/2    &    1 &      1.9532 &         580 &        2.71  &         9245 &      1.9532 &         579 &        2.64  &        12373 &      1.9556 &         577 &        2.62  &        14685 \\ 
   ION    &    0 &      2.0941 &         429 &        2.64  &      -9522   &      2.0889 &         431 &        2.99  &      - 9713       &      2.0883 &         432 &        3.25  &      -9876   \\ 
\end{tabular}
\end{table}

\clearpage

\subsection{Combined potentials}

\begin{figure}[hbtp]
\centering
\includegraphics[width=0.99\textwidth]{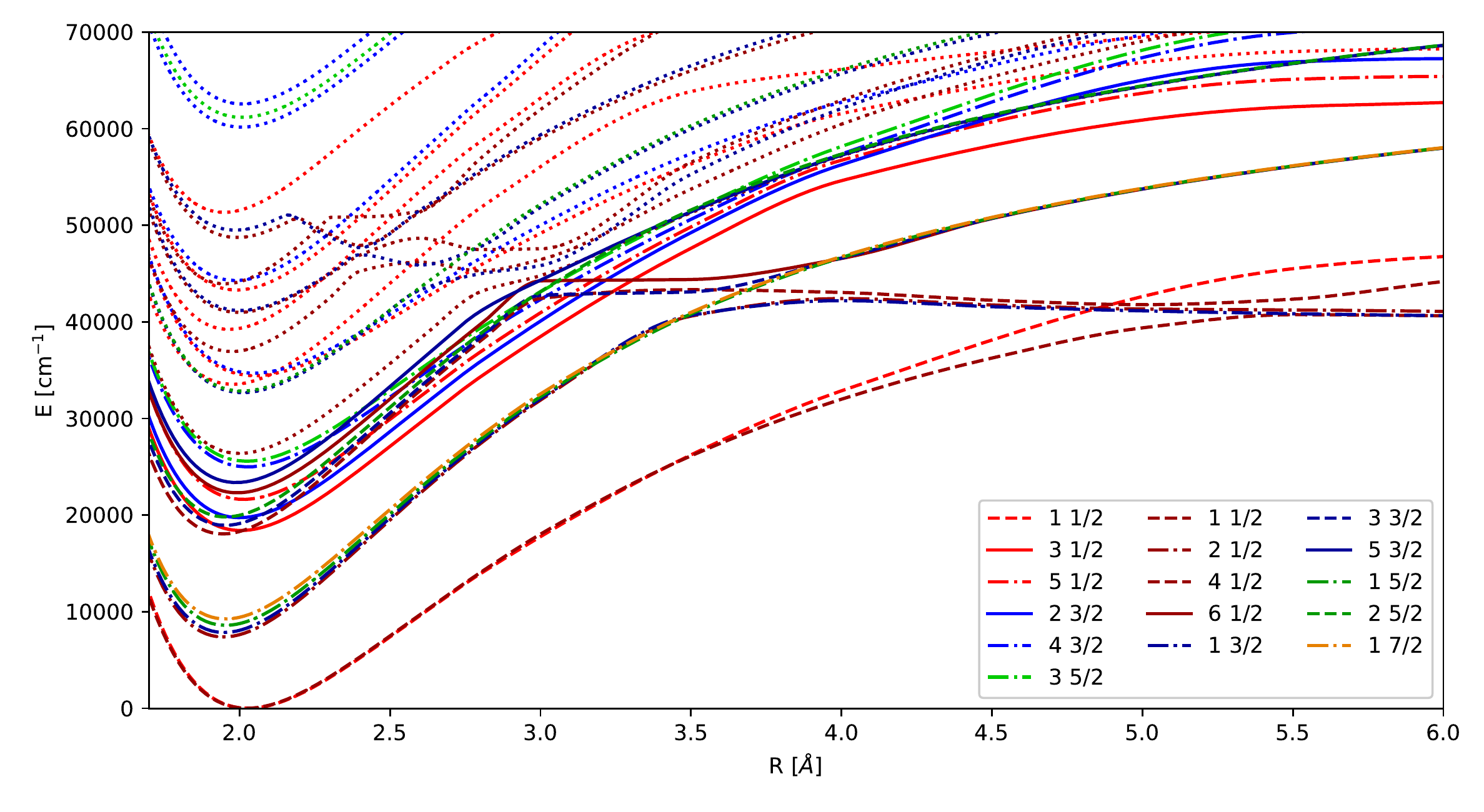}
\caption{EOM-CCSD PECs generated by the use of two different sectors and reference systems for the triple zeta basis sets. }
\label{fig:PES_EOM_v3z}
\end{figure}

\begin{figure}[hbtp]
\centering
\includegraphics[width=0.99\textwidth]{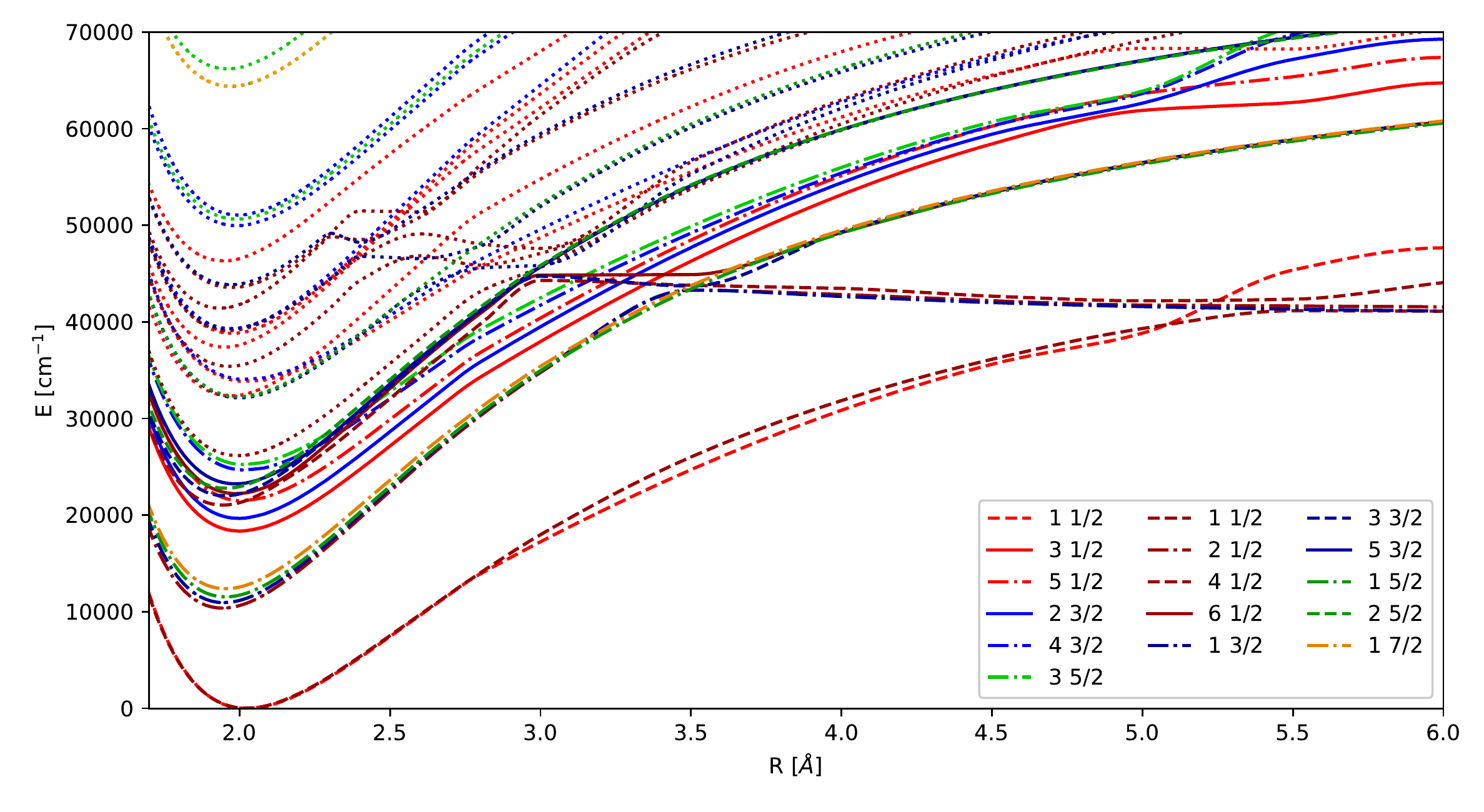}
\caption{EOM-CCSD PECs generated by the use of two different sectors and reference systems for the quadruple zeta basis sets. }
\label{fig:PES_EOM_v4z}
\end{figure}

\begin{figure}[hbtp]
\centering
\includegraphics[width=0.99\textwidth]{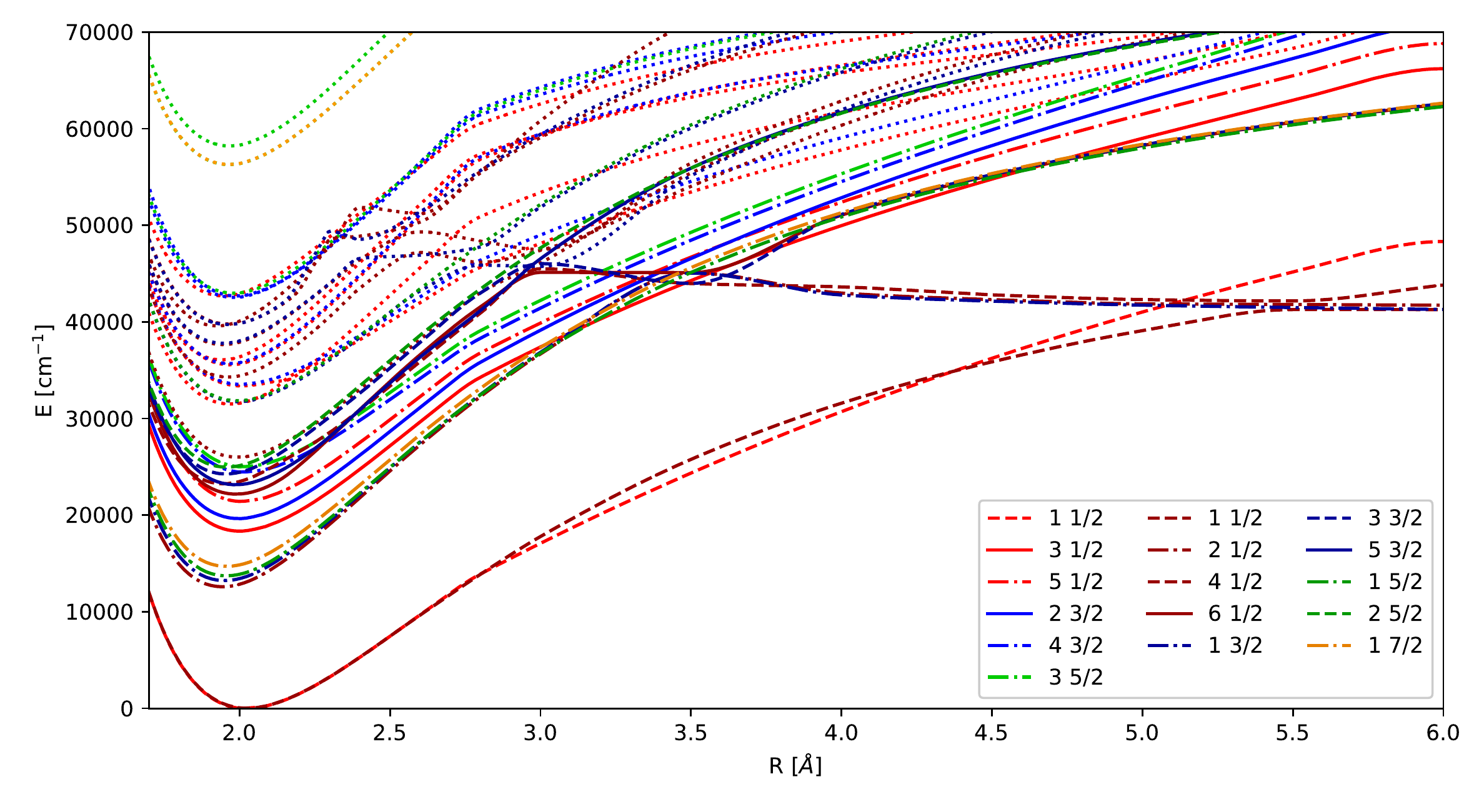}
\caption{EOM-CCSD PECs generated by the use of two different sectors and reference systems for extrapolation to the basis set limit.}
\label{fig:PES_EOM_cbs}
\end{figure}

\clearpage

\section{Frank-Condon factors (FCFs) for different methods}
\label{sec:Spectra}

\begin{figure}[hbtp]
\centering
KRCI \\
\includegraphics[width=\textwidth]{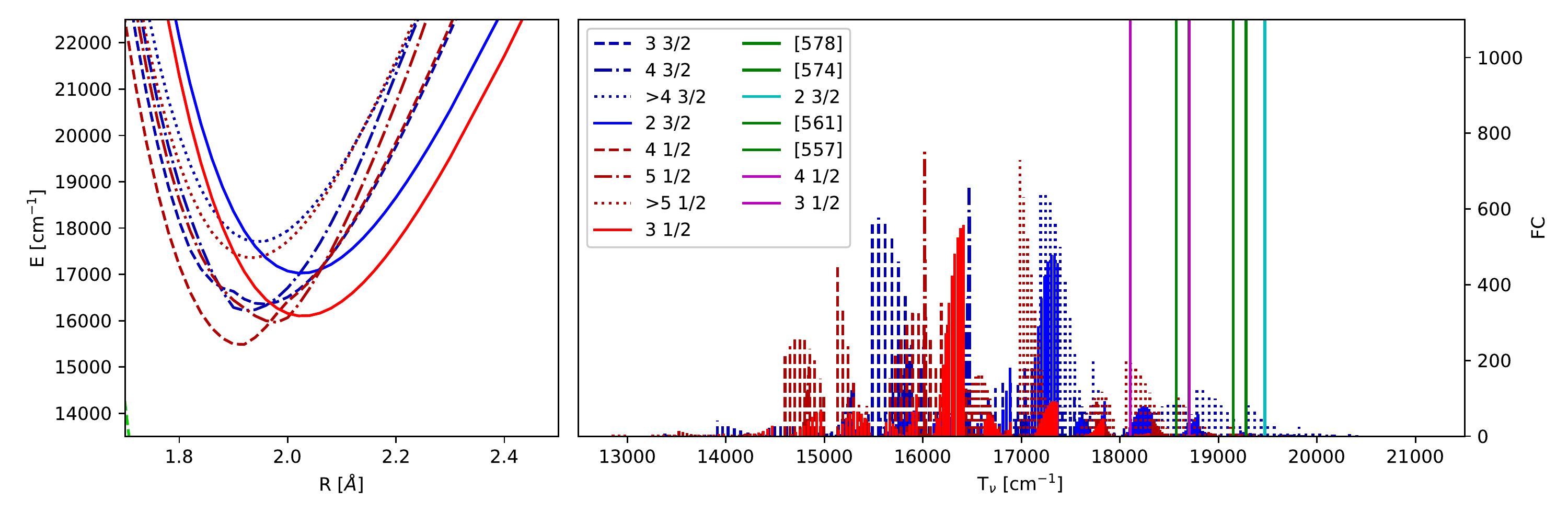} 
\\
EOM-CCSD \\
\includegraphics[width=\textwidth]{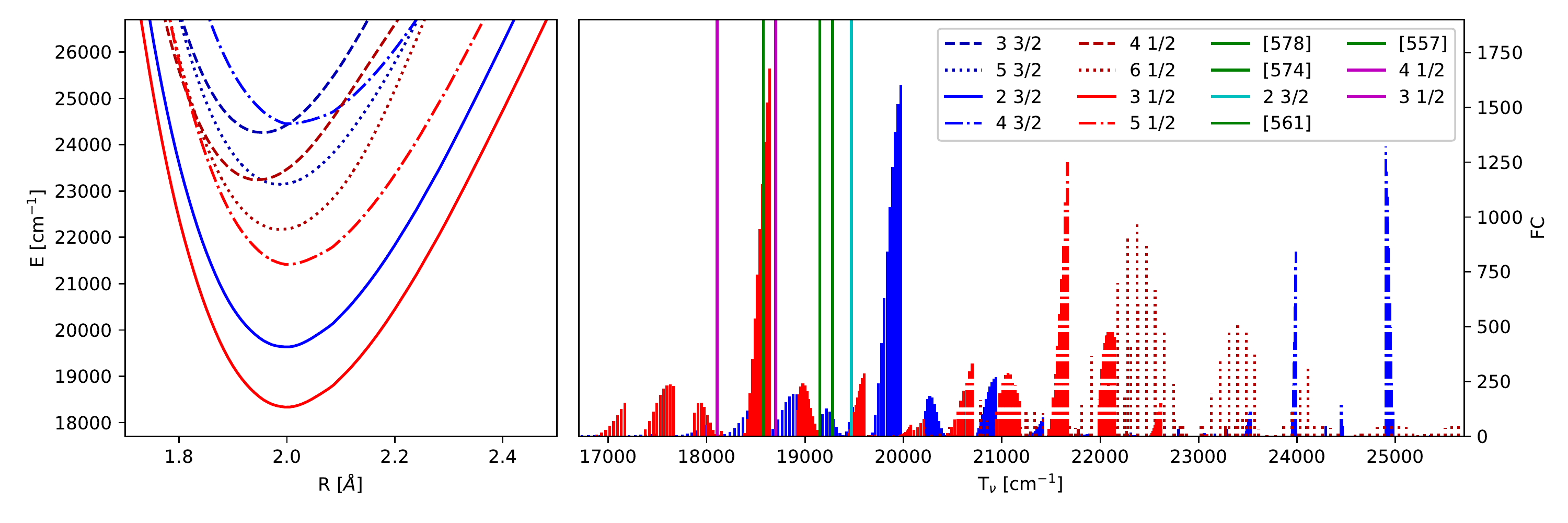}
\\
IHFS-CCSD \\
\includegraphics[width=\textwidth]{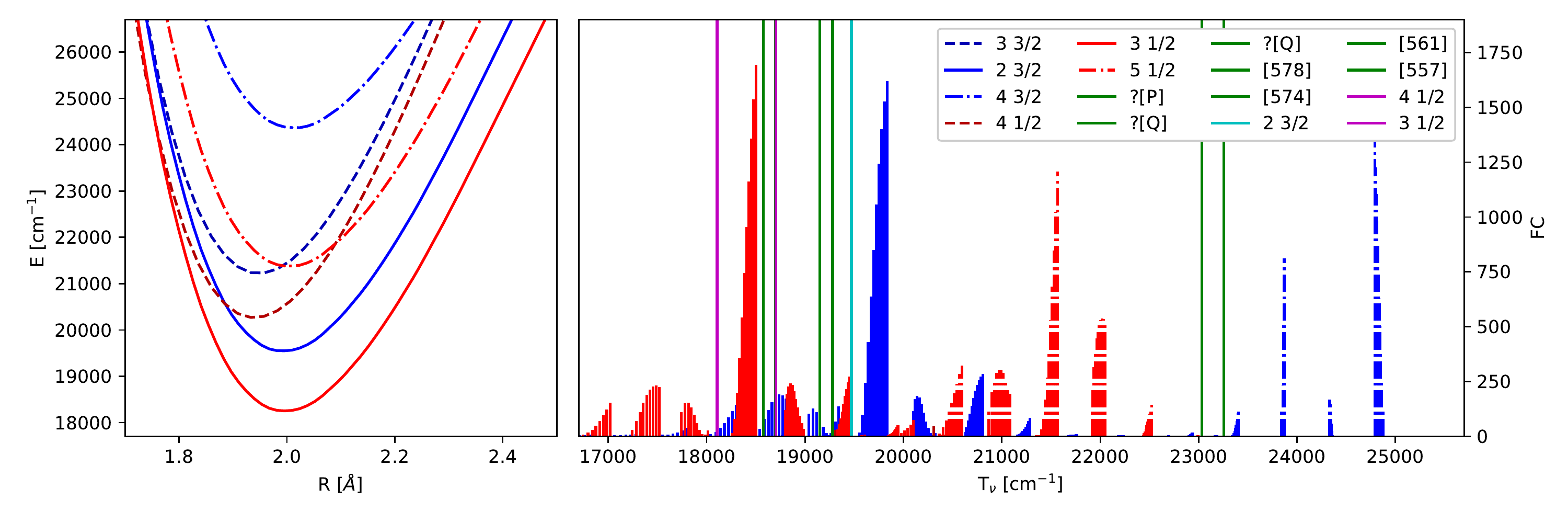}
\caption{Frank-Condon factors obtained by using the potential energy curves on the left, which have been obtained by different ab-intio methods.
The lowest 10 vibrational levels of the ground state were used as well as the lowest 60 vibrational levels of the excited state. The experimental values have been added as straight lines.\cite{Dunfield1995,Lim2017,Smallman2014}. }
\label{fig:compare_methods}
\end{figure}

Frank-Condon factors are shown in figure~\ref{fig:compare_methods} for MRCI, EOM-CCSD and IHFS-CCSD. From it, we observe a marked difference between the configuration interaction and coupled cluster results: first, the equilibrium distance and curvature are quite different, something that, in turn, change the overlap and resulting Frank-Condon factors as well as the progression of the vibrational levels. 

 Additionally, there is more uncertainty with regards to the position of the potential energy curves, as configuration interaction is not size consistent and we observe transition in the range from 14000 to 19000~\wn{} for these states. The states are more dense is this case, which is expected as states of the  Yb($4f^{13}$[$F_{7/2}^\circ$]$5d\sigma_{6s}$)F configuration are included, which are missing in the coupled cluster descriptions. 

The EOM-CCSD and IHFS-CCSD Frank-Condon factors are quite similar to each other, with the exception of the additional $6_{1/2}$ state in EOM-CCSD, probably stemming from the contributions by (2h,1p)/(1h,2p) to the final states. The results for the $3_{1/2}$ and $2_{3/2}$ state are close to the experimentally observed positions, only about 500~\wn{} too high, and show the same splitting. The transition of the $5_{1/2}$ states are also similar in both coupled cluster treatments.

\clearpage

\bibliography{ybf}